\documentstyle[12pt]{article}
\renewcommand{\baselinestretch}{1.2}
\begin{document}
\begin{flushright}
September, 1998
\end{flushright}
\vspace{0mm}
\begin{center}
\large{Chiral Symmetry and the Nucleon Spin Structure Functions}
\end{center}
\vspace{0mm}
\begin{center}
M.~Wakamatsu\footnote{Email \ : \ wakamatu@miho.rcnp.osaka-u.ac.jp}
and T.~Kubota\footnote{Email \ : \ kubota@kern.phys.sci.osaka-u.ac.jp}
\end{center}
\vspace{-2mm}
\begin{center}
Department of Physics, Faculty of Science, \\
Osaka University, \\
Toyonaka, Osaka 560, JAPAN
\end{center}
\vspace{2mm}
\begin{flushleft}
\end{flushleft}
\vspace{0mm}

\begin{center}
\small{{\bf Abstract}}
\end{center}
\vspace{-1mm}
\begin{center}
\begin{minipage}{15.5cm}
\renewcommand{\baselinestretch}{1.0}
\small
\ \ We carry out a systematic investigation of twist-two spin
dependent structure functions of the nucleon within the framework
of the chiral quark soliton model (CQSM) by paying special
attention to the role of chiral symmetry of QCD. The importance
of chiral symmetry is illustrated through the good reproduction of
the recent SLAC data for the neutron spin structure function
$g_1^n (x,Q^2)$. We also observe substantial difference
between the predictions of the longitudinally polarized distribution
functions and those of the transversity distribution functions.
That the chiral symmetry may be responsible for this difference
is seen in the isospin dependence of the corresponding first
moments, i.e. the axial and tensor charges. The CQSM predicts
$g_A^{(0)} / g_A^{(3)} \simeq 0.25$ for the ratio of the
isoscalar to isovector axial charges, while $g_T^{(0)} /
g_T^{(3)} \simeq 0.46$ for the ratio of the isoscalar to
isovector tensor charges, which should be compared with the
prediction $g_A^{(0)} / g_A^{(3)} = g_T^{(0)} / g_T^{(3)} 
= 3 / 5$ of the constituent quark model or of the naive MIT bag model
without proper account of chiral symmetry. Another prominent
prediction of the CQSM is the opposite polarization of the
$\bar{u}$ and $\bar{d}$ antiquarks, thereby indicating the SU(2)
asymmetric sea quark (spin) polarization in the nucleon.
\normalsize
\end{minipage}
\end{center}
\renewcommand{\baselinestretch}{2.00}

\section{Introduction}

\ \ \ \ Undoubtedly, the so-called ``nucleon spin crisis'' caused by
the EMC measurement in 1988 is one of the most exciting topics
in the field of hadron physics [1]. The recent renaissance of nucleon
structure function physics is greatly owing to this
epoch-making finding. Naturally, the physics of nucleon
structure functions has two different aspects. One is a
perturbative aspect, while the other is a nonperturbative aspect.
Because of the asymptotic freedom of QCD, the $Q^2$-evolution of
quark distribution functions can be controlled by the perturbative
QCD at least for large enough $Q^2$ [2].
However, the perturbative QCD is entirely powerless for predicting
distribution functions themselves. Here we need to solve
nonperturbative QCD in some way. Unfortunately, we have no reliable
analytical method for handling this aspect of QCD.
For the present moment, we are then left with two tentative choices.
One is to rely upon lattice QCD, while the other is to use effective
models of QCD. If one takes the first choice, one must first
evaluate infinite towers of moments of distribution functions,
since the direct calculation of distribution functions does not
match this numerical simulation method [3]. Here we take the second
choice, which allows us a direct calculation of quark distribution
functions. Still, there are quite a lot of effective model
of baryons. We advocate that the chiral quark soliton model
(CQSM) is a unique model of baryons which has several appealing
features not possessed by other models of baryons,
especially when applied to the physics
of quark distribution functions. First of all, it is an
effective model of baryons maximally incorporating spontaneous chiral
symmetry breaking of QCD vacuum [4-6]. The nucleon in this model is
a composite of three valence quarks and infinitely many Dirac sea
quarks moving in a slowly rotating M.F. of hedgehog shape.
As a natural consequence, it automatically simulates cloud of pions
surrounding the core of three valence quarks. Nevertheless,
since everything is described in terms of effective
quark fields only, we need not worry about a double
counting of quark and pion degrees of freedom.
(We recall that this kind of double counting occurs, for instance,
in models of hadrons based on the linear-sigma-quark-model type
lagrangian [7,8].) This also means that we do not need to use such an
ambiguous procedure as convoluting the pion structure functions
with {\it pion probability function} (or more precisely a
light-cone momentum distribution of the pion) inside the
nucleon [9-11].

Several group have already attempted to calculate nucleon
structure functions within the CQSM or the Nambu-Jona-Lasinio
(NJL) soliton model. For instance, Weigel et al. investigated
the polarized as well as unpolarized structure functions of
the nucleon under the so-called 
``valence quark approximation'' [12].
This is not an extremely bad approximation, but it is known to
have several unpleasant features. Probably, most serious would
be the violation of positivity condition for the
unpolarized antiquark (or sea quark) distribution functions.
Although such an apparent disaster does not happen for the
spin dependent quark distribution functions, a lesson learned
from the above observation is that a reliable prediction of
{\it antiquark} distributions would not be obtainable unless
incorporating effects of Dirac sea quarks or equivalently
vacuum polarization effects.

More consistent calculation including vacuum polarization
effects have been performed by Diakonov et al. [13,14] and
also by Tanikawa and Saito [15] with different regularization schemes,
but by confining to the isosinglet unpolarized as well as isovector
longitudinally polarized distribution functions, which have
values at the leading order of $1 / N_c$ expansion (or at the
$0$th order of the expansion in the collective angular velocity
$\Omega$ of the hedgehog soliton).
Unfortunately, an abundance of interesting
physics like the physics of ``nucleon spin contents'' is contained
in the next order of $1 / N_c$ expansion [5]. This is easily
understood because the inclusion of $O (\Omega^1)$ terms
is the minimum condition for the collective quantization treatment
of hedgehog solitons to hold. Otherwise,
the nucleon cannot have correct quantum numbers [4-6].

We have recently reported the first calculation of the
$O (\Omega^1)$ contributions to the isovector unpolarized quark
distribution function related to the physics of Gottfried sum [16]
with full inclusion of the vacuum polarization effects [17].
It was shown that the model can explain the excess of the $\bar{d}$
sea over the $\bar{u}$ sea in the proton very naturally [17-19].
However, some of the treatments there were criticized in
a recent paper by Pobylitsa et al. [20]. In the process of obtaining
theoretical quark distribution functions, we need to evaluate
nucleon matrix elements of quark bilinear operators which
are nonlocal in time. Their criticism is that the calculation
in [17]  does not treat this nonlocality in time to the full
extent.

Now the purpose of the present paper is to carry out a systematic
calculation of all the twist-2 spin dependent quark distribution
functions of the nucleon as consistently as possible.
We evaluate both of the $O (\Omega^0)$ and $O (\Omega^1)$
contributions with full inclusion of the vacuum polarization
effects. The above-mentioned nonlocality effects are
also carefully taken into account. We believe that these unique
features of our theoretical analysis would give new and important
information on the nonperturbative aspect of the spin dependent
quark distribution functions including the {\it antiquark}
distributions as well.

The plan of the paper is as follows. For completeness, we give
in sect.2 a precise definition of twist-2 quark distribution
functions which we shall investigate in the present paper.
How to evaluate these quark distribution functions within the
framework of the CQSM is explained in sect.3.
Sect.4 is devoted to the discussion of the numerical results.
We then summarize what we have found in sect.5.

\vspace{4mm}
\section{Definition of quark distribution functions}

\ \ \ Most theoretical analyses of quark distribution functions
of the nucleon are based on a field-theoretical formulation
given by Collins and Soper [21]. As a natural extension, Jaffe and Ji
recently carried out a systematic classification of quark
distribution functions by including chiral-odd distribution
functions which do not appear in the formulas of deep inelastic
scattering cross sections [22]. According to them,
there are nine independent distribution functions, from
twist 2 to twist 4. Here we are interested in the twist-2
distribution functions, which are known to have simple parton model
interpretation. There are three twist-2 distribution functions,
the spin independent (or averaged) distribution $f_1 (x)$, the
longitudinally polarized distribution $g_1 (x)$, and
what is called the transversity distribution $h_1 (x)$.
Following the notation of [22], they are represented as
\begin{eqnarray}
   f_1(x) &=& \frac{1}{\sqrt{2} p^+} \int 
   \frac{d \lambda}{2 \pi} \,e^{i \lambda x}
   <P S | \psi_+^\dagger (0) \psi_+(\lambda n) | P S > ,\\
   g_1 (x) &=& \frac{1}{\sqrt{2} p^+} \int 
   \frac{d \lambda}{2\pi} \,e^{i \lambda x}
   <P S_z | \psi_+^\dagger (0) \gamma_5 \psi_+ (\lambda n) | P S_z > ,\\
   h_1 (x) &=& \frac{1}{\sqrt{2} p^+} \int 
   \frac{d \lambda}{2 \pi} \,e^{i \lambda x}
   <P S_{\bot} | \psi_+^\dagger (0) \gamma_{\bot} \gamma_5 \psi_+ 
   (\lambda n) | P S_{\bot}> ,
\end{eqnarray}
where $p^\mu$ and $n^\mu$ are two light-like (null) vectors, having
the properties,
\begin{equation}
   p^{-} = 0, \ \ \ n^{+} = 0, \ \ \ p^2 = n^2 = 0, \ \ \ p \cdot n = 1 .
\end{equation}
Without loss of generality, one can choose a frame in which the
four-momentum $P^\mu$ of the initial nucleon and the four-momentum
transfer $q^\mu$ from a lepton to a nucleon have the third and
the time components only. In this frame, $p^\mu$ and $n^\mu$ take the
form :
\begin{equation}
   p^\mu = \frac{{\cal P}}{\sqrt{2}} \,(1,0,0,1), \ \ \ 
   n^\mu = \frac{1}{\sqrt{2} {\cal P}} \,(1,0,0,-1) ,
\end{equation}
while $P^\mu$ and $q^\mu$ are represented as
\begin{eqnarray}
   P^\mu &=& p^\mu + \frac{M^2}{2} \,n^\mu , \\
   q^\mu &=& \frac{1}{M_N^2} \left(
   \nu - \sqrt{\nu^2 + M_N^2 Q^2} \,\right) p^\mu
   + \frac{1}{2} \left( \nu + \sqrt{\nu^2 + M_N^2 Q^2} \,\right) 
   n^\mu ,
\end{eqnarray}
with $\nu = P \cdot q$ and $Q^2 = - q^2$. In the above definition of
the twist-2 quark distribution functions, $\psi_{+}$ is a component
of the quark field $\psi$ defined through the decomposition
\begin{equation}
   \psi = (P_{+} + P_{-}) \,\psi = \psi_{+} + \psi_{-} ,
\end{equation}
by the projection operators $P_{\pm} = \frac{1}{2} \gamma^{\mp}
\gamma^{\pm}$ with $\gamma^{\pm} = \frac{1}{\sqrt{2}} (\gamma^0
\pm \gamma^3)$. According to the authors of [22], $\psi_{+}$ is
called the ``good'' component of $\psi$, since it describes an
independent propagating degrees of freedom in the light-cone
quantization scheme [23]. On the other hand, $\psi_{-}$ is
called the ``bad'' component, since it can
be interpreted as quark-gluon composites. It is important to
recognize that only the good component of $\psi$ appears in the
definition of twist-2 quark distribution functions in conformity with
the fact that they have simple parton model interpretation.
In the actual model calculation of these distribution functions,
it is more convenient to rewrite the above expressions with use of
the identities :
\begin{eqnarray}
   P_{+}^2 \ \ \ 
   &=& P_{+} \ = \ \frac{1}{2} \,(1 + \gamma^0 \gamma^3) ,\\
   P_{+} \gamma_5 P_{+} &=& \frac{1}{2} \,(1 + \gamma^0 \gamma^3)
   \gamma_5 ,\\
   P_{+} \gamma_{\perp} \gamma_5 P_{+} &=& \frac{1}{2} \,
   (1 + \gamma^0 \gamma^3) \,\gamma_{\perp} \gamma_5 .
\end{eqnarray}
Since the distribution functions are in principle frame-independent,
it is also convenient to go to the nucleon rest frame, in which one
can set ${\cal P} = M_N / \sqrt{2}$. Now using the change of variable
as
\begin{equation}
   \lambda n^\mu = \lambda \,\frac{1}{M_N} \,
   (1,0,0,-1) \equiv z^\mu ,
\end{equation}
we obtain
\begin{eqnarray}
   z_0 = \frac{\lambda}{M_N}, \ \ \ 
   z_3 = - \frac{\lambda}{M_N} = - z_0, \ \ \ 
   z_{\perp} = 0.
\end{eqnarray}
Noting that
\begin{equation}
   \int_{- \infty}^{\infty} d \lambda \,e^{i \lambda x} \cdots = 
   M_N \int_{- \infty}^{\infty} d z_0 \,
   e^{i \,x \,M_N \,z_0} ,
\end{equation}
we are then led to the following expressions :
\begin{eqnarray}
   f_1 (x) &=& \frac{1}{4 \pi} \,\int \,d z^0 \, 
   e^{\,i \,x \,M_N \,z_0} \nonumber \\
   &\times& < \mbox{\boldmath $P$} = 0, 
   S \,| \,\psi^\dagger (0) (1 + \gamma^0 \gamma^3) 
   \psi (z) \,| \,
   \mbox{\boldmath $P$} = 0, S > |_{z_3 = -z_0, \ z_{\perp} = 0},
   \\
   g_1 (x) &=& \frac{1}{4 \pi} \,\int \,d z^0 \, 
   e^{\,i \,x \,M_N \,z_0} \nonumber \\
   &\times& < \mbox{\boldmath $P$} = 0, S_z \,| \,
   \psi^\dagger (0)
   (1 + \gamma^0 \gamma^3) \gamma_5 \psi (z)
   \,| \,\mbox{\boldmath $P$} = 0, S_z > 
   |_{z_3 = -z_0, \ z_{\perp} = 0}, \\
   h_1 (x) &=& \frac{1}{4 \pi} \,\int \,d z^0 \, 
   e^{\,i \,x \,M_N \,z_0} \nonumber \\
   &\times& < \mbox{\boldmath $P$} = 0, S_{\perp} \,| \,
   \psi^\dagger (0) 
   (1 + \gamma^0 \gamma^3) \gamma_{\perp} \gamma_5
   \psi(z) \,| \,\mbox{\boldmath $P$} = 0, S_{\perp} >|
   _{z_3 = -z_0, \ z_{\perp} = 0} .
\end{eqnarray}
What is left for us now is to evaluate nucleon matrix
elements of quark bilinear operators containing two space-time
coordinates with light-cone distance. How to evaluate these matrix
elements of {\it bilocal} quark operators will be explained in
the next section.

\vspace{4mm}
\section{Theory of quark distribution functions}

\ \ \ As shown in the previous section, the quark distribution
functions of our present interest can generally be represented
in the form :
\begin{equation}
   q (x) \ \ = \ \ \frac{1}{4 \pi} \,\,\int_{-\infty}^\infty
   \,\,d z_0 \,\,e^{\,i \,x \,M_N \,z_0} \,\,
   {\langle N (\mbox{\boldmath $P$} = 0) \,| \, \bar{\psi} (0) \,
   \Gamma \,\psi(z) \,| \,
   N (\mbox{\boldmath $P$} = 0) \rangle} \,\, 
   {|}_{z_3 = - z_0, \,z_\perp = 0} \,\, .
\end{equation}
In the present study, we confine to spin-dependent distribution
functions, so that we are to take
\begin{equation}
   O_a \ = \ (1 + \gamma^0 \gamma^3) \,\gamma_5, \ \ \ 
   \tau_3 \,(1 + \gamma^0 \gamma^3) \gamma_5 ,
\end{equation}
respectively for the isoscalar and isovector parts of the
longitudinally polarized distribution functions, whereas
\begin{equation}
   O_a \ = \ (1 + \gamma^0 \gamma^3) \gamma_{\perp} \gamma_5, \ \ \ 
   \tau_3 \,(1 + \gamma^0 \gamma^3) \,\gamma_{\perp} \gamma_5,
\end{equation}
for the isoscalar and isovector parts of the transversity distributions.
We recall here the fact that, extending the definition of distribution
function $q(x)$ to interval $-1 \le x \le 1$, the relevant antiquark
distributions are given as [14],
\begin{eqnarray}
   \Delta \bar{u} (x) + \Delta \bar{d} (x) &=&
   \Delta u(-x) + \Delta d (-x) \hspace{10mm} (0 < x < 1) , \\
   \Delta \bar{u} (x) - \Delta \bar{d} (x) &=& 
   \Delta u(-x) - \Delta d (-x) \hspace{10mm} (0 < x < 1) ,
\end{eqnarray}
for the longitudinally polarized distributions, while
\begin{eqnarray}
   \ \delta \bar{u} (x) + \delta \bar{d} (x) &=&
   - \,[ \delta u(-x) + \delta d (-x) ] \hspace{10mm} (0 < x < 1) , \\
   \ \delta \bar{u} (x) - \delta \bar{d} (x) &=&
   - \,[ \delta u(-x) - \delta d (-x) ] \hspace{10mm} (0 < x < 1) ,
\end{eqnarray}
for the transversity distributions [22]. As explained in the previous
paper [17], the basis of our analysis is the following path integral
representation of a matrix element of an arbitrary (bilocal) quark
bilinear operator between the nucleon states with definite
momenta :
\begin{eqnarray}
   &\,& \langle N (\mbox{\boldmath $P$}) \,| \,\psi^\dagger (0) \,O \,
   \psi(z) \,| \,N (\mbox{\boldmath $P$})\rangle
   \ \ = \ \ \frac{1}{Z} \,\,\int \,\,d^3 x  \,\,d^3 y \,\,
   e^{\,- \,i \mbox{\boldmath $P$} \cdot \mbox{\boldmath $x$}} \,\,
   e^{\,i \,\mbox{\boldmath $P$} \cdot \mbox{\boldmath $y$}} \,\,
   \int {\cal D} U \nonumber \\
   &\times& \!\!\!\! \int {\cal D}
   \psi \,\,{\cal D} \psi^\dagger \,\,
   J_N (\frac{T}{2}, \mbox{\boldmath $x$}) \,\,\psi^\dagger (0) \,
   O \,\psi(z) \,\,J_N^\dagger (-\frac{T}{2}, \mbox{\boldmath $y$}) \,\,
   \exp \,[\,\,i \int \,d^4 x \,\,\bar{\psi} \,\,
   (\,i \! \not\!\partial \,- \,M
   U^{\gamma_5}) \,\psi \, ] \, , \ \ \ \ \ 
\end{eqnarray}
where
\begin{eqnarray}
   {\cal L} \ \ = \ \ \bar{\psi} \,(\,i \not\!\partial \ - \ 
   M U^{\gamma_5} (x) \,) \,\psi \,\, ,
\end{eqnarray}
with $U^{\gamma_5} (x) = \exp [ \,i \gamma_5 \mbox{\boldmath $\tau$}
\cdot \mbox{\boldmath $\pi$} (x) / f_\pi \,]$ being the basic lagrangian
of the CQSM, and
\begin{equation}
   J_N (x) \ \ = \ \ \frac{1}{N_c !} \,\, 
   \epsilon^{\alpha_1 \cdots \alpha_{N_c}} \,\,
   \Gamma_{J J_3, T T_3}^{\{f_1 \cdots f_{N_c}\}} \,\,
   \psi_{\alpha_1 f_1} (x) \cdots \psi_{\alpha_{N_c} f_{N_c}} (x) \,\, ,
\end{equation}
is a composite operator carrying the quantum numbers $J J_3, T T_3$
(spin, isospin) of the nucleon, where $\alpha_i$ is the color index,
while $\Gamma^{\{ f_1 \cdots f_{N_C} \}}_{J J_3,T T_3}$ is a symmetric
matrix in spin-flavor indices $f_i$. By starting with a stationary
pion field configuration of hedgehog shape $U^{\gamma_5}_0
(\mbox{\boldmath $x$}) = \exp \,[ \,i \gamma_5 \mbox{\boldmath $\tau$}
\cdot \hat{\mbox{\boldmath $r$}} F(r) \,]$, the path integral over
the pion fields $U$ can be done in a saddle point approximation. Next, we
consider two important fluctuations around the static configuration,
i.e. the translational and rotational zero-modes.
To treat the translational zero-modes, we use an approximate
momentum projection procedure of the nucleon state, which amounts to
integrating over all shift $\mbox{\boldmath $R$}$ of the soliton
center-of-mass coordinates [14] :
\begin{equation}
   \langle N (\mbox{\boldmath $P$}) \,| \,
   \psi^\dagger (0) \,O \,\psi (z) \,| \,N
   (\mbox{\boldmath $P$}) \rangle
   \ \longrightarrow \ \int \,d^3 R \,\,
   \langle N (\mbox{\boldmath $P$}) \,| \,
   \psi^\dagger (0,- \mbox{\boldmath $R$}) \,
   O \,\psi (z_0,\mbox{\boldmath $z$} - \mbox{\boldmath $R$}) \,| \,N
   (\mbox{\boldmath $P$}) \rangle \,\, .
\end{equation}
The rotational zero-modes can be treated by introducing a rotating
meson field of the form :
\begin{eqnarray}
   U^{\gamma_5} ( \mbox{\boldmath $x$}, t) \ \ = \ \ A(t) \,\,
   U_0^{\gamma_5} (\mbox{\boldmath $x$}) \,\,A^\dagger (t) \,\, ,
\end{eqnarray}
where $A(t)$ is a time-dependent $SU(2)$ matrix in the isospin space.
Note first the identity
\begin{equation}
   \bar{\psi} \,( \,i \not\!\partial - M A (t) U^{\gamma_5}_0 
   (\mbox{\boldmath $x$}) A^\dagger (t) \,) \,\psi \ \ = \ \ 
   \psi^\dagger_A \,( i \partial_t - H - \Omega \,) \,\psi_A \,\, 
\end{equation}
with
\begin{eqnarray}
   \psi_A \ = \ A^\dagger (t) \,\psi \,\, , \hspace{5mm} 
   H \ = \ \frac{\mbox{\boldmath $\alpha$} \cdot \nabla}{i} 
   \ + \ M \,\beta \,U^{\gamma_5}_0 (\mbox{\boldmath $x$})
   \,\, , \hspace{5mm}
   \Omega \ = \ - \,i \,A^\dagger (t) \,\dot{A} (t) \,\, .
\end{eqnarray}
Here $H$ is a static Dirac Hamiltonian with the background pion
fields $U_0^{\gamma_5} (\mbox{\boldmath $x$})$, playing the role of a
mean field for quarks, while $\Omega = \frac{1}{2} \Omega_a \tau_a$
is the SU(2)-valued angular velocity matrix later to be quantized as
$\Omega_a \rightarrow \hat{J}_a / I$ with $I$ the moment of inertia
of the soliton and $\hat{J}_a$ the angular momentum operator [4-6].
We then introduce a change of quark field variables $\psi \rightarrow
\psi_A$, which amounts to getting on a body-fixed rotating frame.
Denoting $\psi_A$ anew as $\psi$ for notational simplicity, the
nucleon matrix element (25) can then be written as
\begin{eqnarray}
   &\,& \langle N (\mbox{\boldmath $P$}) \,| \,\psi^\dagger (0) \,O \,
   \psi(z) \,   | \,N (\mbox{\boldmath $P$}) \rangle \nonumber \\
   &=& \frac{1}{Z} \,\,\Gamma^{\{ f \}} \,\,{\Gamma^{\{ g \}}}^* \,\,
   \int \,\,d^3 x \,\,d^3 y \,\,
   e^{-i \mbox{\boldmath $P$} \cdot \mbox{\boldmath $x$}} \,\,
   e^{i \mbox{\boldmath $P$} \cdot \mbox{\boldmath $y$}} \,\,
   \int \,\,d^3 R \nonumber \\
   &\times& \int \,\,{\cal D} A \,\,{\cal D} \psi \,\,
   {\cal D} \psi^\dagger
   \,\,\exp \,[ \,\,i \,\int \,d^4 x \,\,\psi^\dagger 
   (\,i \partial_t - H - \Omega) \,\psi \,] \,\,
   \prod^{N_c}_{i = 1} \,\,\,
   [ \,A ( \frac{T}{2}) \,\,\psi_{f_i}
   ( \frac{T}{2}, \mbox{\boldmath $x$} ) \,] \nonumber \\
   &\times& \psi^\dagger (0, - \mbox{\boldmath $R$})
   \,\,A^\dagger (0) \,O \,A(z_0) \,\,
   \psi(z_0,\mbox{\boldmath $z$} - \mbox{\boldmath $R$}) \,\,
   \prod^{N_c}_{j = 1} \,\,\, [ \,\psi_{g_j}^\dagger
   (- \frac{T}{2}, \mbox{\boldmath $y$}) \,\,
   A^\dagger (-\frac{T}{2})] \,\, .
\end{eqnarray}
Now performing the path integral over the quark fields, we obtain
\begin{eqnarray}
   &\,& \langle N (\mbox{\boldmath $P$}) \,
   | \,\psi^\dagger (0) \,O \,\psi (z) \,
   | \,N (\mbox{\boldmath $P$}) \rangle  \nonumber \\
   &=& \frac{1}{Z} \,\,\tilde{\Gamma}^{\{ f \}} \,\,
   {\tilde{\Gamma}}^{{\{ g \}}^\dagger} \,\,N_c \,\,
   \int \,\,d^3 x \,\,d^3 y \,\,e^{\,-i \,\mbox{\boldmath $P$}
   \cdot \mbox{\boldmath $x$}}
   \,\,e^{\,i \mbox{\boldmath $P$} \cdot \mbox{\boldmath $y$}} \,\,
   \int d^3 R \,\,\int {\cal D} A \nonumber \\
   &\times& \!\!\! \Bigl\{ \, 
   {}_{f_1} \langle\frac{T}{2}, \mbox{\boldmath $x$}
   \,| \,\frac{i}
   {\,i \partial_t - H - \Omega} \,| \,0,- \mbox{\boldmath $R$}
   \rangle_{\gamma} \cdot
   {( A^\dagger (0) O_a A(z_0) )}_{\gamma \delta} \cdot
   {}_\delta \langle z_0, \mbox{\boldmath $z$} -
   \mbox{\boldmath $R$} \,| \,\frac{i}
   {\,i \partial_t - H - \Omega} \,
   | - \frac{T}{2}, \mbox{\boldmath $y$} \rangle_{g_1} \nonumber \\
   &-& \!\!\!
   \mbox{Tr} \,\,{\bigl( \,\langle z_0, \mbox{\boldmath $z$} -
   \mbox{\boldmath $R$} \,| \,
   \frac{i}{i \partial_t - H - \Omega} \,| \,
   0, - \mbox{\boldmath $R$} \rangle 
   A^\dagger (0) O_a A (z_0) \,\bigr)} \,\,\,
   {}_{f_1} \langle \frac{T}{2},\mbox{\boldmath $x$} \,| \,
   \frac{i}{\,i \partial_t - H - \Omega} \,| - \frac{T}{2},
   \mbox{\boldmath $y$}\rangle_{g_1} \,\, \Bigr\} \nonumber \\
   &\times& \prod^{N_c}_{j = 2} \,\,\,[ \,{}_{f_j}
   \langle \frac{T}{2}, \mbox{\boldmath $x$}
   \,| \,\frac{i}{\,i \partial_t - H - \Omega} \,
   | -\frac{T}{2}, \mbox{\boldmath $y$}\rangle_{g_j} \,] \,\cdot \,
   \exp \,[\,\mbox{Sp} \log \,(\,i \partial_t - H - \Omega) \, ] \,\, ,    
\end{eqnarray}
with $\tilde{\Gamma}^{\{f\}} = \Gamma^{\{f\}} \,
{[A(\frac{T}{2})]}^{N_c}$ etc. Here $\mbox{Tr}$ is to be taken over
spin-flavor indices. Assuming a slow rotation of the
hedgehog soliton, we can make use of an expansion in $\Omega$.
(Since $\Omega$ is known to be an $O (1 / N_c)$ quantity, this
perturbative expansion in $\Omega$ can also be taken as a $1 / N_c$
expansion. For an effective action, this gives
\begin{equation}
   \mbox{Sp} \log \,
   ( \,i \partial_t - H - \Omega) \ \ = \ \ 
   \mbox{Sp} \log \,(\,i \partial_t - H)
   \ + \ \,i \,\,\,\frac{1}{2} \,\,I \,\,
   \int \,\,\Omega_a^2 \,\,d t \,\, . 
\end{equation}
The second term here is essentially the action of a rigid rotor,
which plays the role of the evolution operator in the space of
collective coordinates. We also use the expansion of the
single quark propagator as
\begin{eqnarray}
   &\,& \!\!\!\!\!
   {}_{f_1} \langle \frac{T}{2}, \mbox{\boldmath $x$} \,| \,
   \frac{i}{i \partial_t - H - \Omega} \,| \,0,
   - \mbox{\boldmath $R$} \rangle {}_\gamma
   \ \ = \ \ 
   {}_{f_1} \langle \frac{T}{2}, \mbox{\boldmath $x$} \,| \,
   \frac{i}{i \partial_t - H }
   \,| \,0, - \mbox{\boldmath $R$} \rangle {}_\gamma \nonumber \\
   &\,& - \ \ \int \,\,d z_0^\prime \,\,
   d ^3 z^\prime \,\,{}_{f_1}
   \langle \frac{T}{2}, \mbox{\boldmath $x$} \,| \,
   \frac{i}{i \partial_t - H} \,
   | \,z_0^\prime, \mbox{\boldmath $z$}^\prime \rangle {}_\alpha \,\,
   \cdot i \,\Omega_{\alpha \beta} (z_0^\prime) \cdot
   {}_\beta \langle z_0^\prime, \mbox{\boldmath $z$}^\prime \,| \,
   \frac{i}{i \partial_t - H} \,| \,0, - \mbox{\boldmath $R$} 
   \rangle {}_\gamma \ \ \ \ \ \nonumber \\
   &\,& + \hspace{65mm} \dots \ \ \ \ \ . 
\end{eqnarray}
An important suggestion made in a recent paper by Pobilytsa et al.
[20] is that one must also take account of the nonlocality (in time)
of the operator $A^\dagger (0) O_a A(z_0)$. Expanding this operator
around $0$ or $z_0$, one respectively obtains
\begin{eqnarray}
   A^\dagger (0) O_a A(z_0) &=& A^\dagger (0) O_a A(0) 
   \ \,\,\,+ \ \,\,\,
   z_0 A^\dagger (0) O_a \dot{A} (0) \ \,\,\,+ \ \,\,\,\cdots , \\
   \mbox{or} \ \ \ \ \ \ 
   A^\dagger (0) O_a A(z_0) &=& A^\dagger (z_0) O_a A(z_0) \ - \ z_0
   \dot{A}^\dagger (z_0) O_a A (z_0) \ + \ \cdots .
\end{eqnarray}
Since both choices are known to lead to the same answer [20],
it is convenient to use a symmetrized form in the following
manipulation. This amounts to performing the following replacement :
\begin{eqnarray}
   A^\dagger (0) O_a A(z_0) &\longrightarrow&
   A^\dagger O_a A \ + \ \frac{1}{2} \,z_0 \,
   (A^\dagger O_a A \,\, A^\dagger \dot{A} - 
   \dot{A}^\dagger A \,\, A^\dagger O_a A) , \nonumber \\
   &=& \tilde{O}_a \ + \ i \,z_0 \,\frac{1}{2} \,
   \{ \Omega , \tilde{O}_a \} ,
\end{eqnarray}
in the process of collective quantization of the rotational motion.
Here we have introduced the notation
\begin{equation}
   \tilde{O}_a \ \equiv \ A^\dagger O_a A ,
\end{equation}
for saving space. Eq.(38) means that the nonlocality of the operator
$A^\dagger (0) O_a A(z_0)$ causes a rotational correction
{\it proportional} to the collective angular velocity $\Omega$.
After taking all these into account, we are then led to a perturbative
series in $\Omega$, which is also regarded as a $1 / N_c$ expansion :
\begin{eqnarray}
   &\,& \langle N (\mbox{\boldmath $P$}) | 
   \psi^\dagger (0) O_a \psi (z)
   | N (\mbox{\boldmath $P$}) \rangle \nonumber \\
   &=& {\langle N (\mbox{\boldmath $P$}) | 
   \psi^\dagger (0) O_a \psi (z)
   | N (\mbox{\boldmath $P$}) \rangle}^{\Omega^0} \ + \ 
      {\langle N (\mbox{\boldmath $P$}) | 
   \psi^\dagger (0) O_a \psi (z)
   | N (\mbox{\boldmath $P$}) \rangle}^{\Omega^1} \ + \ \cdots ,
\end{eqnarray}
where
\begin{eqnarray}
   &\,& \!\!\!\!\! {\langle N (\mbox{\boldmath $P$}) 
   \,| \,\psi^\dagger (0) \,
   O \,\psi (z) \,| \,
   N (\mbox{\boldmath $P$}) \rangle}^{\Omega^0}  \nonumber \\
   &=& \frac{1}{Z} \,\,{\,\tilde{\Gamma}}^{\{ f \}} \,\,
   {\tilde{\Gamma}}^{{\{ g \}}^\dagger} \,\,N_c \,\, 
   \int \,\,d^3 x \,\,d^3 y  \,\,e^{\,-i \mbox{\boldmath $P$} 
   \cdot \mbox{\boldmath $x$}} 
   \,\,e^{\,i \mbox{\boldmath $P$} \cdot \mbox{\boldmath $y$}}
   \,\,\int \,\,d^3 R \,\,\int \,\,{\cal D} A \,\,
   {(\tilde{O}_a)}_{\gamma \delta} \nonumber \\
   &\times& \!\!\! \Bigl[ \,{}_{f_1}
   \langle\frac{T}{2}, \mbox{\boldmath $x$} \,| \, 
   \frac{i}{i \partial_t - H} \,| \,
   0, -\mbox{\boldmath $R$} \rangle_{\gamma} \cdot
   {}_ {\delta} \langle z_0, 
   \mbox{\boldmath $z$} - \mbox{\boldmath $R$}
   \,| \,\frac{i}{i \partial_t - H} \,|\, - 
   \frac{T}{2}, \mbox{\boldmath $y$} \rangle {}_{g_1} \nonumber \\
   &-& \!\!\! {}_{\delta} \langle z_0, 
   \mbox{\boldmath $z$} - \mbox{\boldmath $R$} \,| \,
   \frac{i}{i \partial_t - H} \,| \,
   0, - \mbox{\boldmath $R$} \rangle_{\gamma}
   \cdot {}_{f_1} \langle \frac{T}{2}, \mbox{\boldmath $x$} \,| \,
   \frac{i}{i \partial_t - H} \,| \,- \frac{T}{2}, 
   \mbox{\boldmath $y$} \rangle {}_{g_1} \,\Bigr]  \nonumber \\
   &\times& \prod^{N_c}_{j = 2} \,\,\,\bigl[\, {}_{f_j}
   \langle \frac{T}{2}, \mbox{\boldmath $x$}
   \,| \,\frac{i}{i \partial_t - H} \,| \,
   - \frac{T}{2}, \mbox{\boldmath $y$} \rangle_{g_j} \,\bigr] 
   \,\cdot \,
   \,\,\,\exp \,[\,\,\mbox{Sp} \log (\,i \partial_t - H)
   \,+ \,i \,\frac{I}{2} 
   \,\int \Omega_a^2 \,d t \,\,]  \,\, ,
\end{eqnarray}
and
\begin{eqnarray}
   &\,& \langle N (\mbox{\boldmath $P$}) \,| \,\psi^\dagger (0) \,
   O \,\psi (z) \,| \,
   N (\mbox{\boldmath $P$}) \rangle^{\Omega^1}  \nonumber \\
   &=& \frac{1}{Z} \,\,{\,\tilde{\Gamma}}^{\{ f \}} \,\,
   {\tilde{\Gamma}}^{{\{ g \}}^\dagger} \,\,N_c \,\, 
   \int \,\,d^3 x \,\,d^3 y  \,\,e^{\,-i \mbox{\boldmath $P$} 
   \cdot \mbox{\boldmath $x$}} 
   \,\,e^{\,i \mbox{\boldmath $P$} \cdot \mbox{\boldmath $y$}}
   \,\,\int \,\,d^3 R \,\,\int \,\,{\cal D} {\cal A} \nonumber \\
   &\times& \!\!\! \Biggl\{ \,
   \int d^3 z^{\prime} \,\,d z_0^\prime \,\,\,\,
   i \,\Omega_{\alpha \beta} (z_0^\prime) \,\,
   {(A^\dagger (0) \Omega_a A(z_0))}_{\gamma \delta} \nonumber \\
   &\times& \!\!\!\! 
   \Bigl[ \,{}_{f_1}
   \langle\frac{T}{2}, \mbox{\boldmath $x$} \,| \, 
   \frac{i}{i \partial_t - H} \,| \,
   z_0^\prime, \mbox{\boldmath $z$}^\prime \rangle_{\alpha} \cdot
   {}_ {\beta} \langle z_0^\prime, \mbox{\boldmath $z$}^\prime 
   \,| \,\frac{i}{i \partial_t - H} \,|\, 0, - \mbox{\boldmath $R$} 
   \rangle_\gamma  \cdot {}_\delta \langle z_0,
   \mbox{\boldmath $z$} - \mbox{\boldmath $R$}
   \,| \,\frac{i}{i \partial_t - H} \,| \,- 
   \frac{T}{2}, \mbox{\boldmath $y$} \rangle_{g_1} \ \ \ \ \nonumber \\
   &+& \!\!\!\! {}_{f_1} \langle \frac{T}{2}, 
   \mbox{\boldmath $x$} \,| \,
   \frac{i}{i \partial_t - H} \,| \,
   0, - \mbox{\boldmath $R$} \rangle_{\gamma}
   \cdot {}_{\delta} \langle z_0, \mbox{\boldmath $z$} - 
   \mbox{\boldmath $R$} \,| \,
   \frac{i}{i \partial_t - H} \,| \,z_0^\prime, 
   \mbox{\boldmath $z$}^\prime \rangle_{\alpha} \cdot {}_{\beta}
   \langle z_0^\prime, 
   \mbox{\boldmath $z$}^\prime \,| \,\frac{i}{i \partial_t -H} | 
   - \frac{T}{2},
   \mbox{\boldmath $y$}\rangle_{g_1} \ \ \ \ \nonumber \\
   &-& \!\!\!\! {}_{f_1} \langle \frac{T}{2}, \mbox{\boldmath $x$}
   \,|\, \frac{i}{i \partial_t - H} \,| \,
   - \frac{T}{2}, \mbox{\boldmath $y$} \rangle_{g_1} \cdot
   {}_{\delta} \langle z_0, \mbox{\boldmath $z$} - 
   \mbox{\boldmath $R$} \,| 
   \,\frac{i}{i \partial_t - H}
   \, | \,  z_0^\prime, \mbox{\boldmath $z$}^\prime \rangle_{\alpha}
   \cdot {}_{\beta} \langle z_0^\prime, 
   \mbox{\boldmath $z$}^\prime \,| \, 
   \frac{i}{i \partial_t - H} \,| \, 0, - \mbox{\boldmath $R$} 
   \rangle_{\gamma} \,\Bigr] \ \ \ \ \nonumber \\
   &+& i \,z_0 \,\frac{1}{2} \,
   {\{ \Omega , \tilde{O}_a \}}_{\gamma \delta} \nonumber \\
   &\times& \!\!\!\! 
   \Bigl[ \,{}_{f_1}
   \langle\frac{T}{2}, \mbox{\boldmath $x$} \,| \, 
   \frac{i}{i \partial_t - H} \,| \,
   0, -\mbox{\boldmath $R$} \rangle_{\gamma} \cdot
   {}_ {\delta} \langle z_0, 
   \mbox{\boldmath $z$} - \mbox{\boldmath $R$}
   \,| \,\frac{i}{i \partial_t - H} \,|\, - 
   \frac{T}{2}, \mbox{\boldmath $y$} \rangle {}_{g_1} \nonumber \\
   &-& \!\!\!\! {}_{\delta} \langle z_0, 
   \mbox{\boldmath $z$} - \mbox{\boldmath $R$} \,| \,
   \frac{i}{i \partial_t - H} \,| \,
   0, - \mbox{\boldmath $R$} \rangle_{\gamma}
   \cdot {}_{f_1} \langle \frac{T}{2}, \mbox{\boldmath $x$} \,| \,
   \frac{i}{i \partial_t - H} \,| \,- \frac{T}{2}, 
   \mbox{\boldmath $y$} \rangle {}_{g_1} \,\Bigr] \,\Biggr\} \nonumber \\
   &\times& \prod^{N_c}_{j = 2} \,\,\,\bigl[\, {}_{f_j}
   \langle \frac{T}{2}, \mbox{\boldmath $x$}
   \,| \,\frac{i}{i \partial_t - H} \,| \,
   - \frac{T}{2}, \mbox{\boldmath $y$} \rangle_{g_j} \,\bigr] \,\cdot \,
   \,\,\,\exp \,[\,\,\mbox{Sp} \log (\,i \partial_t - H)
   \,+ \,i \,\frac{I}{2} 
   \,\int \Omega_a^2 \,d t \,\,]  \,\, .
\end{eqnarray}
Let us first discuss the leading $O (\Omega^0)$ term. As usual [4,5],
we introduce the eigenstates $| m \rangle$ and the associated
eigenenergies $E_m$ of the static Dirac hamiltonian $H$, satisfying
\begin{equation}
   H \,| m \rangle \ = \ E_m \,| m \rangle .
\end{equation}
This enables us to write down a spectral representation of the
single quark Green function as follows :
\begin{eqnarray}
   {}_{\alpha} \langle\mbox{\boldmath $x$}, t \,|\,
   \frac{i}{i \partial_t - H} \,| \,
   \mbox{\boldmath $x$}^\prime, t^\prime\rangle_{\beta}
   &=& \theta(t - t^\prime) \,\,\sum_{m>0} \,\,
   e^{\,- i E_m (t - t^\prime)} \,\,
   {}_{\alpha} \langle\mbox{\boldmath $x$}\,| \,m \rangle 
   \langle m \,| \,\mbox{\boldmath $x$}^\prime \rangle_{\beta}
   \nonumber \\
   &-& \theta(t^\prime - t) \,\,\sum_{m<0} \,\, 
   e^{\,-i E_m (t - t^\prime)} \,\,
   {}_{\alpha} \langle\mbox{\boldmath $x$} \,| \,m \rangle
   \langle m \,| \,\mbox{\boldmath $x$}^\prime\rangle_{\beta} \,\, . \ \
\end{eqnarray}
Using this equation together with the relation
\begin{equation}
   \langle \mbox{\boldmath $z$} - \mbox{\boldmath $R$} \, | \, 
   \ \ = \ \ \langle -\mbox{\boldmath $R$} \,| \,\,
   e^{\,i \mbox{\boldmath $p$} \cdot \mbox{\boldmath $z$}} \,\, ,
\end{equation}
we can perform the integration over $\mbox{\boldmath $R$}$ in (41). 
The resultant expression is then put into (18) to carry out the
integration over $z_0$. We then arrive at a formula, which provides
us with a theoretical basis for evaluating the zeroth order
contributions in $\Omega$ to quark distribution functions of the
nucleon :
\begin{equation}
   q (x ; \Omega^0) \ \ = \ \ \int \,
   \Psi_{J_3 T_3}^{{(J)}^*} [\xi_A] \,\,\,O^{(0)} [\xi_A] \,\,\,
   \Psi_{J_3 T_3}^{(J)} [\xi_A] \,\,d \xi_A ,
\end{equation}
where
\begin{equation}
   \Psi_{J_3 T_3}^{(J)} [\xi_A] 
   \ \ = \ \ \sqrt{\frac{2J+1}{8 \pi^2}} \,\,
   {(-1)}^{T + T_3} \,\,D_{-T_3 J_3}^{(J)} (\xi_A) \,\, ,
\end{equation}
are wave functions, describing the collective rotational motion
of the hedgehog soliton, while
\begin{equation}
   O^{(0)} [\xi_A] \ \ = \ \ M_N \,\frac{N_c}{2} \,
   \Bigl( \,\sum_{n \leq 0} - \sum_{n > 0} \,\Bigr) \,
   \langle n | \,\tilde{O}_a \delta (x M_N - E_n - p^3) \,
   | n \rangle .
\end{equation}
Using the identity
\begin{equation}
   \Bigl( \,\sum_{n \leq 0} + \sum_{n > 0} \,\Bigr) \,
   \langle n | \,\tilde{O}_a \delta (x M_N - E_n - p^3) \,
   | n \rangle \ = \ 0 ,
\end{equation}
eq.(48) can be expressed in either of the following two forms :
\begin{eqnarray}
   O^{(0)} [\xi_A] &=& \ \,\,\,\,M_N \,N_c \,
   \sum_{n \leq 0} \, \langle n | \,\tilde{O}_a 
   \delta (x M_N - E_n - p^3) \, | n \rangle \nonumber \\
   &=& - \,M_N \,N_c \,
   \sum_{n > 0} \, \langle n | \,\tilde{O}_a 
   \delta (x M_N - E_n - p^3) \, | n \rangle ,
\end{eqnarray}
i.e., as a sum over the occupied states or as a sum over the
nonoccupied states. As was emphasized in [14], it is better to use the
first form for $x > 0$, whereas the second form for $x < 0$,
for the purpose of numerical calculation.

Next we turn to the $O (\Omega^1)$ contribution.
In writing down (42), we have retained the time arguments $0, z_0$
and $z'_0$ in $A^\dagger$, $A$ and $\Omega$,
since we have to pay attention to the time order of these
collective space operators,
which do not generally commute after collective quantization of
the rotational zero-energy modes. In the previous paper [17],
motivated by the physical picture that the time-scale of deep
inelastic-scattering processes is much shorter than that of collective
rotational motion of the soliton, we dropped special time-order
diagrams in which the Coriolis coupling $\Omega$ between the
collective rotational motion and the intrinsic quark motion operates
in the time interval between $z_0$ and $0$. However, this procedure
was criticized by Pobylitsa et al. in a recent paper [20].
According to the them,
there is little reason to assume approximate degeneracy of $0$ and
$z_0$ in $A^\dagger (0) O_a A(z_0)$, since the deep-inelastic
scattering processes are not necessarily short distance phenomena.
Taking this nonlocality in time arguments more seriously, one
should retain all the possible time-order diagrams. In doing so,
we must pay attention to the time order of collective space
operators $A$ and $\Omega$. By ordering these operators according
to their time orders, we are led to the replacement :
\begin{eqnarray}
   &\,& \!\!\!\!\! \Omega_{\alpha \beta} (z'_0) \,
   {(A^\dagger (0) O_a A(z_0))}_{\gamma \delta} \nonumber \\
   &\longrightarrow& [ \theta (z'_0, 0, z_0) + \theta (z'_0, z_0, 0) ]
   \,\Omega_{\alpha \beta} \,\tilde{O}_{\gamma \delta} \,
   \ + \ [ \theta (0, z_0, z'_0) + \theta (z_0, 0, z'_0) ] \,
   \tilde{O}_{\gamma \delta} \Omega_{\alpha \beta}  \nonumber \\
   &+& \ \theta(0, z'_0, z_0) \,{(O_a)}_{\gamma^\prime \delta^\prime}
   \,A^\dagger_{\gamma \gamma^\prime} \,
   \Omega_{\alpha \beta} \,A_{\delta^\prime \delta}
   \ \,\,+ \ \,\,\theta(z_0, z'_0, 0) \,
   {(O_a)}_{\gamma^\prime \delta^\prime} \,
   A_{\delta^\prime \delta} \,\Omega_{\alpha \beta} \,
   A^\dagger_{\gamma \gamma^\prime} .
\end{eqnarray}
Here the third and the fourth terms are new ones discarded in the
treatment of [17]. In order to handle these somewhat peculiar terms,
we first recall the rule of collective quantization :
\begin{equation}
   \Omega \ = \ \frac{1}{2} \,\Omega_a \tau_a
   \longrightarrow \frac{1}{2 I} \,J_a \tau_a ,
\end{equation}
where $J_a$ is the total angular momentum operator satisfying the
commutation relations (CR) as follows :
\begin{eqnarray}
   \,[ J_a, J_b ] &=& i \,\epsilon_{a b c} \,J_c , \\
   \,[ J_a, A ] &=& \,\,\,\frac{1}{2} \,A \,\tau_a , \\
   \,[ J_a, A^\dagger ] &=& - \frac{1}{2} \,\tau_a \,A^\dagger .
\end{eqnarray}
Using these CR, one can show that
\begin{eqnarray}
   {(O_a)}_{\gamma^\prime \delta^\prime} \,
   A^\dagger_{\gamma \gamma^\prime} \,
   \,\Omega_{\alpha \beta}
   \,A_{\delta^\prime \delta}
   &=& \frac{1}{2 I} \,{(\tau_c)}_{\alpha \beta} \,
   {(O_a)}_{\gamma^\prime \delta^\prime} \,
   A^\dagger_{\gamma \gamma^\prime} \,J_c A_{\delta^\prime \delta}
   \nonumber \\
   &=& \frac{1}{2 I} \,{(\tau_c)}_{\alpha \beta} \,
   {(O_a)}_{\gamma^\prime \delta^\prime} \,
   A^\dagger_{\gamma \gamma^\prime} \,
   \,\bigl[\, \frac{1}{2} \,{(A \tau_c)}_{\delta^\prime \delta}
   + A_{\delta^\prime \delta} \,J_c \bigr] \nonumber \\
   &=& \frac{1}{2 I} \,{(\tau_c)}_{\alpha \beta}
   \,\bigl[\, \frac{1}{2} \,{(A^\dagger O_a A \tau_c)}_{\gamma \delta}
   + {(A^\dagger O_a A)}_{\gamma \delta} \,J_c \bigr] ,
\end{eqnarray}
where we have used (54). Similarly, by using (55), one
may obtain an alternative expression
\begin{eqnarray}
   &\,& {(O_a)}_{\gamma^\prime \delta^\prime} \,
   A^\dagger_{\gamma \gamma^\prime} \,
   \Omega_{\alpha \beta}
   \,A_{\delta^\prime \delta} \ = \ 
   \frac{1}{2 I} \,{(\tau_c)}_{\alpha \beta}
   \,\bigl[\, \frac{1}{2} \,{(\tau_c A^\dagger O_a A)}_{\gamma \delta}
   + J_c \,{(A^\dagger O_a A)}_{\gamma \delta} \bigr] .
\end{eqnarray}
In the following manipulation, we find it convenient to take an
average of these two expressions as
\begin{eqnarray}
   {(O_a)}_{\gamma^\prime \delta^\prime} \,
   A^\dagger_{\gamma \gamma^\prime} \,
   \Omega_{\alpha \beta} \,A_{\delta^\prime \delta}
   &=& \frac{1}{8 I} \,{(\tau_c)}_{\alpha \beta} \,
   [ \,{(A^\dagger O_a A \tau_c)}_{\gamma \delta} + 
   {(\tau_c A^\dagger O_a A)}_{\gamma \delta} \,] \nonumber \\
   &+& \frac{1}{4 I} \,{(\tau_c)}_{\alpha \beta} \,
   [\, {(A^\dagger O_a A )}_{\gamma \delta} J_c + 
   J_c {(A^\dagger O_a A)}_{\gamma \delta} \,] .
\end{eqnarray}
Now we must treat two cases separately. The first is the case in
which the operator $O_a$ contains an isospin factor $\tau_a$ as
\begin{equation}
   O_a = \tau_a \bar{O} .
\end{equation}
In this case, using the relation $A^\dagger O_a A = D_{ab} \tau_b
\bar{O}$, we can rewrite as
\begin{eqnarray}
   &\,& {(A^\dagger O_a A \tau_c)}_{\gamma \delta} + 
   {(\tau_c A^\dagger O_a A)}_{\gamma \delta}
   \ = \ D_{ab} \,{\Bigl((\tau_b \tau_c + \tau_c \tau_b) 
   \bar{O}\Bigr)}_{\gamma
   \delta} = 2 \,D_{ac} \,{(\bar{O})}_{\gamma \delta} . \ \ \ 
\end{eqnarray}
On the other hand, if $O_a$ contains no isospin factor as
\begin{equation}
   O_a = \bar{O} ,
\end{equation}
we obtain
\begin{equation}
   {(A^\dagger O_a A \tau_c)}_{\gamma \delta} + 
   {(\tau_c A^\dagger O_a A)}_{\gamma \delta} \ = \ 
   2 \,{(\tau_c \bar{O})}_{\gamma \delta} .
\end{equation}
Unifying the two cases, we can then write as
\begin{eqnarray}
   &\,& {(O_a)}_{\gamma^\prime \delta^\prime} \,
   A^\dagger_{\gamma \gamma^\prime} \,
   \Omega_{\alpha \beta} A_{\delta^\prime \delta}
   \ = \ \frac{1}{4 I} \,{(\tau_c)}_{\alpha \beta}
   \left\{ \begin{array}{c}
      D_{ac} \bar{O}_{\gamma \delta} \\
      {(\tau_c \bar{O})}_{\gamma \delta}
   \end{array}
   \right\} +
   \frac{1}{2} \,\{ \Omega_{\alpha \beta},
   {(A^\dagger O_a A)}_{\gamma \delta} \}_{+} .
\end{eqnarray}
A similar manipulation for the fourth term in (51) leads to
\begin{eqnarray}
   &\,& {(O_a)}_{\gamma^\prime \delta^\prime} 
   A_{\delta^\prime \delta}
   \Omega_{\alpha \beta} A^\dagger_{\gamma \gamma^\prime}
   \ = \ - \,\frac{1}{4 I} \,{(\tau_c)}_{\alpha \beta}
   \left\{ \begin{array}{c}
      D_{ac} \bar{O}_{\gamma \delta} \\
      {(\tau_c \bar{O})}_{\gamma \delta}
   \end{array}
   \right\} +
   \frac{1}{2} \,\{ \Omega_{\alpha \beta},
   {(A^\dagger O_a A)}_{\gamma \delta} \}_{+} .
\end{eqnarray}
Retaining all these possible time order diagrams, the
$O(\Omega^1)$ contribution to the distribution function now becomes
\begin{eqnarray}
   &\,& {\langle N (\mbox{\boldmath $P$}) \,| \,\psi^\dagger (z) \,
   O \,\psi (0) \,| \,
   N (\mbox{\boldmath $P$}) \rangle}^{\Omega^1}  \nonumber \\
   &=& \frac{1}{Z} \,\,{\,\tilde{\Gamma}}^{\{ f \}} \,\,
   {\tilde{\Gamma}}^{{\{ g \}}^\dagger} \,\,N_c \,\, 
   \int \,\,d^3 x \,\,d^3 y  \,\,e^{\,-i \mbox{\boldmath $P$} 
   \cdot \mbox{\boldmath $x$}} 
   \,\,e^{\,i \mbox{\boldmath $P$} \cdot \mbox{\boldmath $y$}}
   \,\,\int \,\,d^3 R \,\,\int \,\,{\cal D} A \nonumber \\
   &\times& 
   \Biggl\{ \,i \int d^3 z^\prime \,d z_0^\prime \nonumber \\
   &\times& 
   \biggl( \,
   [ \theta (z^\prime_0, 0, z_0) + \theta (z^\prime_0, z_0, 0) ]
   \,\Omega_{\alpha \beta} \tilde{O}_{\gamma \delta} \ + \ 
   [ \theta (0, z_0, z^\prime_0) + \theta (z_0, 0, z^\prime_0) ]
   \,\tilde{O}_{\gamma \delta} \Omega_{\alpha \beta} \nonumber \\
   &+& \theta(0, z^\prime_0, z_0) \,
   \Bigl[ \,\frac{1}{2} {\{ \Omega_{\alpha \beta}, 
   \,\tilde{O}_{\gamma \delta} \}}_{+} + \frac{1}{4 I}
   {(\tau_c)}_{\alpha \beta} 
   \left\{ \begin{array}{c}
      D_{ac} \bar{O}_{\gamma \delta} \\
      {(\tau_c \bar{O})}_{\gamma \delta}
   \end{array} \right\} 
   \Bigr] \nonumber \\
   &+& \theta(z_0, z^\prime_0, 0) \,
   \Bigl[ \,\frac{1}{2} {\{ \Omega_{\alpha \beta}, 
   \tilde{O}_{\gamma \delta} \}}_{+} - \frac{1}{4 I}
   {(\tau_c)}_{\alpha \beta} 
   \left\{ \begin{array}{c}
      D_{ac} \bar{O}_{\gamma \delta} \\
      {(\tau_c \bar{O})}_{\gamma \delta}
   \end{array} \right\} \Bigr] \biggr)
   \nonumber \\
   &\times& \!\!\!\! \Bigl[ \,{}_{f_1}
   \langle\frac{T}{2}, \mbox{\boldmath $x$} \,| \, 
   \frac{i}{i \partial_t - H} \,| \,
   z_0^\prime, \mbox{\boldmath $z$}^\prime \rangle_{\alpha} \cdot
   {}_ {\beta} \langle z_0^\prime, \mbox{\boldmath $z$}^\prime 
   \,| \,\frac{i}{i \partial_t - H} \,|\, 0, - \mbox{\boldmath $R$}
   \rangle_\gamma  \cdot {}_\delta \langle z_0, \mbox{\boldmath $z$} - 
   \mbox{\boldmath $R$}
   \,| \,\frac{i}{i \partial_t - H} \,| \,- 
   \frac{T}{2}, \mbox{\boldmath $y$} \rangle_{g_1} \ \ \ \ \nonumber \\
   &+& \!\!\!\! {}_{f_1} \langle \frac{T}{2}, 
   \mbox{\boldmath $x$} \,| \,
   \frac{i}{i \partial_t - H} \,| \,
   0, - \mbox{\boldmath $R$} \rangle_{\gamma}
   \cdot {}_{\delta} \langle z_0, \mbox{\boldmath $z$} - 
   \mbox{\boldmath $R$} \,| \,
   \frac{i}{i \partial_t - H} \,| \,z_0^\prime, 
   \mbox{\boldmath $z$}^\prime \rangle_{\alpha} \cdot {}_{\beta}
   \langle z_0^\prime, 
   \mbox{\boldmath $z$}^\prime \,| \,\frac{i}{i \partial_t -H} | 
   - \frac{T}{2},
   \mbox{\boldmath $y$}\rangle_{g_1} \ \ \ \ \nonumber \\
   &-& \!\!\!\! {}_{f_1} \langle \frac{T}{2}, \mbox{\boldmath $x$}
   \,|\, \frac{i}{i \partial_t - H} \,| \,
   - \frac{T}{2}, \mbox{\boldmath $y$} \rangle_{g_1} \cdot
   {}_{\delta} \langle z_0, \mbox{\boldmath $z$} - 
   \mbox{\boldmath $R$} \,|    \,\frac{i}{i \partial_t - H}
   \, | \,  z_0^\prime, \mbox{\boldmath $z$}^\prime\rangle_{\alpha}
   \cdot {}_{\beta} \langle z_0^\prime, 
   \mbox{\boldmath $z$}^\prime \,| \, 
   \frac{i}{i \partial_t - H} \,| \, 0, - \mbox{\boldmath $R$}
   \rangle_{\gamma} \Bigr] \ \ \ \ \nonumber \\
   &+& i \,z_0 \,\frac{1}{2} \,
   \{ \Omega, \tilde{O}_a \}_{\gamma \delta}
   \nonumber \\
   &\times& \!\!\!\! \Bigl[ \,{}_{f_1}
   \langle\frac{T}{2}, \mbox{\boldmath $x$} \,| \, 
   \frac{i}{i \partial_t - H} \,| \,
   0, -\mbox{\boldmath $R$} \rangle_{\gamma} \cdot
   {}_ {\delta} \langle z_0, 
   \mbox{\boldmath $z$} - \mbox{\boldmath $R$}
   \,| \,\frac{i}{i \partial_t - H} \,|\, - 
   \frac{T}{2}, \mbox{\boldmath $y$} \rangle {}_{g_1} \nonumber \\
   &-& \!\!\!\! {}_{\delta} \langle z_0, 
   \mbox{\boldmath $z$} - \mbox{\boldmath $R$} \,| \,
   \frac{i}{i \partial_t - H} \,| \,
   0, - \mbox{\boldmath $R$} \rangle_{\gamma}
   \cdot {}_{f_1} \langle \frac{T}{2}, \mbox{\boldmath $x$} \,| \,
   \frac{i}{i \partial_t - H} \,| \,- \frac{T}{2}, 
   \mbox{\boldmath $y$} \rangle {}_{g_1} \,\Bigr] \Biggr\}
   \nonumber \\
   &\times& \prod^{N_c}_{j = 2} \,\,\,[\, {}_{f_j}
   \langle \frac{T}{2}, \mbox{\boldmath $x$}
   \,| \,\frac{i}{i \partial_t - H} \,| \,
   - \frac{T}{2}, \mbox{\boldmath $y$} \rangle_{g_j} \,] \,\cdot \,
   \,\,\,\exp \,[\,\,\mbox{Sp} \log (\,i \partial_t - H)
   \,+ \,i \,\frac{I}{2} 
   \,\int \Omega_a^2 \,d t \,\,]  \,\, ,
\end{eqnarray}
After stating all the delicacies inherent in the structure function
problem, we can now proceed in the same way as [17] and [24]. Using
the spectral representation of the single quark Green function (44)
together with the relation (45),
we can perform the integration over $\mbox{\boldmath $R$}, 
\mbox{\boldmath $z$}^\prime$, and $z'_0$.
The resultant expression is then put into (18) to carry out the
integration over $z_0$. We then arrive at a formula, which gives
a theoretical basis for evaluating the $O (\Omega^1)$
contributions to quark distribution functions of the
nucleon :
\begin{equation}
   q (x ; \Omega^1) \ \ = \ \ \int \,
   \Psi_{J_3 T_3}^{{(J)}^*} [\xi_A] \,\,\,O^{(1)} [\xi_A] \,\,\,
   \Psi_{J_3 T_3}^{(J)} [\xi_A] \,\,d \xi_A ,
\end{equation}
where
\begin{equation}
   O^{(1)} [\xi_A] \ \ = \ \ O^{(1)}_A \ + \ O^{(1)}_B \ + \ 
   O^{(1)}_{B^\prime} \ + \ O^{(1)}_C ,
\end{equation}
with
\begin{eqnarray}
   O^{(1)}_A &=& M_N \,\frac{N_c}{4} \,
   ( \sum_{m > 0, n \leq 0} - \sum_{n > 0, m \leq 0} )
   \,\frac{1}{E_m - E_n} \nonumber \\
   &\times& \!\! [ \,\langle n \,| \,\tilde{O}_a
   (\delta_n + \delta_m) \,| \,m \rangle \,\,
   \langle m \,| \,\Omega \,| \,n \rangle  + 
   \langle n \,| \,\Omega \,| \,m \rangle \,\,
   \langle m \,| \, \tilde{O}_a \,(\delta_n + \delta_m) \,
   | \,m \rangle \, ] , \\
   O^{(1)}_B &=& M_N \,\frac{N_c}{4} \,
   ( \sum_{m \leq 0, n \leq 0} - \sum_{n > 0, m > 0} )
   \,\frac{1}{E_m - E_n} \nonumber \\
   &\times& \!\! [ \,\langle n \,| \,\tilde{O}_a
   (\delta_n - \delta_m) \,| \,m \rangle \,\,
   \langle m \,| \,\Omega \,| \,n \rangle
   + \,\,\langle n \,| \,\Omega \,| \,m \rangle \,\,
   \langle m \,| \, \tilde{O}_a \,(\delta_n - \delta_m) \,
   | \,m \rangle \, ] , \\
   O^{(1)}_{B^\prime} &=& M_N \, \frac{N_c}{8 I} \,
   ( \sum_{m \leq 0, n \leq 0} - \sum_{n > 0, m > 0} )
   \,\frac{1}{E_m - E_n} \nonumber \\
   &\,& \hspace{35mm} \times \, 
   \langle n \,| \,\tau_c \,| \,m \rangle \,\,
   \langle m \,| \,
   \left\{ \begin{array}{c}
     D_{ac} \bar{O} \\
     \tau_c \bar{O}
   \end{array} \right\} \, (\delta_n - \delta_m) \,
   | \,n \rangle , \\
   O^{(1)}_C &=& \frac{d}{d x} \,\frac{N_c}{4} \,
   ( \sum_{n \leq 0} - \sum_{n > 0} ) \,
   \langle n \,| \, \{ \tilde{O}_a, \Omega \} \,\delta_n \,| \,n
   \rangle .
\end{eqnarray}
In the above equations, we have used the notation
\begin{equation}
   \delta_m \ \equiv \ \delta (x M_N - E_m - p^3), \ \ \mbox{and} \ \ 
   \delta_n \ \equiv \ \delta (x M_N - E_n - p^3) .
\end{equation}
for saving space. Here $O^{(1)}_A$ is the contribution from the
diagram in which $z^\prime_0$ is later (or earlier) than both
of $0$ or $z_0$. As was emphasized in [17], this term contains
transitions between the occupied and nonoccupied single quark
levels so that it is not in conflict with the Pauli principle.
On the other hand, $O^{(1)}_B$ and $O^{(1)}_{B^\prime}$ are the
contributions from diagrams in which $z^\prime_0$ lies between
$0$ and $z_0$. Although these terms appear to contain
Pauli-violating transitions between the occupied levels themselves
or the nonoccupied ones, we take here the viewpoint advocated in
[20] that there is no compulsory reason to drop them since
we are here dealing with operators which are non-local in time.
Finally, $O^{(1)}_C$ is the $O (\Omega^1)$ contribution resulting
from the nonlocality of the operator $A^\dagger (0) O_a A(z_0)$,
i.e. the second term of (38). In deriving $O^{(1)}_C$, use has been
made of the identity,
\begin{equation}
   \frac{1}{2 \pi} \int_{- \infty}^{\infty} dz_0 \,\,i \,z_0 \,
   e^{i \,(x M_N - E_n - p^3) \,z_0} \ = \ 
   \frac{1}{M_N} \,\frac{\partial}{\partial x} \,
   \delta (x M_N - E_n - p^3) .
\end{equation}

As will become clear shortly, it is convenient to treat
$O^{(1)}_A$ and $O^{(1)}_B$ in a combined way. To see it, first
note that, after a simple change of summation indices, $O^{(1)}_A$
can be rewritten as
\begin{eqnarray}
   O^{(1)}_A &=& M_N \,\frac{N_c}{4} \,
   \Bigl\{ \sum_{m > 0, n \leq 0} \frac{1}{E_m - E_n} \,
   [ \langle n | \tilde{O}_a \delta_n | m \rangle
   \langle m | \Omega | n \rangle + \langle n | \Omega | m \rangle
   \langle m | \tilde{O}_a \delta_n | n \rangle ] \nonumber \\
   &\,& \hspace{10mm} - \ 
   \sum_{m \leq 0, n > 0} \frac{1}{E_m - E_n} \,
   [ \langle m | \tilde{O}_a \delta_n | n \rangle
   \langle n | \Omega | m \rangle + \langle m | \Omega | n \rangle
   \langle n | \tilde{O}_a \delta_n | m \rangle ] \Bigr\} .
\end{eqnarray}
From now on, we treat the two cases separately. First, assume
that the relevant operator $O_a$ contains an isospin factor
$\tau_a$ in such a form as $O_a = \tau_a \bar{O}$.
In this case, in view of the
relations $\tilde{O}_a = A^\dagger O_a A = D_{ab} \tau_b \bar{O}$
and $\Omega = \frac{1}{2 I} J_c \tau_c$, we must carefully treat
the noncommutativity of the two collective space operators
$D_{ab}$ and $J_c$. By keeping the order of $D_{ab}$ and $J_c$,
$O^{(1)}_A$ can generally be divided into two pieces [24] as
\begin{eqnarray}
   O^{(1)}_A &=& M_N \,\frac{N_c}{4 I} \,
   \frac{1}{2} \,\{ D_{ab}, J_c \}_{+} \nonumber \\
   &\times& \Bigl\{ \sum_{m > 0, n \leq 0} \frac{1}{E_m - E_n} \,
   [ \langle n | \tau_b \bar{O} \delta_n | m \rangle
   \langle m | \tau_c | n \rangle + \langle n | \tau_c | m \rangle
   \langle m | \tau_b \bar{O} \delta_n | n \rangle ] \nonumber \\
   &-& \,\,\,\sum_{m \leq 0, n > 0} \frac{1}{E_m - E_n} \,
   [ \langle n | \tau_b \bar{O} \delta_n | m \rangle
   \langle m | \tau_c | n \rangle + \langle n | \tau_c | m \rangle
   \langle m | \tau_b \bar{O} \delta_n | n \rangle ] \Bigr\} 
   \nonumber \\
   &+& M_N \,\frac{N_c}{4 I} \,
   \,\frac{1}{2} \,[ D_{ab}, J_c ] \nonumber \\
   &\times& \Bigl\{ \sum_{m > 0, n \leq 0} \frac{1}{E_m - E_n} \,
   [ \langle n | \tau_b \bar{O} \delta_n | m \rangle
   \langle m | \tau_c | n \rangle - \langle n | \tau_c | m \rangle
   \langle m | \tau_b \bar{O} \delta_n | n \rangle ] \nonumber \\
   &-& \,\,\,\sum_{m \leq 0, n > 0} \frac{1}{E_m - E_n} \,
   [ \langle n | \tau_b \bar{O} \delta_n | m \rangle
   \langle m | \tau_c | n \rangle - \langle n | \tau_c | m \rangle
   \langle m | \tau_b \bar{O} \delta_n | n \rangle ] \Bigr\} ,
\end{eqnarray}
which contains symmetric and antisymmetric combinations of
the two collective space operators $D_{ab}$ and $J_c$.
On the other hand, it can be easily verified that $O^{(1)}_B$
term contains symmetric combination only :
\begin{eqnarray}
   O^{(1)}_B &=& M_N \,\frac{N_c}{4 I} \,
   \,\frac{1}{2} \,\{ D_{ab}, J_c \}_{+} \nonumber \\
   &\times& \Bigl\{ \sum_{m > 0, n \leq 0} \frac{1}{E_m - E_n} \,
   [ \langle n | \tau_b \bar{O} \delta_n | m \rangle
   \langle m | \tau_c | n \rangle + \langle n | \tau_c | m \rangle
   \langle m | \tau_b \bar{O} \delta_n | n \rangle ] \nonumber \\
   &-& \,\,\,\sum_{m \leq 0, n > 0} \frac{1}{E_m - E_n} \,
   [ \langle n | \tau_b \bar{O} \delta_n | m \rangle
   \langle m | \tau_c | n \rangle + \langle n | \tau_c | m \rangle
   \langle m | \tau_b \bar{O} \delta_n | n \rangle ] \Bigr\} .
\end{eqnarray}
Combining $O^{(1)}_A$ and $O^{(1)}_B$ terms, we then obtain for
the isovector case
\begin{equation}
   O^{(1)}_A \ + \ O^{(1)}_B \ = \ 
   O^{(1)}_{\{A,B \}} \ + \ O^{(1)}_{[A, B]} ,
\end{equation}
where
\begin{eqnarray}
   &\,& O^{(1)}_{\{A, B \}} = M_N \,\frac{N_c}{4 I} \,
   \,\frac{1}{2} \,\{ D_{ab}, J_c \}_{+} \,
   \left( \sum_{m > 0, n \leq 0} - \sum_{m \leq 0, n > 0} +
   \sum_{m \leq 0, n \leq 0} - \sum_{m > 0, n > 0} 
   \right) , \\
   &\,& \hspace{33mm} \times \,
   \frac{1}{E_m - E_n} \,
   \left[ \langle n | \tau_b \bar{O} \delta_n | m \rangle
   \langle m | \tau_c | n \rangle + \langle n | \tau_c | m \rangle
   \langle m | \tau_b \bar{O} \delta_n | n \rangle 
   \right] \nonumber \\
   &\,& O^{(1)}_{[A, B]} = M_N \,\frac{N_c}{4 I} \,
   \,\frac{1}{2} \,
   [ D_{ab}, J_c ] \,
   \left( 
   \sum_{m > 0, n \leq 0} + \sum_{m \leq 0, n > 0} 
   \right) \nonumber \\
   &\,& \hspace{33mm} \times \,
   \frac{1}{E_m - E_n} \,
   \left[ \langle n | \tau_b \bar{O} \delta_n | m \rangle
   \langle m | \tau_c | n \rangle - \langle n | \tau_c | m \rangle
   \langle m | \tau_b \bar{O} \delta_n | n \rangle 
   \right] .
\end{eqnarray}

The situation is much simpler for isoscalar operators $O_a = \bar{O}$.
Since $\tilde{O}_a = A^\dagger O_a A = A^\dagger \bar{O} A = \bar{O}$,
we have only to replace both of $D_{ab}$ and $\tau_b$ by $1$
in the above manipulation, thereby leading to
\begin{eqnarray}
   O^{(1)}_{\{ A, B \}} &=& M_N \,\frac{N_c}{4 I} \,J_c 
   \left( \sum_{m > 0, n \leq 0} - \sum_{m \leq 0, n > 0} +
   \sum_{m \leq 0, n \leq 0} - \sum_{m > 0, n > 0} 
   \right) \nonumber \\
   &\,& \hspace{20mm} \times \,\frac{1}{E_m - E_n} \,
   \left[ 
   \langle n | \bar{O} \delta_n | m \rangle
   \langle m | \tau_c | n \rangle + \langle n | \tau_c | m \rangle
   \langle m | \bar{O} \delta_n | n \rangle
   \right] , \\
   O^{(1)}_{[ A, B ]} &=& 0 .
\end{eqnarray}
One notices that only the symmetric combination of the matrix elements
survives for this isoscalar case. This should be contrasted to the isovector
case in which either of the symmetric part or the antisymmetric
part survives, depending on the symmetry property of the relevant
single quark matrix elements appearing in (78) and (79).
As we shall discuss later, the symmetric part contributes to the
isoscalar unpolarized distribution function $\Delta u(x) + \Delta d(x)$
and $\delta u(x) + \delta d(x)$ at the $O(\Omega^1)$,
whereas the antisymmetric part plays an important
role in the $O(\Omega^1)$ term of the isovector polarized
distribution functions $\Delta u(x) - \Delta d(x)$ or
$\delta u(x) - \delta d(x)$ [24].

Now we shall investigate the case of our interest in more detail
for obtaining explicit formulas, which can be used for numerical
calculation of polarized distribution functions of the nucleon.

\subsection{$\Delta u(x) + \Delta d(x)$}

The relevant operator in this case is
\begin{equation}
   \tilde{O}_a \ = \ A^\dagger (1 + \gamma^0 \gamma^3) \gamma_5 A
   \ = \ (1 + \gamma^0 \gamma^3) \gamma^5 .
\end{equation}
Since the $O (\Omega^0)$ contribution to $\Delta u(x) + \Delta d(x)$
vanishes due to the hedgehog symmetry,
the leading contribution to this distribution function
arises from the $O (\Omega^1)$ terms. Due to the symmetry property
of the relevant single quark matrix elements, only the symmetric
combination of $O^{(1)}_A + O^{(1)}_B$ survives. The total
$O (\Omega^1)$ term therefore consists of three pieces,
$O^{(1)}_{\{A, B\}}$, $O_{B^\prime}$ and $O^{(1)}_C$.
Using the general formulas obtained so far, the contributions of these
three terms to $\Delta u(x) + \Delta d(x)$ are given as
\begin{eqnarray}
   &\,& {[ \Delta u(x) + \Delta d(x) ]}^{(1)}_{\{ A, B \}} \ = \ 
   {\langle J_3 \rangle}_{p \uparrow} \cdot M_N \,\frac{N_c}{4 I} \,
   \,( \sum_{m = all, n \leq 0} - 
   \sum_{m = all, n > 0} ) \frac{1}{E_m - E_n} \nonumber \\
   &\,& \hspace{20mm} 
   \times \,[ \langle n | (1 + \gamma^0 \gamma^3) \gamma_5 \delta_n
   | m \rangle \langle m | \tau_3 | n \rangle + 
   \langle n | \tau_3 | m \rangle 
   \langle m | (1 + \gamma^0 \gamma^3) \gamma_5 \delta_n
   | n \rangle ] , \\
   &\,& {[ \Delta u(x) + \Delta d(x) ]}^{(1)}_{B^\prime} \ = \ 
   {\langle 1 \rangle}_{p \uparrow} \cdot M_N \,\frac{N_c}{8 I} \,
   \, ( \sum_{m \leq 0, n \leq 0} - 
   \sum_{m >, n > 0} ) \frac{1}{E_m - E_n} \nonumber \\
   &\,& \hspace{56mm} 
   \times \,\langle n | \tau_c | m \rangle 
   \langle m | \tau_c (1 + \gamma^0 \gamma^3) \gamma_5 
   (\delta_n - \delta_m)    | n \rangle , \\
   &\,& {[ \Delta u(x) + \Delta d(x) ]}^{(1)}_{C} \ = \ 
   {\langle J_3 \rangle}_{p \uparrow} \cdot \frac{d}{d x} \,
   M_N \,\frac{N_c}{4 I} \,( \sum_{n \leq 0} - \sum_{n > 0} ) \,
   \langle n | \tau_3 (1 + \gamma^0 \gamma^3) \gamma_5 \delta_n
   | n \rangle .
\end{eqnarray}
In the above equations, ${\langle {\cal O} \rangle}_{p \uparrow}$
denotes a matrix element of a collective space operator ${\cal O}$
with respect to the proton in the spin up state along the $z$-axis,
i.e.
\begin{equation}
   {\langle {\cal O} \rangle}_{p \uparrow} \ = \ \int
   \Psi^{(\frac{1}{2})}_{\frac{1}{2} \frac{1}{2}} [\xi_A]
   \,{\cal O} \,\Psi^{(\frac{1}{2})}_{\frac{1}{2} \frac{1}{2}}
   [\xi_A] \,d \xi_A  \ = \ 
   \langle p, S_z = 1 / 2 | {\cal O} | 
   p, S_z = 1 / 2 \rangle .
\end{equation}
In deriving (83), we have used the relation
\begin{eqnarray}
   {\langle {\{ \tilde{O}_a, \Omega \}}_{+} \rangle}_{p \uparrow} &=&
   {\langle \,{\{ (1 + \gamma^0 \gamma^3) \gamma_5, \frac{1}{2}
   J_c \tau_c \}}_{+} \,\rangle}_{p \uparrow} 
   \ = \ {\langle J_3 \rangle}_{p \uparrow} \cdot \tau_3 \,
   (1 + \gamma^0 \gamma^3) \,\gamma_5 .
\end{eqnarray}
One may notice that the collective space operator contained in the
term ${[ \Delta u(x) + \Delta d(x)]}^{(1)}_{B^\prime}$ is $1$ and it
is different from $J_3$ contained in other two terms.
The appearance of this term seems to be inconsistent, since it does not
change sign in contrast to the other two terms when the direction of
the proton spin is reversed. Fortunately, it can be shown that
this potentially dangerous term vanishes identically due to the
symmetry of the double sum of the single quark matrix element :
\begin{equation}
   {[ \Delta u(x) + \Delta d(x)]}^{(1)}_{B^\prime} = 0 .
\end{equation}
We are then left with the two terms, i.e.
${[ \Delta u(x) + \Delta d(x)]}^{(1)}_{\{ A, B \}}$ and
${[ \Delta u(x) + \Delta d(x)]}^{(1)}_C$, which both have required
state dependence. For the purpose of numerical calculation, it is
convenient to rewrite the above two terms slightly further.
Using the argument given in appendix of [15], we can prove the
identity :
\begin{eqnarray}
   &\,& ( \sum_{m = all, n \leq 0} + \sum_{m = all, n > 0} )
   \,\frac{1}{E_m - E_n} \nonumber \\
   &\,& \hspace{1mm} \times \,\,\bigl[ 
   \langle n | (1 + \gamma^0 \gamma^3) \gamma_5 \delta_n | m \rangle
   \langle m | \tau_3 | n \rangle + 
   \langle n | \tau_3 | m \rangle
   \langle m | (1 + \gamma^0 \gamma^3) \gamma_5 \delta_n | n \rangle
   \bigr]
   = 0 .
\end{eqnarray}
As a consequence, ${[ \Delta u(x) + \Delta d(x) ]}^{(1)}_{\{A, B\}}$ can
be expressed either of the following two forms :
\begin{eqnarray}
   [ \Delta u(x) \!\!\! &+& \!\!\!
   \Delta d(x) ]^{(1)}_{\{A, B \}} \nonumber \\
   &=& \ \,{\langle J_3 \rangle}_{p \uparrow} \cdot M_N \,\frac{N_c}{4 I}
   \sum_{m = all, n \leq 0} \frac{1}{E_m - E_n} \nonumber \\
   &\times& [
   \langle n | (1 + \gamma^0 \gamma^3) \gamma_5 \delta_n | m \rangle
   \langle m | \tau_3 | n \rangle + 
   \langle n | \tau_3 | m \rangle
   \langle m | (1 + \gamma^0 \gamma^3) \gamma_5 \delta_n | n \rangle ]
   \nonumber \\
   &=& - \,{\langle J_3 \rangle}_{p \uparrow} \cdot M_N \,\frac{N_c}{4 I}
   \sum_{m = all, n > 0} \frac{1}{E_m - E_n} \nonumber \\
   &\times& [
   \langle n | (1 + \gamma^0 \gamma^3) \gamma_5 \delta_n | m \rangle
   \langle m | \tau_3 | n \rangle + 
   \langle n | \tau_3 | m \rangle
   \langle m | (1 + \gamma^0 \gamma^3) \gamma_5 \delta_n | n \rangle ] .
\end{eqnarray}
As advocated in [20], it is convenient to use the first expression
given as a sum over the occupied states for the numerical calculation of
distribution functions in the region $x > 0$, while to use the second
one given as sum over the non-occupied states when $x < 0$, since one
can thus avoid vacuum subtraction, i.e. subtraction of the corresponding
sums over vacuum levels (with $U = 1$). Following [20], we also separate
the $E_m = E_n$ contribution from the above sum over the single quark
levels. This can be done by noting the identities,
\begin{eqnarray}
   &\,& \sum_{m \leq 0, n \leq 0} \frac{1}{E_m - E_n} \,
   \langle n | (1 + \gamma^0 \gamma^3) \gamma_5 \delta_n | m \rangle
   \langle m | \tau_3 | n \rangle + 
   \langle n | \tau_3 | m \rangle
   \langle m | (1 + \gamma^0 \gamma^3) \gamma_5 \delta_n | n \rangle ]
   \nonumber \\
   &\,& \hspace{10mm} \ = \ 
   \frac{1}{2} \sum_{m \leq 0, n \leq 0} \frac{1}{E_m - E_n} \,
   \bigl[ \langle n | (1 + \gamma^0 \gamma^3) \gamma_5 
   (\delta_n - \delta_m)| m \rangle
   \langle m | \tau_3 | n \rangle \nonumber \\
   &\,& \hspace{47mm} \ + \ 
   \langle n | \tau_3 | m \rangle
   \langle m | (1 + \gamma^0 \gamma^3) \gamma_5 
   (\delta_n - \delta_m) | n \rangle ] ,
\end{eqnarray}
and
\begin{eqnarray}
   \lim_{E_m \longrightarrow E_n} \!\!\!\!\!\! &\,& \!\!\!\!\!\!
   \frac{\delta (x M_N - E_n - p^3) -
   \delta (x M_N - E_m - p^3)}{E_m - E_n} \nonumber \\
   &=& \delta^\prime (x M_N - E_n - p^3) 
   \ = \ \frac{1}{M_N} \,\frac{d}{d x} \,\delta (x M_N - E_n - p^3) .
\end{eqnarray}
From (90), we can then readily obtain
\begin{eqnarray}
   &\,& \!\!\! {[ \Delta u(x) + \Delta d(x) ]}^{(1)}_{\{ A, B \}} 
   \nonumber \\
   &=& {\langle J_3 \rangle}_{p \uparrow} 
   \cdot M_N \,\frac{N_c}{4 I} \,
   \sum_{\stackrel{\scriptstyle m = all, n \leq 0}{(E_m \neq E_n)}} \,
   \frac{1}{E_m - E_n} \nonumber \\
   &\,& \hspace{5mm} \times \,
   \left[ \langle n | (1 + \gamma^0 \gamma^3) \gamma_5 
   \delta_n | m \rangle
   \langle m | \tau_3 | n \rangle + 
   \langle n | \tau_3 | m \rangle
   \langle m | (1 + \gamma^0 \gamma^3) \gamma_5 
   \delta_n | n \rangle 
   \right] \nonumber \\
   &+& {\langle J_3 \rangle}_{p \uparrow} \cdot
   \frac{d}{d x} \,\frac{N_c}{4 I}
   \sum_{\stackrel{\scriptstyle m \leq 0, n \leq 0}{(E_m = E_n)}} \,
   \frac{1}{E_m - E_n} \nonumber \\
   &\,& \!\!\!\!\!\! \times \,
   \left[ \langle n | (1 + \gamma^0 \gamma^3) \gamma_5 
   (\delta_n - \delta_m)| m \rangle
   \langle m | \tau_3 | n \rangle + 
   \langle n | \tau_3 | m \rangle
   \langle m | (1 + \gamma^0 \gamma^3) \gamma_5 
   (\delta_n - \delta_m) | n \rangle 
   \right] ,
\end{eqnarray}
and a corresponding expression given as sums over non-occupied levels.
The remaining term ${[ \Delta u(x) + \Delta d(x)]}^{(1)}_C$ can similarly
be expressed in either of the two equivalent forms as
\begin{eqnarray}
   {[ \Delta u(x) + \Delta d(x)]}^{(1)}_C &=&
   \,\,\,\,\,\,{\langle J_3 \rangle}_{p \uparrow} \cdot \frac{d}{d x}
   \,\frac{N_c}{4 I} \,\sum_{n \leq 0} \,
   \langle n | \tau_3 (1 + \gamma^0 \gamma^3) 
   \gamma_5 \delta_n | n \rangle
   \nonumber \\
   &=& - \,{\langle J_3 \rangle}_{p \uparrow} \cdot \frac{d}{d x}
   \,\frac{N_c}{4 I} \,\sum_{n > 0} \,
   \langle n | \tau_3 (1 + \gamma^0 \gamma^3) 
   \gamma_5 \delta_n | n \rangle .
\end{eqnarray}
Inserting the complete set of single quark states into the first expression
and separating the $E_m \neq E_n$ and $E_m = E_n$ terms in this sum,
we obtain
\begin{eqnarray}
   {[ \Delta u(x) + \Delta d(x)]}^{(1)}_C &=&
   {\langle J_3 \rangle}_{p \uparrow} \cdot \frac{d}{d x}
   \,\frac{N_c}{4 I} \,
   \sum_{\stackrel{\scriptstyle m = all, n \leq 0}{(E_m \neq E_n)}}
   \langle n | \tau_3 | m \rangle \langle m |
   (1 + \gamma^0 \gamma^3) \gamma_5 \delta_n | n \rangle
   \nonumber \\
   &+& {\langle J_3 \rangle}_{p \uparrow} \cdot \frac{d}{d x}
   \,\frac{N_c}{4 I} \,\,
   \sum_{\stackrel{\scriptstyle m \leq 0, n \leq 0}{(E_m = E_n)}}
   \langle n | \tau_3 | m \rangle \langle m |
   (1 + \gamma^0 \gamma^3) \gamma_5 \delta_n | n \rangle ,
\end{eqnarray}
and a corresponding expression given as sums over non-occupied states.
Just as argued in [15] for the case of the unpolarized distribution
function $u(x) - d(x)$, $E_m = E_n$ contribution in the double sums in
${[\Delta u(x) + \Delta d(x)]}^{(1)}_{\{A, B \}}$ and
${[ \Delta u(x) + \Delta d(x)]}^{(1)}_C$ precisely cancel each other.
After regrouping the terms
in such a way that this cancellation occurs at the level of analytical
expressions, the $O (\Omega^1)$ contribution to the distribution
function $\Delta u(x) + \Delta d(x)$ can finally be written in the
following form :
\begin{eqnarray}
   &\,& {[ \Delta u(x) + \Delta d(x) ]}^{(1)}
   \ = \ 
   {[ \Delta u(x) + \Delta d(x) ]}^{(1)}_{\{ A, B \}^\prime}
   \ + \ 
   {[ \Delta u(x) + \Delta d(x) ]}^{(1)}_{C^\prime}
\end{eqnarray}
where
\begin{eqnarray}
   &\,& \!\!\! 
   {[ \Delta u(x) + \Delta d(x) ]}^{(1)}_{\{ A, B \}^\prime}
   \nonumber \\
   &=& \,\,{\langle J_3 \rangle}_{p \uparrow} \cdot 
   M_N \,\frac{N_c}{2 I} \,
   \sum_{\stackrel{\scriptstyle m = all, n \leq 0}{(E_m \neq E_n)}} 
   \frac{1}{E_m - E_n} 
   \langle n | (1 + \gamma^0 \gamma^3) \gamma_5 
   \delta_n | m \rangle
   \langle m | \tau_3 | n \rangle , \\
   &=& \!\!\! - \,{\langle J_3 \rangle}_{p \uparrow} \cdot
   \,M_N \,\frac{N_c}{2 I} \,
   \sum_{\stackrel{\scriptstyle m = all, n > 0}{(E_m \neq E_n)}} 
   \frac{1}{E_m - E_n} 
   \langle n | (1 + \gamma^0 \gamma^3) \gamma_5 
   \delta_n | m \rangle
   \langle m | \tau_3 | n \rangle ,
\end{eqnarray}
and
\begin{eqnarray}
   {[ \Delta u(x) + \Delta d(x)]}^{(1)}_{C^\prime}
   &=& \,\,\,\,\,
   {\langle J_3 \rangle}_{p \uparrow} \cdot \frac{d}{d x} \,
   \frac{N_c}{4 I} \,
   \sum_{\stackrel{\scriptstyle m = all, n \leq 0}{(E_m \neq E_n)}}
   \langle n | \tau_3 | m \rangle \langle m |
   (1 + \gamma^0 \gamma^3) \gamma_5 \delta_n | n \rangle ,
   \\
   &=& - \,{\langle J_3 \rangle}_{p \uparrow} \cdot \frac{d}{d x} \,
   \frac{N_c}{4 I} \,
   \sum_{\stackrel{\scriptstyle m = all, n > 0}{(E_m \neq E_n)}}
   \langle n | \tau_3 | m \rangle \langle m |
   (1 + \gamma^0 \gamma^3) \gamma_5 \delta_n | n \rangle .
\end{eqnarray}
These expression will be used in the numerical calculation.

\subsection{$\Delta u(x) - \Delta d(x)$}

The relevant operator for the isovector longitudinally polarized
distribution function is
\begin{equation}
   \tilde{O}_{a = 3} \ = \ A^\dagger \tau_3 (1 + \gamma^0 \gamma^3)
   \gamma_5 A \ = \ D_{3 b} \,\tau_b \,
   (1 + \gamma^0 \gamma^3) \gamma_5 .
\end{equation}
The main contribution to this distribution function comes from the
$0$th order term in $\Omega$. A simple manipulation gives
\begin{eqnarray}
   {[ \Delta u(x) - \Delta d(x)]}^{(0)} &=&
   \,\,\,\,\,
   {\langle D_{33} \rangle}_{p \uparrow} \cdot M_N \,N_c \,
   \sum_{n \leq 0} \,\langle n | \tau_3 (1 + \gamma^0 \gamma^3)
   \gamma_5 \delta_n | n \rangle \nonumber \\
   &=& - \,{\langle D_{33} \rangle}_{p \uparrow} \cdot M_N \,N_c \,
   \sum_{n > 0} \,\langle n | \tau_3 (1 + \gamma^0 \gamma^3)
   \gamma_5 \delta_n | n \rangle .
\end{eqnarray}
The $O (\Omega^1)$ contribution to $\Delta u(x) - \Delta d(x)$ is
much far complicated. It generally consists of 4 terms,
$O^{(1)}_A$, $O^{(1)}_B$, $O^{(1)}_{B^\prime}$ and $O^{(1)}_C$.
As was already mentioned, the symmetric part of the sum of
$O^{(1)}_A$ and $O^{(1)}_B$ vanishes for this particular operator,
owing to the symmetry of the single quark matrix elements.
Using the familiar commutation relation
\begin{equation}
   [ J_c, D_{3 b} ] \ = \ i \,\epsilon_{c b e} \,D_{3 e} ,
\end{equation}
the antisymmetric part of $O^{(1)}_A + O^{(1)}_B$ becomes
\begin{eqnarray}
   &\,& \!\!\!\!\!\! {[\Delta u(x) - \Delta d(x)]}^{(1)}_{[A,B]}
   \nonumber \\
   &=& {\langle D_{33} \rangle}_{p \uparrow} \cdot M_N \,
   \frac{N_c}{8 I} \,\,i \,\epsilon_{3 c b} \,\sum_{m > 0, n \leq 0}
   \frac{1}{E_m - E_n} \nonumber \\
   &\times& \!\!\!\! [ \langle n | \tau_c | m \rangle \langle m |
   \tau_b (1 + \gamma^0 \gamma^3) \gamma_5 (\delta_n + \delta_m) 
   | n \rangle + 
   \langle n | \tau_b
   (1 + \gamma^0 \gamma^3) \gamma_5 (\delta_n + \delta_m) 
   | m \rangle \langle m | \tau_c | n \rangle ] \ \ \ \nonumber \\
   &=& - \,{\langle D_{33} \rangle}_{p \uparrow} \cdot M_N \,
   \frac{1}{I} \,\frac{N_c}{2} \,\sum_{m > 0, n \leq 0}
   \frac{1}{E_m - E_n} \nonumber \\
   &\times& \!\!\!\! [ \langle n | \tau_{+1} | m \rangle \langle n |
   \tau_{+1} (\gamma_5 + \Sigma_3) \frac{\delta_n + \delta_m}{2}
   | m \rangle - \langle n | \tau_{-1} | m \rangle 
   \langle n | \tau_{-1} (\gamma_5 + \Sigma_3) 
   \frac{\delta_n + \delta_m}{2} | m \rangle ] ,
\end{eqnarray}
with the standard definition $\tau_{\pm} = \mp (\tau_1 \pm
i \tau_2) / \sqrt{2}$ and $\Sigma_3 = \gamma^0 \gamma^3 \gamma_5$.
Next, from (70) with the case of isovector operator, we find that
\begin{eqnarray}
   {[ \Delta u(x) - \Delta d(x) ]}_{B^\prime}^{(1)}
   &=& {\langle D_{33} \rangle}_{p \uparrow} \cdot
   M_N \,\frac{N_c}{8 I} \,\left( \sum_{m \leq 0, n \leq 0} - 
   \sum_{m > 0, n > 0} \right) \nonumber \\
   &\,& \hspace{5mm} \times \,
   \frac{1}{E_m - E_n} \,
   \langle n | \tau_3 | m \rangle \langle m | 
   (1 + \gamma^0 \gamma^3) \gamma_5 (\delta_n - \delta_m) 
   | n \rangle .
\end{eqnarray}
One should notice that the state dependence of this somewhat
peculiar contribution is nothing different from that of the main term,
which implies that there is no reason for this term to vanish.
In fact, the single quark matrix element appearing in the above
double sum is essentially the same as that appearing in the expression
for ${[ \Delta u(x) + \Delta d(x)]}^{(1)}$.

The last but potentially important contribution comes from the
nonlocality correction term $O^{(1)}_C$. First note that
\begin{eqnarray}
   {\{ \tilde{O}_{a=3}, \Omega \}}_{+} &=& 
   \Bigl\{ D_{3b} \tau_b (1 + \gamma^0 \gamma^3) \gamma_5,
   \,\frac{1}{2 I} J_c \tau_c \Bigr\}_{+} \nonumber \\
   &=& \frac{1}{2 I} \,
   \left ( \,\frac{1}{2} \{ D_{3b}, J_c \}_{+}
   \left[ 
   \,\tau_b (1 + \gamma^0 \gamma^3) \gamma_5 \tau_c +
   \,\tau_c \tau_b (1 + \gamma^0 \gamma^3) \gamma_5 
   \right] \right. \nonumber \\
   &\,& \hspace{6mm} + \left. \frac{1}{2}
   \,[ D_{3b}, J_c ] \,\,\,
   \left[ \tau_b (1 + \gamma^0 \gamma^3) \gamma_5 \tau_c -
   \tau_c \tau_b (1 + \gamma^0 \gamma^3) \gamma_5 
   \right] \right)
   \nonumber \\
   &=& \frac{1}{2 I} 
   \left( \{ D_{3c}, J_c \}_{+}
   (1 + \gamma^0 \gamma^3) \gamma_5 \ + \ 
   i \,\epsilon_{bce} \,[ D_{3b}, J_c ] \,
   \tau_e \,(1 + \gamma^0 \gamma^3) \gamma_5 
   \right) .
\end{eqnarray}
The first term of the above quation does not contribute, since
\begin{eqnarray}
   \sum_{n \leq 0} \,\langle n | (1 + \gamma^0 \gamma^3)
   \gamma_5 \delta_n | n \rangle = \sum_{n > 0} \,
   \langle n | (1 + \gamma^0 \gamma^3) \gamma_5 \delta_n |
   n \rangle = 0 .
\end{eqnarray}
Simplifying the second term by using the CR (103), we finally
obtain
\begin{eqnarray}
   &\,& {[ \Delta u(x) - \Delta d(x)]}^{(1)}_C
   = - \,{\langle D_{33} \rangle}^{(1)}_{p \uparrow}
   \cdot \frac{d}{d x}
   \,\frac{N_c}{4 I} \,\left( \sum_{n \leq 0} - \sum_{n > 0} \right)
   \,\langle n | \tau_3 (1 + \gamma^0 \gamma^3) \gamma^5
   \delta_n | n \rangle . \ \ \ 
\end{eqnarray}
For the same reason as before, it is convenient to
consider these term in a combined way. To this end, we first
rewrite (108) by inserting a complete set of single quark states
and by separating the $E_m = E_n$ contributions from the
resultant double sum. The result can be expressed in two
alternative forms as
\begin{eqnarray}
   &\,& {[ \Delta u(x) - \Delta d(x)]}^{(1)}_C \nonumber \\
   &=& \!\!\! - \,{\langle D_{33} \rangle}_{p \uparrow} \cdot 
   \frac{d}{dx} \, \frac{N_c}{4 I} \,
   \left( 
   \sum_{\stackrel{\scriptstyle m = all, n \leq 0}{(E_m \neq E_n)}}
   + \sum_{\stackrel{\scriptstyle m \leq 0, n \leq 0}{(E_m = E_n)}}
   \right)
   \langle n | \tau_3 | m \rangle \langle m |
   (1 + \gamma^0 \gamma^3) \gamma_5  \delta_n | n \rangle
   \nonumber \\
   &=& \,{\langle D_{33} \rangle}_{p \uparrow} \cdot
   \frac{d}{dx} \,\frac{N_c}{4 I} 
   \left( 
   \sum_{\stackrel{\scriptstyle m = all, n > 0}{E_m \neq E_n}}
   + \sum_{\stackrel{\scriptstyle m > 0, n > 0}{E_m = E_n}} \right)
   \langle n | \tau_3 | m \rangle \langle m |
   (1 + \gamma^0 \gamma^3) \gamma_5  \delta_n | n \rangle .
\end{eqnarray}
To rewrite the $B^\prime$ term, we first separate $E_m = E_n$
contributions in the double sum of (105) as
\begin{eqnarray}
   &\,& {[ \Delta u(x) - \Delta d(x)]}^{(1)}_{B^\prime} 
   \nonumber \\
   &=& {\langle D_{33} \rangle}_{p \uparrow} \cdot
   \frac{d}{dx} \,M_N \,\frac{N_c}{2 I} \,
   \left( 
   \sum_{\stackrel{\scriptstyle m \leq 0, n \leq 0}{(E_m \neq E_n)}} -
   \sum_{\stackrel{\scriptstyle m > 0, n > 0}{(E_m \neq E_n)}}
   \right) \nonumber \\
   &\,& \hspace{30mm} \times \,
   \frac{1}{E_m - E_n} \,
   \langle n | \tau_3 | m \rangle 
   \langle m |
   (1 + \gamma^0 \gamma^3) \gamma_5  \delta_n | n \rangle ,
   \nonumber \\
   &+& \!\!\! {\langle - D_{33} \rangle}_{p \uparrow} \cdot
   \,\frac{d}{dx} \,\frac{N_c}{4 I} \,
   \left( 
   \sum_{\stackrel{\scriptstyle m \leq 0, n \leq 0}{(E_m = E_n)}} 
   \! - \!
   \sum_{\stackrel{\scriptstyle m > 0, n > 0}{(E_m = E_n)}} \right)
   \langle n | \tau_3 | m \rangle \langle m |
   (1 + \gamma^0 \gamma^3) \gamma_5  \delta_n | n \rangle .
\end{eqnarray}
Next, we notice the identity
\begin{eqnarray}
   0 &=& \sum_{\stackrel{\scriptstyle m = all, n = all}{(E_m \neq E_n)}}
   \frac{1}{E_m - E_n} \langle n | \tau_3 | m \rangle
   \langle m | (1 + \gamma^0 \gamma^3) \gamma_5 \delta_n | n
   \rangle \nonumber \\
   &+& \frac{1}{2 M_N} \,\frac{d}{dx} \,
   \sum_{\stackrel{\scriptstyle m = all, n = all}{(E_m = E_n)}}
   \langle n | \tau_3 | m \rangle
   \langle m | (1 + \gamma^0 \gamma^3) \gamma_5 \delta_n | n
   \rangle .
\end{eqnarray}
Using this identity, (110) can be rewritten in either of the
following two forms :
\begin{eqnarray}
   [ \Delta u(x) \!\!\! &-& \!\!\! \Delta d(x)]^{(1)}_{B^\prime}
   \nonumber \\
   &=& {\langle D_{33} \rangle}_{p \uparrow} \cdot
   M_N \,\frac{N_c}{4 I} \,
   \left( 2 
   \sum_{\stackrel{\scriptstyle m \leq 0, n \leq 0}{(E_m \neq E_n)}} +
   \sum_{m > 0, n \leq 0}  + \sum_{m \leq 0, n > 0} \right) \nonumber \\
   &\,& \hspace{40mm} \times \frac{1}{E_m - E_n} \,
   \langle n | \tau_3 | m \rangle \langle m |
   (1 + \gamma^0 \gamma^3) \gamma_5  \delta_n | n \rangle ,
   \nonumber \\
   &+& \,{\langle D_{33} \rangle}_{p \uparrow} \cdot
   \frac{d}{dx} \,\frac{N_c}{4 I} \,
   \sum_{\stackrel{\scriptstyle m \leq 0, n \leq 0}{(E_m = E_n)}}
   \langle n | \tau_3 | m \rangle \langle m |
   (1 + \gamma^0 \gamma^3) \gamma_5  \delta_n | n \rangle ,
   \nonumber \\
   &=& - {\langle D_{33} \rangle}_{p \uparrow} \cdot
   M_N \frac{N_c}{4 I} 
   \left( 2 
   \sum_{\stackrel{\scriptstyle m > 0, n > 0}{(E_m \neq E_n)}} +
   \sum_{m > 0, n \leq 0}  + \sum_{m \leq 0, n > 0}
   \right) \nonumber \\
   &\,& \hspace{40mm} \times \frac{1}{E_m - E_n}
   \, \langle n | \tau_3 | m \rangle \langle m |
   (1 + \gamma^0 \gamma^3) \gamma_5  \delta_n | n \rangle ,
   \nonumber \\
   &\,& \hspace{3mm} 
   - \,{\langle D_{33} \rangle}_{p \uparrow} \cdot \frac{d}{dx}
   \frac{N_c}{4 I} 
   \sum_{\stackrel{\scriptstyle m > 0, n > 0}{(E_m = E_n)}}
   \langle n | \tau_3 | m \rangle \langle m |
   (1 + \gamma^0 \gamma^3) \gamma_5  \delta_n | n \rangle ,
\end{eqnarray}
Comparing (109) and (112), one notices that the $E_m = E_n$ pieces
in the double sums cancels precisely between $B^\prime$ and $C$ terms.
(This is true for both of the occupied and nonoccupied
expressions.) After some manipulation by taking care of this
cancellation, the sum of these two terms can finally be expressed as
\begin{eqnarray}
   [ \Delta u(x) \!\!\! &-& \!\!\! \Delta d(x)]^{(1)}_{B^\prime + C}
   \nonumber \\
   &=& \,\,\,{\langle D_{33} \rangle}_{p \uparrow} \cdot
   M_N \,\frac{N_c}{2 I} \nonumber \\
   &\times& \Bigl\{ 
   \sum_{\stackrel{\scriptstyle m = all, n \leq 0}{(E_m \neq E_n)}}
   \frac{1}{E_m - E_n} \,
   \langle n | \tau_3 | m \rangle \langle m |
   (1 + \gamma^0 \gamma^3) \gamma_5  \delta_n | n \rangle ,
   \nonumber \\
   &-& \frac{1}{2 M_N} \,\frac{d}{dx}
   \sum_{\stackrel{\scriptstyle m = all, n \leq 0}{(E_m \neq E_n)}}
   \langle n | \tau_3 | m \rangle \langle m |
   (1 + \gamma^0 \gamma^3) \gamma_5  \delta_n | n \rangle ,
   \nonumber \\
   &-& \sum_{m > 0, n \leq 0} \frac{1}{E_m - E_n} \,
   \langle n | \tau_3 | m \rangle \langle m |
   (1 + \gamma^0 \gamma^3) \gamma_5
   \frac{\delta_n + \delta_m}{2} | n \rangle \Bigr\}
   \nonumber \\
   &=& - \,{\langle D_{33} \rangle}_{p \uparrow} \cdot
   M_N \,\frac{N_c}{2 I} \nonumber \\
   &\times& \Bigl\{ 
   \sum_{\stackrel{\scriptstyle m = all, n > 0}{(E_m \neq E_n)}}
   \frac{1}{E_m - E_n} \,
   \langle n | \tau_3 | m \rangle \langle m |
   (1 + \gamma^0 \gamma^3) \gamma_5  \delta_n | n \rangle ,
   \nonumber \\
   &-& \frac{1}{2 M_N} \,\frac{d}{dx}
   \sum_{\stackrel{\scriptstyle m = all, n > 0}{(E_m \neq E_n)}}
   \langle n | \tau_3 | m \rangle \langle m |
   (1 + \gamma^0 \gamma^3) \gamma_5  \delta_n | n \rangle ,
   \nonumber \\
   &-& \sum_{m \leq 0, n > 0} \frac{1}{E_m - E_n} \,
   \langle n | \tau_3 | m \rangle \langle m |
   (1 + \gamma^0 \gamma^3) \gamma_5
   \frac{\delta_n + \delta_m}{2} | n \rangle \Bigr\} .
\end{eqnarray}
For numerical calculation, we shall use the first form for
$x > 0$, while the second form for $x < 0$.

\subsection{$\delta u(x) + \delta d(x)$}

Since the evaluation of the transversity distribution can be
done in a completely parallel way as the longitudinally
polarized distribution functions, we shall show below only the
final results. The $O (\Omega^0)$ contribution to
$\delta u(x) + \delta d(x)$ vanishes, i.e.
\begin{equation}
   {[ \delta u(x) + \delta d(x)]}^{(0)} \ = \ 0 .
\end{equation}
The $O (\Omega^1)$ contribution consists of two pieces as
\begin{equation}
   {[ \delta u(x) + \delta d(x)]}^{(1)} \ = \ 
   {[ \delta u(x) + \delta d(x)]}^{(0)}_{\{A, B\}^\prime} \ + \ 
   {[ \delta u(x) + \delta d(x)]}^{(0)}_{C^\prime} ,
\end{equation}
where
\begin{eqnarray}
   [ \delta u(x) \!\!\! &+& \!\!\!
   \delta d(x) ]^{(1)}_{\{ A, B \}^\prime}
   \nonumber \\
   &=& \,\,{\langle J_x \rangle}_{p S_x} \cdot M_N \,\frac{N_c}{2 I} \,
   \sum_{\stackrel{\scriptstyle m = all, n \leq 0}{(E_m \neq E_n)}} 
   \frac{1}{E_m - E_n} 
   \langle n | (\gamma^1 \gamma_5 - i \gamma^2)
   \delta_n | m \rangle
   \langle m | \tau_1 | n \rangle , \nonumber \\
   &=& \!\!\! - \,{\langle J_x \rangle}_{p S_x} \cdot 
   M_N \,\frac{N_c}{2 I}
   \sum_{\stackrel{\scriptstyle m = all, n > 0}{(E_m \neq E_n)}}
   \frac{1}{E_m - E_n} 
   \langle n | (\gamma^1 \gamma_5 - i \gamma^2)
   \delta_n | m \rangle
   \langle m | \tau_1 | n \rangle , \ \ 
\end{eqnarray}
and
\begin{eqnarray}
   &\,& {[ \Delta u(x) + \Delta d(x)]}^{(1)}_{C^\prime} \nonumber \\
   &=& \,\,\,\,\,\,
   {\langle J_x \rangle}_{p S_x} \cdot \frac{d}{d x} \,
   \frac{N_c}{4 I} \,
   \sum_{\stackrel{\scriptstyle m = all, n \leq 0}{(E_m \neq E_n)}}
   \langle n | \tau_1 | m \rangle \langle m |
   (\gamma^1 \gamma_5 - i \gamma^2) \delta_n | n \rangle
   \nonumber \\
   &=& - \,{\langle J_x \rangle}_{p S_x} \cdot \frac{d}{d x} \,
   \frac{N_c}{4 I} 
   \sum_{\stackrel{\scriptstyle m = all, n > 0}{(E_m \neq E_n)}}
   \langle n | \tau_1 | m \rangle \langle m |
   (\gamma^1 \gamma_5 - i \gamma^2) \delta_n | n \rangle
\end{eqnarray}
Here ${\langle J_x \rangle}_{p S_x}$ is defined by
\begin{equation}
   {\langle J_x \rangle}_{p S_x} \ = \ \langle p S_x | J_x |
   p S_x \rangle .
\end{equation}

\subsection{$\delta u(x) - \delta d(x)$}

The $O (\Omega^0)$ contribution is given by
\begin{eqnarray}
   {[ \delta u(x) - \delta d(x)]}^{(0)}
   &=& \,\,\,\,\,{\langle D_{31} \rangle}_{p S_x} \cdot M_N \,N_c \,
   \sum_{n \leq 0} \,\langle n | \tau_3 
   (\gamma^1 \gamma_5 - i \gamma^2) \delta_n | n \rangle \nonumber \\
   &=& \! - \,{\langle D_{31} \rangle}_{p S_x} \cdot M_N \,N_c \,
   \sum_{n > 0} \,\langle n | \tau_3 
   (\gamma^1 \gamma_5 - i \gamma^2) \delta_n | n \rangle .
\end{eqnarray}
The $O (\Omega^1)$ contribution consists of two pieces as
\begin{equation}
   {[ \delta u(x) - \delta d(x) ]}^{(1)} \ = \ 
   {[ \delta u(x) - \delta d(x) ]}^{(1)}_{[A,B]} \ + \ 
   {[ \delta u(x) - \delta d(x) ]}^{(1)}_{B^\prime + C}
\end{equation}
where
\begin{eqnarray}
   &\,& \!\!\! {[ \delta u(x) - \delta d(x)]}^{(1)}_{[A,B]} \nonumber \\
   &=& {\langle D_{31} \rangle}_{p S_x} \cdot M_N \,
   \frac{N_c}{8 I} \,\,i \,\epsilon_{3 c b} \,\sum_{m > 0, n \leq 0}
   \frac{1}{E_m - E_n} \nonumber \\
   &\times& \!\!\!\! [ \langle n | \tau_c | m \rangle \langle m |
   \tau_b (\gamma^1 \gamma_5 \! - \! i \gamma^2)
   (\delta_n + \delta_m) 
   | n \rangle + 
   \langle n | \tau_b
   (\gamma^1 \gamma_5 - i \gamma^2) (\delta_n + \delta_m) 
   | m \rangle \langle m | \tau_c | n \rangle ] \ \ \ 
   \nonumber \\
   &=& \!\! - \,{\langle D_{31} \rangle}_{p S_x} \cdot M_N \,
   \frac{1}{I} \,\frac{N_c}{2} \,\sum_{m > 0, n \leq 0}
   \frac{1}{E_m - E_n} \nonumber \\
   &\times& \!\!\!\!\! [ \langle n | \tau_{+1} | m \rangle \langle n |
   \tau_{+1} (\gamma^1 \gamma_5 \! - \! i \gamma^2)
   \frac{\delta_n + \delta_m}{2}
   | m \rangle - \langle n | \tau_{-1} | m \rangle 
   \langle n | \tau_{-1} (\gamma^1 \gamma_5 - i \gamma^2)
   \frac{\delta_n + \delta_m}{2} | m \rangle ] , \ \ \ 
\end{eqnarray}
and
\begin{eqnarray}
   [ \delta u(x) \!\!\! &-& \!\!\!
   \delta d(x)]^{(1)}_{B^\prime + C} \nonumber \\
   &=& {\langle D_{31} \rangle}_{p S_x} \cdot
   M_N \,\frac{N_c}{2 I} \nonumber \\
   &\times& \Bigl\{ 
   \sum_{\stackrel{\scriptstyle m = all, n \leq 0}{(E_m \neq E_n)}}
   \frac{1}{E_m - E_n} \,
   \langle n | \tau_3 | m \rangle \langle m |
   (\gamma^1 \gamma_5 - i \gamma^2) \delta_n | n \rangle
   \nonumber \\
   &-& \frac{1}{2 M_N} \,\frac{d}{dx}
   \sum_{\stackrel{\scriptstyle m = all, n \leq 0}{(E_m \neq E_n)}}
   \langle n | \tau_3 | m \rangle \langle m |
   (\gamma^1 \gamma_5 - i \gamma^2) \delta_n | n \rangle
   \nonumber \\
   &-& \sum_{m > 0, n \leq 0} \frac{1}{E_m - E_n} \,
   \langle n | \tau_3 | m \rangle \langle m |
   (\gamma^1 \gamma_5 - i \gamma^2)
   \frac{\delta_n + \delta_m}{2} | n \rangle \Bigr\} ,
   \nonumber \\
   &=& - \,{\langle D_{31} \rangle}_{p S_x} \cdot
   M_N \, \frac{N_c}{2 I} \nonumber \\
   &\times& \Bigl\{ 
   \sum_{\stackrel{\scriptstyle m = all, n > 0}{(E_m \neq E_n)}}
   \frac{1}{E_m - E_n} \,
   \langle n | \tau_3 | m \rangle \langle m |
   (\gamma^1 \gamma_5 - i \gamma^2) \delta_n | n \rangle ,
   \nonumber \\
   &-& \frac{1}{2 M_N} \,\frac{d}{dx}
   \sum_{\stackrel{\scriptstyle m = all, n > 0}{(E_m \neq E_n)}}
   \langle n | \tau_3 | m \rangle \langle m |
   (\gamma^1 \gamma_5 - i \gamma^2) \delta_n | n \rangle ,
   \nonumber \\
   &-& \sum_{m \leq 0, n > 0} \frac{1}{E_m - E_n} \,
   \langle n | \tau_3 | m \rangle \langle m |
   (\gamma^1 \gamma_5 - i \gamma^2)
   \frac{\delta_n + \delta_m}{2} | n \rangle \Bigr\} .
\end{eqnarray}

\vspace{4mm}
\section{Numerical results and discussion}

\ \ \ \ \ Before showing the results of numerical calculations, we
briefly discuss the parameters of our effective
model specified by the lagrangian (26).
Fixing $f_{\pi}$ to its physical value , i.e. 
$f_{\pi} = 93 \,\mbox{MeV}$, only one parameters of the model is
the constituent quark mass M, which plays the role of the coupling
constant between the pion and the effective quark 
fields. There is some argument based on the instanton picture of the
QCD vacuum that the value of this mass parameter should not be extremely
far from $350 \,\mbox{MeV}$ [25]. Phenomenological analyses of various static
baryon observables based on this model prefer a slightly larger value of
$M$ between $350 \,\mbox{MeV}$ and $425 \,\mbox{MeV}$ [5,6].
In the present analysis, we use the value $M = 375 \,\mbox{MeV}$ favored
from analyses of various static observables of baryons.
Actually the model contains ultraviolet divergences so that it must be
regularized by introducing some physical cutoff.
In the case of static nucleon observables, most frequently used
regularization scheme is the one based
on Schwinger's proper-time representation [5,6].
Unfortunately, how to generalize this regularization scheme in the evaluation
of nucleon structure functions is an open problem. For evaluating
quark distribution functions, Diakonov et al. then proposed to use
the so-called Pauli-Villars regularization scheme, which they claim
has several nice properties as compared with the energy cutoff scheme
like the proper-time regularization scheme [14].
The basic idea of this regularization scheme is very simple.
Using the derivative (gradient) expansion, one can evaluate the
effective meson action corresponding to the original effective quark
lagrangian (26) as
\begin{eqnarray}
   S_{eff}^M [U] &=& - \,i \,N_c \,\mbox{Sp} [ i \not\!\partial - 
   M e^{i \gamma_5 
   \mbox{\boldmath $\tau$} \cdot \mbox{\boldmath $\pi$}/ f_{\pi}} ]
   \nonumber \\
   &=& \,\,\frac{4 N_c}{f_{\pi}^2} \,I_2 (M) \cdot \frac{1}{2} \,
   ( \partial_{\mu} \mbox{\boldmath $\pi$})^2 \ + \ \cdots \ .
\end{eqnarray}
Here the coefficient of the pion kinetic term given by
\begin{equation}
   I_2 (M) \ \equiv \ i \int \frac{d^4 k}{(2 \pi)^2} 
   \frac{M^2}{(k^2 - M^2)^2},
\end{equation}
contains logarithmic divergence. Clearly, this divergence can be 
removed by introducing a regularized action  $S_{eff}^{reg}$ by 
\begin{equation}
   S_{eff}^{reg} \ \equiv \ S_{eff}^M - 
   {\left( \frac{M}{M_{PV}}\right)}^2 \,S_{eff}^{M_{PV}} .
\end{equation}
Here $S_{eff}^{M_{PV}}$ denotes the effective meson action obtained
from $S_{eff}^M$ by replacing the dynamical quark mass M with the
Pauli-Villars mass $M_{PV}$. In fact, this replaces $I_2 (M)$ with 
\begin{equation}
   I_2^{reg} \ \equiv \ I_2 (M) - 
   {\left( \frac{M}{M_{PV}} \right)}^2 \,I_2 (M_{PV}) 
   \ = \ \frac{M^2}{16 \pi^2} \log \frac{M_{PV}^2}{M^2} ,
\end{equation}
which is clearly finite.
Demanding further that the pion kinetic term in $S_{eff}^{reg}$ has 
the correct normalization, one obtain
\begin{equation}
   \frac{N_c}{4 \pi^2} \,M^2 \,\log \frac{M_{PV}^2}{M^2} = f_{\pi}^2 .
\end{equation}
For $M = 375 \,\mbox{MeV}$, for instance, this gives
$M_{PV} \simeq 562 \,\mbox{MeV}$.
Other observables like quark distribution functions,
which contains logarithmic divergence, can similarly be regularized as 
\begin{equation}
   {\langle O \rangle}^{reg} \ \equiv \ 
   {\langle O \rangle}^M - {\left( \frac{M}{M_{PV}} \right)}^2 
   {\langle O \rangle}^{M_{PV}} .
\end{equation}
For the sake of consistency, a soliton solution should also be obtained
in the same regularization scheme.
The startingpoint of soliton construction is the mean field equation
\begin{equation}
   {\langle \bar{\psi} \psi \rangle}_r^{reg'} \sin F (r) 
   \ = \ {\langle \bar{\psi} \, i \,\gamma_5 \mbox{\boldmath $\tau$} 
   \psi \rangle}_r^{reg'} \cos F(r) ,
\end{equation}
obtained under the assumption of the static hedgehog configuration
\begin{equation}
   \hat{\mbox{\boldmath $\pi$}} (\mbox{\boldmath $r$}) 
   = f_{\pi} \,\hat{\mbox{\boldmath $r$}} \,F (r) .
\end{equation}
Here ${\langle \bar{\psi} \psi \rangle}_r^{reg'}$ and
${\langle \bar{\psi} \, i \,\gamma_5 \mbox{\boldmath $\tau$}
\psi \rangle}_r^{reg'}$ are
the regularized scalar and pseudoscalar densities in
the Pauli-Villars subtraction scheme :
\begin{eqnarray}
   {\langle \bar{\psi} \psi \rangle}_r^{reg'} \ \ &\equiv& \ \ 
   {\langle \bar{\psi} \psi \rangle}_r^M 
   \ - \ \left( \frac{M}{M_{PV}} \right) 
   {\langle \bar{\psi} \psi \rangle}_r^{M_{PV}} \\
   {\langle \bar{\psi} \gamma_5 \mbox{\boldmath $\tau$} 
   \psi \rangle}_r^{reg'}
   &\equiv& {\langle \bar{\psi} i \gamma_5 
   \mbox{\boldmath $\tau$} \psi \rangle}_r^M 
   \ - \ \left( \frac{M}{M_{PV}} \right) {\langle \bar{\psi} i \gamma_5 
   \mbox{\boldmath $\tau$} \psi \rangle}_r^{M_{PV}} .
\end{eqnarray}
Recently, self-consistent solutions of this equation of motion
has been obtained in [26] with use of the Kahana-Ripka basis [27].
(Essentially the same equation was solved in [28] within the  
framework of the Nambu-Jona Lasinio model with an {\it ad hoc}
nonlinear constraint.) 
However, one should use this regularization scheme with some care. 
In fact, it is known that the single subtraction is not enough to get 
rid of linear divergences, for instance, contained in the 
expression of the vacuum quark condensate [28], which implies that 
${<\bar{\psi} \psi>}_r^{reg'}$ and 
${< \bar{\psi} i \gamma_5 \mbox{\boldmath $\tau$} \psi>}_r^{reg'}$  
also contain convergences. Why could the authors of refs.[26,28]
obtain self-consistent solutions then?
The reason is in the way of solving the equation of motion (129) in
the nonlinear model. Given an appropriate initial form of $F(r)$,
one can evaluate ${\langle \bar{\psi} \psi \rangle}_r^{reg'}$ and
${\langle \bar{\psi} \gamma_5 \mbox{\boldmath $\tau$} \psi
\rangle}_r^{reg'}$ by using the Kahana-Ripka plane-wave
basis as long as the box size D and the maximum momentum
$k_{max}$ are finite. 
A new $F(r)$ can then be obtained from
\begin{equation}
   F(r) = \mbox{arctan}
   \left( \frac{ {< \bar{\psi} i \gamma_5 
   \mbox{\boldmath $\tau$} 
   \psi > }_r^{reg'}}{{<\bar{\psi} 
   \psi >}_r^{reg'}}
   \right) .
\end{equation}
As $k_{max}$ increases, both of ${<\bar{\psi} \psi>}_r^{reg'}$ and 
${<\bar{\psi} i \gamma_5 \mbox{\boldmath $\tau$} 
\psi>}_r^{reg'}$ tend to 
diverge. We numerically find that both quantifies increases at 
the same rate as $k_{max}$ increases so that the resultant $F(r)$ 
is quite insensitive to the value of $k_{max}$ for large enough
$k_{max}$. This is the reason why stable soliton solutions could be
found in the above mentioned single-subtraction Pauli-Villars
regularization scheme.
The existence of finite energy soliton could also be inferred from 
the derivative expansion analysis of the nonlinear lagrangian (123)
with vanishing current quark masses. Nevertheless, one
should keep in mind that it is not a completely satisfactory scheme
in the sense that its predictions for some special quantities like
the vacuum quark condensate contain divergences.
For obtaining satisfactory answers also for these special quantities,
the single-substraction Pauli-Villars scheme is not enough.
We found that more sophisticated Pauli-Villars scheme with 
three substraction meets this requirement, and that its
self-consistent solutions are only slightly different from 
those of the naive single-subtraction scheme, except when discussing
some special quantities as pointed  out above. (This analysis
will be reported elsewhere.)
Considering the fact that the calculation of the structure functions
are very time-consuming, we shall then use here the 
single-subtraction Pauli-Villars scheme, keeping in mind that some
particular observables are out of the application of this regularization
scheme. Finally, as for the nucleon mass, we prefer to using the
theoretical soliton mass $M_N \simeq 1102 \,\mbox{MeV}$ rather
than the physical mass, since it respects the energy-momentum
sum rule at the energy scale of the model.

For evaluating quark distribution functions at the $O (\Omega^1)$,
we must perform infinite double sums over all the single-quark
orbitals which are eigenstates of the static Hamiltonian $H$ given
by (31). As far as static nucleon observables, a numerical technique
for carrying out such double sums was established in [5].
On the other hand, several new subtleties arising in the evaluation
of quark distribution functions have been explained in [17].
In the actual numerical calculation, the expression of each
physical quantity is divided into two pieces, i.e. the contribution
of what we call the valence quark level (it is the lowest energy
eigenstate of the static Dirac hamiltonian $H$, which emerges from
the positive energy continuum) and that of the Dirac sea quarks (or
the vacuum polarization contribution) as explained in [17].
The regularization is introduced into the latter part only.

Now we start to show the results of our numerical calculation for
polarized quark distribution functions of the nucleon. 
Shown in Fig.1 are the $O(\Omega^0)$ contributions to the isovector 
longitudinally polarization distribution functions 
$\Delta u (x) - \Delta d (x)$ (solid curves) and 
$\Delta \bar{u} (x) - \Delta \bar{d} (x)$ (dashed curves), 
which was first calculated by Diakonov et al. [14].
Here Fig.1(a) represents the contributions of the discrete valence
quark level, while Fig.1(b) is the vacuum polarization contribution
to the same quantities. Sum of these two contributions are shown
in Fig.1(c).
As shown by Diakonov et al., the $O (\Omega^0)$ vacuum
polarization contributions to $\Delta u(x) - \Delta d(x)$ and 
$\Delta \bar{u} (x) - \Delta \bar{d} (x)$ are fairly large.
(The large and positive longitudinal polarization of the isovector
combination of the antiquark distributions seems to be a
characteristic prediction of
the CQSM, which can in principle be tested by the improved
phenomenological analyses of polarized parton distribution functions
in the near future. More detailed
discussion on this point will be given after finishing the
evaluation of the $O (\Omega^1)$ contribution to the same
distribution function as well as that of the isoscalar
longitudinally polarized distribution functions.)

Next, we show in Fig.2 the $O (\Omega^1)$ contribution to the
same distribution functions $\Delta u(x) - \Delta d(x)$ and
$\Delta \bar{u} (x) - \Delta \bar{d} (x)$.
Fig.2(a), 2(b) and 2(c) respectively stand for the
$O (\Omega^1)$ contributions of the discrete valence quark level, 
those of the Dirac sea quarks (or the vacuum polarization
contributions), and their sums.
One sees that the  $O (\Omega^1)$ contributions to the isovector 
longitudinally polarized distribution function are far from
negligible as compared with the leading $O (\Omega^0)$ contributions.
This could be expected since the first moment of this
distribution functions gives the isovector axial coupling constant
of the nucleon.
\begin{equation}
   g_A^{(3)} \ = \  \int_0^1 \,\{ [ \Delta u(x) - \Delta d(x)] 
   \ + \ [ \Delta \bar{u}(x) - \Delta \bar{d}(x)] \} \,d x ,
\end{equation}
while we already know from the previous analyses that the
$O (\Omega^1)$ contribution to $g_A^{(3)}$ is large enough to resolve
the longstanding $g_A$ problem in the hedgehog soliton model [29,30,24].
Adding this $O(\Omega^1)$ contribution to the leading $O(\Omega^0)$
contribution, we obtain final answers for
$\Delta u(x) - \Delta d(x)$
and $\Delta \bar{u} (x) - \Delta \bar{d} (x)$, which will be shown 
later together with the final answer for the isoscalar longitudinal 
distribution functions $\Delta u(x) + \Delta d(x)$ and
$\Delta \bar{u} (x) + \Delta \bar{d}(x)$.
Before showing those, we give in Fig.3 the result for
the $O (\Omega^1)$ contributions to the isoscalar longitudinally 
polarized distribution functions, $\Delta u (x) + \Delta d(x)$
and $\Delta \bar{u}(x) + \Delta \bar{d}(x)$. (We recall that 
there is no $O (\Omega^0)$ contribution to these distribution functions.)
Fig.3(a), 3(b) and 3(c) respectively stand for the contributions of 
the discrete valence quark level, those of the vacuum polarization
contributions, and their sums. One sees that the vacuum polarization
contributions to the distribution functions $\Delta u (x) + 
\Delta d (x)$ and $\Delta \bar{u} (x) + \Delta \bar{d} (x)$ are
much smaller than those of the corresponding isovector distributions
$\Delta u (x) - \Delta d (x)$ and $\Delta \bar{u} (x) - 
\Delta \bar{d} (x)$. Now we show in Fig.4 the final answers for
$\Delta u(x) - \Delta d(x)$ and $\Delta \bar{u}(x) - \Delta \bar{d}(x)$,
which are the sums of the $O (\Omega^0)$ and $O (\Omega^1)$
contributions, in comparison with
the final answers for $\Delta u(x) + \Delta d(x)$ and
$\Delta \bar{u}(x) + \Delta \bar{d}(x)$ arising from
the $O (\Omega^1)$ terms alone.
We observe quite a big difference between the isovector distributions 
and the isoscalar one. The overall magnitude of $\Delta u (x) + 
\Delta d (x)$ is much smaller than that of $\Delta u (x) - \Delta d (x)$,
which denotes that $u$-quark is positively polarized, while the $d$-quark
is negatively polarized to the direction of proton spin.

At this stage, it may be interesting to compare our theoretical
predictions for the longitudinally polarized quark distribution
functions with some of the semi-phenomenological parametrization.
The parametrization given by Gl\"{u}ck, Reya, Stratmann and Vogelsang is
especially convenient for the purpose of handy comparison [31], since the
normalization point ($Q_{init}^2 \simeq 0.34 \,\mbox{GeV}^2$) of
their parametrization is fairly close to the energy scale of our
effective quark model ($M_{PV}^2 \simeq 0.32 \,\mbox{GeV}^2$.
Fig.5 shows this comparison.
The filled squares in Fig.5(a) and Fig.5(b) stand for the GRSV
parametrizations for the quark distribution functions
$x ( \Delta u(x) + \Delta \bar{u} (x) + \Delta d(x) + 
\Delta \bar{d} (x))$ and $x ( \Delta u(x) + \Delta \bar{u} (x) - 
\Delta d(x) - \Delta \bar{d} (x))$, respectively.
Of the two theoretical curves in
each figure, the solid curve is the answer of the present
calculation, whereas the dashed curve is obtained by using
the old treatment used in [17], which amounts to dropping
some of the nonlocality effects in time.
One observes that the nonlocality corrections newly introduced in
the present analysis are quite important especially for the
isoscalar distribution $x ( \Delta u(x) + \Delta \bar{u} (x) + 
\Delta d(x) + \Delta \bar{d} (x))$, while it is less important for the
isovector distribution $x ( \Delta u(x) + \Delta \bar{u} (x) - 
\Delta d(x) - \Delta \bar{d} (x))$. (This is probably because the
nonlocality correctios appearing at the $O (\Omega^1)$ are masked
by the dominant $O (\Omega^0)$ contribution in the case of
isovector polarized distribution functions.)
By comparing the two theoretical curves
for $x ( \Delta u(x) + \Delta \bar{u} (x) + \Delta d(x) + \Delta
\bar{d} (x))$ with the corresponding GRSV parametrization, one
finds that the new treatment leads to a better agreement.
Especially impressive is that the new treatment reproduces the
negative sign of the GRSV distribution function in the smaller $x$
region, although one should not forget the fact that the GRSV
parametrizations are not experimental data themselves.
We point out that the most important factor leading to this
qualitative difference between the old and new treatments of the quark
distribution functions is the nonlocality correction arising from
the second term of (38), i.e. the proper account of nonlocality in
time of the operator $A^\dagger (0) O_a A(z_0)$.

Turning back to Fig.4, let us inspect the theoretical predictions
for the antiquark distributions in more detail. An interesting
feature is that, in most region of $x$, $\Delta \bar{u} (x) - 
\Delta \bar{d} (x) > 0$ and $\Delta \bar{u} (x) + 
\Delta \bar{d} (x) < 0$ with the relation
$| \Delta \bar{u} (x) - \Delta \bar{d} (x) | \gg 
| \Delta \bar{u} (x) + \Delta \bar{d} (x) |$. This denotes that
$\bar{d}$ is strongly polarized in the opposite direction to
the proton spin, while $\bar{u}$ is weakly polarized in the
same direction to the proton spin. This appears to be a prominent
prediction of the CQSM, which is worthy of special mention.
In fact, it sharply contradicts the assumption of SU(2) symmetric
sea quark polarization $\Delta \bar{u} (x) = \Delta \bar{d} (x)$,
which is frequently used in semi-phenomenological analyses of
parton distributions.
The isospin symmetric polarization is also assumed in the analysis
by Gl\"{u}ck, Reya, Stirling and Vogt [31]. We compare in Fig.6
our prediction for the $x \Delta \bar{u} (x)$ and 
$x \Delta \bar{d} (x)$ with the GRSV parametrization, which assumes
that $x \Delta \bar{u} (x) = x \Delta \bar{d} (x) \,
( \equiv x \Delta \bar{q} (x) )$. Naturally, one finds
qualitative difference between the theoretical distributions and
the GRSV parametrization. Still, it is interesting to see that
the average of the two theoretical distributions $x \Delta \bar{u} (x)$
and $x \Delta \bar{d} (x)$ is not extremely different from the
corresponding GRSV parametrization $x \Delta \bar{q} (x)$.
As for the unpolarized distribution functions,
the breakdown of the assumption of SU(2) symmetric sea has already
been confirmed by the NMC measurement [16]. By the same token, there
is no compelling reason to believe that the spin dependent
antiquark (sea quark) distributions are isospin symmetric.
In fact, our previous analyses based on the same model shows that
the isospin asymmetry of the unpolarized sea quark
distributions can be explained very naturally as
combined effects of two ingredients, i.e. the apparently
existing flavor asymmetry of valence quark numbers in the nucleon
and the spontaneous chiral symmetry breaking of QCD vacuum [17-19].
It is just the same mechanism that is responsible for the
opposite longitudinal polarization of the $\bar{u}$ and $\bar{d}$
quarks.

The above-mentioned fairly big difference between the isovector
and isoscalar longitudinally polarized distribution functions
manifests itself also in their first moments, i.e. the isovector
and isoscalar axial charges given as
\begin{eqnarray}
   g_A^{(3)} &=& \int_0^1 \,\{ [ \Delta u(x) - \Delta d(x)] 
   \ + \ [ \Delta \bar{u}(x) - \Delta \bar{d}(x)] \} \,d x 
   \ \simeq \ 1.41 ,\\
   g_A^{(0)} &=& \int_0^1 \,\{ [ \Delta u(x) + \Delta d(x) ] 
   \ + \ [ \Delta \bar{u}(x) + \Delta \bar{d}(x)] \} \,d x
   \ \simeq \ 0.35 .
\end{eqnarray}
The resultant large isovector axial charge and small isoscalar 
(flaver-singlet) one seem to be qualitatively consistent
with the observation.
Especially interesting here is the flavor-singlet axial charge
identified with the quark spin content of the nucleon.
In the context of the CQSM, this quantity was first investigated
in [5] with use of the self-consistent soliton solution obtained
in the proper-time regularization scheme. The value of
$g_A^{(0)} = \langle \Sigma_3 \rangle$ obtained there ranges from
$0.4 \sim 0.5$ corresponding to the variation of the
dynamical quark mass $M$ from $425 \,\mbox{MeV}$ to $375 \,
\mbox{MeV}$. One may notice that the value $g_A^{(0)} \simeq
0.35$ obtained in the present calculation is a little smaller
than the previous one.
The cause of this difference can be traced back to the qualitative
change of the self-consistent soliton solution obtained in the new
regularization scheme. As a general trend, the Pauli-Villars
regularization scheme cuts off high momentum components
more weakly than the energy-cutoff scheme like the proper-time
one, thereby leading to soliton solutions with stronger
distortion. Incidentally, owing to the nucleon spin sum rule
$\langle L_3 \rangle + \frac{1}{2} \langle \Sigma_3 \rangle
= \frac{1}{2}$ proved in [5], the rest of the nucleon spin is
carried by the orbital angular momentum of the effective quark
fields. (Naturally, this is true only at low $Q^2$ corresponding
to the energy scale of our effective model. It will be shown
later that an increasing portion of the nucleon spin is
carried by gluons as $Q^2$ increases.) A soliton with stronger
distortion gives larger orbital angular momentum, and consequently
smaller quark spin fraction [5].

The characteristic feature of the above theoretical prediction,
i.e. larger isovector charge and smaller isoscalar one seems
also consistent with the idea of $N_c$ counting
or $1 / N_c$ expansion of QCD. For understanding it, we just recall
the fact that the collective angular velocity $\Omega$ scales as
$1 /N_c$, so that the leading contributions to the isovector and
isoscalar polarized distribution functions are respectively of
the $O (N_c^1)$ and $O (N_c^0)$.
The detailed comparison of the theoretical first moments with the
corresponding experimental data will be given later
after taking account of the scale dependence of them.

Now we show the results of our numerical calculation for
transversity distributions. Fig.7 shows the $O(\Omega^0)$
contributions to the isovector transversity distribution
functions $\delta u(x) - \delta d(x)$ (solid curves) and 
$\delta \bar{u}(x) - \delta \bar{d}(x)$ (dashed curves).
Here Fig.7(a) stands for the contributions of the discrete valence 
level, while 7(b) is the vacuum polarization contributions to the 
same quantities. The sum of these two contributions are shown
in Fig.7(c).
One finds that the vacuum polarization contributions to these 
distribution functions are insignificant.
Next, we show in Fig.8 the $O (\Omega^1)$ contributions to the same 
transversity distribution functions $\delta u(x) - \delta d(x)$ and 
$\delta \bar{u}(x) - \delta \bar{d}(x)$.
Also for these $O (\Omega^1)$ terms, the
vacuum polarization contributions are very small as compared with 
the contributions of  the discrete valence quark level.
However, we emphasized that the valence level contribution at the 
$O (\Omega^1)$ is far from small as compared with the leading
$O (\Omega^0)$ contributions, and should not be discarded.
Shown in Fig.9 are the theoretical isoscalar transversity
distributions resulting at the $O (\Omega^1)$.
One sees that vacuum polarization contributions to the isoscalar 
transversity distributions is also rather small.
The final predictions of the CQSM for $\delta u(x) - \delta d(x)$ and 
$\delta \bar{u}(x) - \delta \bar{d}(x)$, which are the sums of the 
$O(\Omega^0)$ and $O(\Omega^1)$ contributions are shown in Fig.10,
in comparison with the final answer for $\delta u(x) + \delta d(x)$
and $\delta \bar{u}(x) + \delta \bar{d}(x)$ arising from
$O (\Omega^1)$ terms.
One again sees that the magnitudes of the isoscalar distributions
are much smaller than those of the isovector distributions. 
Remember the similar observation made for the longitudinally
polarized distribution functions. To see it in more detail, 
we find that the ratio of the isoscalar to isovector distribution
are much smaller for the longitudinally polarized distribution than
for the transversity one.
We shall come back later to this point when discussing 
the corresponding first moments of these spin dependent quark
distribution functions.

Roughly speaking, the quark distribution functions evaluated here
corresponds to the energy scale of the order of the Pauli-Villars cutoff
mass  $M_{PV} \simeq 0.56 \,\mbox{GeV}$. The $Q^2$ evolution must be taken
into account in some way before comparing them with the observed nucleon
structure functions at high $Q^2$. Recently, Saga group provided a
Fortran program, which gives numerical solution of DGLAP
evolution equations at the next-to-leading order (NLO) 
for the polarized as well as unpolarized structure functions 
of the nucleon [32-34].  We shall make use of their Fortran programs 
to evaluate the polarized distribution functions at large $Q^2$ [33,34].
The question here is what value we should take for the 
initial energy scale of this $Q^2$ evolution. 
Since the use of perturbative QCD below $1 \,\mbox{GeV}$
is anyhow questionable, one may take this initial energy scale
$Q_{init}^2$ as an adjustable parameter, 
which would be fixed by adjusting the observed structure functions
at high energy region.        
Here we have tried to see the effect of variation of $Q_{init}^2$ in
a small range of $Q^2$ around the model energy scale of $M_{PV}^2 
\simeq (0.56 \,\mbox{GeV})^2$. The value $Q_{init}^2 = 
{(0.5 \,\mbox{GeV})}^2 = 0.25 \,\mbox{GeV}^2$
obtained from this analysis will be used throughout the following
investigation. Before showing the results of $Q^2$ evolution, we want to
make a short comment. One notices from the figures 
given so far, the distribution functions evaluated in our effective
model have unphysical tails beyond $x > 1$, although they are not
so significant. These unphysical tails of the theoretical distribution
functions come from an approximate nature of our treatment of the soliton
center-of-motion (as well as the collective rotational motion), 
which is essentially nonrelativistic. A simple procedure to 
remedy this defect was proposed by Jaffe based on the $(1 + 1)$
dimensional bag model [35] and recently reinvestigated by Gamberg et al. 
within the context of the NJL soliton model [36].
According to the latter authors [36], the effect of Lorentz contraction 
can effectively be taken into account by first evaluating the
distribution functions in the soliton rest frame (as we are doing here)
and then by using a simple analytical transformation that preserves
first moments of distribution functions,
as far as the $O (\Omega^0)$ contributions to the distribution
functions are concerned. Such a simple relation may not be expected 
however if we consider the rotational motion of the soliton, which 
are anyhow three dimensional. 
In fact, a comparison with the corresponding phenomenological
distribution functions seems to indicate that the above procedure based on
the $(1+1)$ dimensional dynamics tends to overestimate the effect of
Lorentz contraction. In the present investigation, we therefore
decided not to use their procedure. Still we want distribution functions
which vanish outside the range $0 < x < 1$ so that we can use the
$Q^2$-evolution Fortran program provided by Saga group [33,34].
Since the unphysical tails
of our theoretical distributions are rather small in magnitude, we
are to use a simple cutoff procedure as follows. That is, we
obtain modified distribution functions, which can be used as
input distributions of the above Fortran program, from the original
theoretical distribution functions by multiplying
the $x$-dependent cutoff factor $(1 - x^{10})$. (This special
cutoff factor is invented from the requirement that only the tails
of the distribution functions are modified.)
Fig.11 illustrates the effect of this tentative cutoff procedure.
The solid curve here is the theoretical distribution
function $\Delta u(x) - \Delta d(x)$ given as a sum of the
$O (\Omega^0)$ and $O (\Omega^1)$ contributions. We point out that
this distribution function is the worst case in the sense that the
tail beyond $x = 1$ is most significant as compared with the other
distributions. The dashed curve in the same figure is obtained by
using the above cutoff procedure. One sees that it leaves
the distribution function for $x \leq 0.7$ almost
intact. Naturally, this cutoff procedure alters the values of
integrals of the distribution functions, i.e. the first moments.
However, it turns out that the reduction is less than
$2 \,\%$ even in the above worst case. We therefore expect that the
tentative nature of the above procedure hardly affects
the following qualitative analyses of scale dependence of the quark
distribution functions.

For the sake of comparison, we have carried out a similar 
evolution procedure also for the initial distributions given 
by the MIT bag model. The distribution functions of the (naive) 
MIT bag model are already known and they are given analytically
as follows [22]. The isoscalar longitudinally polarized distribution
functions is given by
\begin{eqnarray}
   \Delta u(x) + \Delta d(x) &=&
   \frac{(M_N R) \,\omega_1}{2 \pi (\omega_1 - 1) j_0^2 (\omega_1)} \,
   \biggl\{ \int_{y_{min}}^{\infty} d y y \,
   \Bigl[\, t_0^2 (\omega_1, y) \nonumber \\
   &+& 2 \,t_0 (\omega_1, y) \,t_1 (\omega_1, y) 
   \left( \frac{y_{min}}{y} \right)
   + t_1^2 (\omega_1, y) \left( 2 
   {\left( \frac{y_{min}}{y} \right)}^2 - 1 \right) 
   \Bigr] \biggr\} ,
\end{eqnarray}
whereas the isoscalar transversity distribution functions is given as
\begin{eqnarray}
   \delta u(x) + \delta d(x) &=&
   \frac{(M_NR) \,\omega_1}{2 \pi (\omega_1 - 1) j_0^2 (\omega_1)}
   \biggl\{ \int_{y_{min}}^{\infty} dy y \,
   \Bigl[ t_0^2 (\omega_1, y) \nonumber \\
   &+& 2 \,t_0 (\omega_1, y) \,t_1 \left( \omega_1, y \right)
   (\frac{y_{min}}{y})
   + t_1^2 \left( \omega_1, y \right)
   (\frac{y_{min}}{y})^2 \,\Bigr] \biggr\} .
\end{eqnarray}
On the other hand, the isovector distribution functions are 
simply related to the isoscalar ones as 
\begin{eqnarray}
   \Delta u(x) - \Delta d(x) &=& \frac{5}{3} \,\,
   [\, \Delta u(x) + \Delta d(x) \,] , \\
   \delta u(x) - \delta d(x) &=& \frac{5}{3} \,\,
   [\,\delta u(x) + \delta d(x) \,] .
\end{eqnarray}
In eqs. (137) and (138), $M_N$ and $R$ respectively stard for the
nucleon mass and the bag radius, while $\omega_n$ is the nth root of
the bag eigenvalue equation as 
\begin{equation}
   \tan \omega_n = - \,\frac{\omega_n}{\omega_n - 1} ,
\end{equation}
and $y_{min} = | x M_N R - \omega_1|$. The function $t_l (\omega_n,y)$
is defined by
\begin{equation}
   t_l (\omega_n, y) = \int_0^1 \,j_l (u \omega_n) j_l (u y) 
   \,u^2 du .
\end{equation}
The bag radius $R$ is only one free parameter of this simple model.
In the numerical calculation, we adopt the value used 
by Jaffe and Ji [22], i.e.
\begin{equation}
M_N R = 4.0 \,\omega_1 ,
\end{equation}
where $\omega_1 \simeq 2.043$ is the lowest (dimensionless) eigenvalue
of the bag equation.

To get a rough idea about the scale dependence, we show in Fig.12
and Fig.13 the theoretical polarized quark distribution functions
before and after $Q^2$-evolution.
Here $\Delta u(x)$ and $\delta u(x)$ in Fig.12(a) respectively stand
for the longitudinal and transversity distributions for $u$-quark.
In our model, the difference between the two distributions are sizable
even at the initial low energy scale. A comparison with the existing
and yet-to-be-obtained high energy data must be done with care, since
the way of evolution of these two distributions are pretty different
and the deference between the two becomes larger and larger as
$Q^2$ increases. A general trend is a rapid growth of small $x$
component of the longitudinally polarized distribution due to the
coupling with gluons. A similar tendency is also observed for the
corresponding $d$-quark distributions shown in Fig.12(b).
We can also give some predictions for the polarized antiquark
distribution functions. As one can see in Fig.13, even the signs
are different for the longitudinal and transversity distributions.
(This is the case for both of $\bar{u}$ and $\bar{d}$ quarks.)
The twist-2 spin dependent distribution functions were calculated
by several authors based on various effective models of baryons
[38,39,12]. As for the polarized quark distribution functions,
the predictions of various models give more or less similar shape
of distributions assuming that they take account of the dominant
nature of the valence quark contribution as well as the effects of
pion cloud in some effective way. The situation is quite different
for the polarized {\it antiquark} distributions. The transversity
distribution functions for the antiquarks have, for instance, been
evaluated by Barone et al. within the chiral chromodielectric
model [38]. Comparing their predictions for $\delta \bar{u} (x)$ and
$\delta \bar{d} (x)$ with ours shown in Fig.13, we find that
their model gives $\delta \bar{u} (x) > 0$, while ours does
$\delta \bar{u} (x) < 0$. The shapes of $\delta \bar{u} (x)$ and
$\delta \bar{d} (x)$ are also quite different in both
models. In consideration of the fact that
the polarized antiquark distributions are quite
sensitive to the detailed dynamics of the model, it is very
important to get precise phenomenological information for them.

Next we show in Fig.14(a) the theoretical predictions for the proton 
structure function $g_1^p (x,Q^2)$ at $Q^2 = 5 \,\mbox{GeV}^2$ in
comparison with the corresponding experimental data given by
E143 collaboration [40]. The theoretical curves 
are obtained as follows. Starting with the initial distributions
$\Delta u(x) + \Delta d(x), \Delta \bar{u}(x) + \Delta \bar{d}(x)$
and $\Delta u(x) - \Delta d(x), \Delta \bar{u}(x) - 
\Delta \bar{d}(x)$ or equivalently $\Delta u(x), \Delta \bar{u}(x)$
and $\Delta d(x), \Delta \bar{d}(x)$ given at
$Q_{init}^2 \simeq 0.25 \,\mbox{GeV}^2$
(we assume $\Delta s(x) = \Delta \bar{s}(x) = 0$ and $\Delta g(x) = 0$
at this energy scale), we solve the NLO evolution equation to obtain
the distribution functions at $Q^2 = 5 \,\mbox{GeV}^2$.
These distribution functions are then convoluted with the relevant 
quark and gluon coefficient functions at the NLO within the framework 
of perturbative QCD. These procedures have been carried out for 
the initial distribution given by the CQSM and also by the MIT bag model.
The solid and dashed curves in Fig.14(a) respectively stand for the 
prediction of the CQSM and that of the MIT bag model. 
A remarkable feature of the CQSM as compared with the MIT bag model 
is the enhancement of the structure function at small $x$ region, i.e. 
large sea quark components.
One also observes that a clear peak of $g_1^p (x,Q^2)$ around 
$x \simeq 0.3$ predicted by the MIT bag model (a relativistic valence
quark model) is not seen in the experimental structure function.
On the other hand, one can say that the prediction of the CQSM 
reproduces qualitative feature of the observed structure function 
in the whole range of $x$.
Fig.14(b) shows the theoretical prediction of the CQSM (solid curve) 
and that of the MIT bag model (dashed curve) for the neutron
spin structure function $g_1^n (x,Q^2)$ in comparison 
with the E154 data [41]. One clearly sees that the neutron spin 
structure function $g_1^n (x,Q^2)$ predicted by the MIT bag model
is negligibly small in magnitude even after evolution.
We recall that at the 
initial energy scale the naive MIT bag model predict $g_1^n (x) = 0$,
which is a necessary consequence of a model that does not properly
incorporate chiral symmetry.
On the other hand, the prediction of the CQSM for $g_1^n (x,Q^2)$ is
seen to be large and negative especially in the small $x$ region
in good agreement with the experimental observation.
Then, this agreement may be regarded as a
manifestation of the importance 
of chiral symmetry in the physics of high-energy deep-inelastic 
scattering.

  As is widely known, the simplest but the most important quantities 
characterizing the quark distribution functions are the associated 
first moments. Here we are interested in the first moments of the 
longitudinally polarized distribution functions and of the 
transversity ones, which are respectively called the axial and 
tensor charges defined as 
\begin{eqnarray}
   g_A^{(3)} &=& \int_0^1 \,\{ [ \Delta u(x) - \Delta d(x) ] 
   + [ \Delta \bar{u}(x) - \Delta \bar{d}(x) ] \} \,dx ,\\ 
   g_A^{(0)} &=& \int_0^1 \,\{ [ \Delta u(x) + \Delta d(x) ]
   + [ \Delta \bar{u}(x) + \Delta \bar{d}(x) ] \} \,dx ,\\
   g_T^{(3)} &=& \int_0^1 \,\{ [ \delta u(x) - \delta d(x) ]
   - [ \delta \bar{u}(x) - \delta \bar{d}(x) ] \} \,dx ,\\
   g_T^{(0)} &=& \int_0^1 \,\{ [ \delta u(x) + \delta d(x) ]
   - [ \delta \bar{u}(x) + \delta \bar{d}(x) ] \} \,dx .
\end{eqnarray}
Before discussing the prediction of the CQSM for these quantities,
it may be instructive to remember some basic properties of those.
(We recall that the first calculation of the tensor charge
in the CQSM was given in [42].)
As emphasized by Jaffe and Ji [22], there is a remarkable 
difference between the axial and tensor charges originating from 
the charge conjugation properties of the relevant operators. 
For each flavor, the tensor charge counts the number of valence
quarks (quarks {\it minus} antiquarks) of opposite transversity.
Consequently, the sea quarks do not contribute to the tensor charge.
(This does not necessarily means vanishing transverse
polarization of antiquarks, however.) On the other hand, the axial
charge counts the number of quarks {\it plus} antiquarks of opposite
helicity. In fact, by rewriting (146) and (147) as
\begin{eqnarray}
   g_A^{(3)} &=& \int_0^1 \,\{ [ \Delta u(x) - \Delta d(x) ] 
   - [ \Delta \bar{u}(x) - \Delta \bar{d}(x) ] \} \,d x 
   + 2 \int_0^1 [ \Delta \bar{u}(x) - \Delta \bar{d}(x) ] \,d x ,
   \ \ \\
   g_A^{(0)} &=& \int_0^1 \,\{ [ \Delta u(x) + \Delta d(x) ] 
   - [ \Delta \bar{u}(x) + \Delta \bar{d}(x) ] \} \,d x 
   + 2 \int_0^1 [ \Delta \bar{u}(x) + \Delta \bar{d}(x) ] \,d x ,
\end{eqnarray}
the first and the second terms of the above equation can respectively 
be interpreted as valence and sea quark contributions in the parton model.
Since the sea quark degrees of freedom is absent in the nonrelativistic
framework, the difference between the axial and tensor charges is 
purely relativistic. Still, one must clearly distinguish two types 
of relativistic effect. The one is dynamical effects, which generate 
sea quark polarization.  The other is kinematical effects, which
make a difference between the axial and tensor charges even though 
the sea quark degrees of freedom are totally neglected. The existence
of this latter effect can readily be convinced by comparing 
the prediction of two ``valence quark models'', i.e. 
the nonrelativistic (constituent) quark model and the MIT bag model.
In fact, the non-relativistic quark model predicts
\begin{eqnarray}
   g_A^{(3)} &=& g_T^{(3)} \ \ = \ \ \frac{5}{3}, \\
   g_A^{(0)} &=& g_T^{(0)} \ \ = \ \ 1,
\end{eqnarray}
while the prediction of the MIT bag model is given by
\begin{eqnarray}
   g_A^{(3)} &=& \frac{5}{3} \cdot 
   \int (f^2 - \frac{1}{3} g^2) \,r^2 d r,   
   \ \ \ \ \ \ g_T^{(3)} = \frac{5}{3} \cdot 
   \int (f^2 + \frac{1}{3} g^2 ) \,r^2 d r ,\\
   g_A^{(0)} &=& 1 \cdot \int (f^2 - \frac{1}{3} g^2) \,r^2 d r, 
   \ \ \ \ \ \ \,\,g_T^{(0)} = 1 \cdot 
   \int (f^2 + \frac{1}{3} g^2 ) \,r^2 d r ,
\end{eqnarray}
where $f$ and $g$ are upper and lower components of the lowest energy 
quark wave functions. For a typical bag radius 
$R \simeq 4.0 \,\omega_1 / M_N$, which was used before, this gives
\begin{eqnarray}
   g_A^{(3)} &\simeq& 1.06, \ \ \ \ \ g_T^{(3)} \ \simeq \ 1.34 ,
   \\
   g_A^{(0)} &\simeq& 0.64, \ \ \ \ \ g_T^{(0)} \ \simeq \ 0.84 .
\end{eqnarray}
As is obvious from eqs. (152) and (153), the splittings of the axial
and tensor charges are due to the different sign of the lower
component (p-wave) contributions [22].
One should however notice that there is one interesting 
feature shared by both of the nonrelativistic quark model and 
the MIT bag model. The predictions of the both models for the ratio 
of the isoscalar to isovector axial charges as well as the ratio 
of the isoscalar to isovector tensor charges are just the same :
\begin{equation}
   g_A^{(0)} / g_A^{(3)} \ = \ g_T^{(0)} / g_T^{(3)} \ = \ 3/5 .
\end{equation}
Although there is no experimental information yet for the tensor 
charges, the above prediction for the ratio of the two axial charges 
obviously contradicts the EMC observation. 
\begin{table}[h]
\vspace{5mm}
\caption{The theoretical predictions for the isovector and
isoscalar axial charges as well as the corresponding tesnsor charges.
The predictions of the MIT bag model and those of the lattice QCD [43]
are also shown together with some experimental data [44,45].}
\vspace{2mm}
\newcommand{\lw}[1]{\smash{\lower2.ex\hbox{#1}}}
\begin{center}
\begin{tabular}{|c|c|c|c|c|} \hline
 \, & \ CQSM \ & \ MIT-bag \ & $\mbox{Lattice QCD}$ [43] & 
 Experiment \\ \hline\hline
 \smash{\lower2.ex\hbox{$g_A^{(3)}$}} & \smash{\lower2.ex\hbox{1.41}} & \smash{\lower2.ex\hbox{1.06}} & \smash{\lower2.ex\hbox{0.99}} &
 1.254 $\pm$ 0.006 [44] \\
 & & & & ($Q^2$-indep.) \\ \hline
 \smash{\lower2.ex\hbox{$g_A^{(0)}$}} & \smash{\lower2.ex\hbox{0.35}} & \smash{\lower2.ex\hbox{0.64}} & \smash{\lower2.ex\hbox{0.18}} &
 0.31 $\pm$ 0.07 [45] \\
 & & & & ($Q^2$ = 10 \,$\mbox{GeV}^2$) \\ \hline\hline
 $g_T^{(3)}$ & 1.22 & 1.34 & 1.07 & -- \\ \hline
 $g_T^{(0)}$ & 0.56 & 0.80 & 0.56 & -- \\ \hline\hline
 $g_A^{(0)} / g_A^{(3)}$ & 0.25 & 0.60 & 0.18 & 0.24
 \\ \hline
 $g_T^{(0)} / g_T^{(3)}$ & 0.46 & 0.60 & 0.52 & -- \\ \hline
\end{tabular}
\end{center}
\end{table}
Now we shall argue that the 
above prediction may be interpreted as showing the limitation of
simple valence quark models, which fail to properly incorporate chiral
symmetry of QCD. To convince it, the predictions of the NRQM and 
the MIT bag model are compared with those of the CQSM in table 1,
which maximally incorporate chiral symmetry. For the sake of reference,
the predictions of the lattice QCD are also shown [43]. (Here we
have omitted the errors of the lattice QCD calculation, for simplicity.)
We first point out that the predictions of the CQSM for the above
ratios, i.e. 
\begin{equation}
   g_A^{(0)} / g_A^{(3)} \ \simeq \ 0.25 , \ \ \ \ \ 
   g_T^{(0)} / g_T^{(3)} \ \simeq \ 0.46 ,
\end{equation}
strongly deviate from the above predictions of the two valence quark 
models. What is remarkable here is that the CQSM predicts very small 
isoscalar axial charge in consistent with the EMC observation.
(More meaningful comparison should be made after taking account of 
the scale dependence of this quantity.) Its prediction for the 
isovector axial charge is also qualitatively consistent with the 
experimental value determined from the neutron beta decay. 
(The deviation from the experimental value is only about 11 $\%$.)
The lattice gauge theory also predicts a very small isoscalar axial 
charge $g_A^{(0)} \simeq 0.18$. However, this prediction may not be
taken as a final one since it largely underestimates the isovector 
axial charge. At any rate, one can observe qualitative similarities 
between the predictions of the CQSM and those of the lattice QCD.
Both predicts quite a small number for the ratio of the isoscalar 
to isovector axial charges as compared with the prediction 
$g_A^{(0)} / g_A^{(3)} = 0.6$ of the NRQM or the MIT bag model.
On the other hand, the predictions of both models for the ratio 
of the isoscalar to isovector charges is not extremely different from 
the prediction $g_T^{(0)} / g_T^{(3)} = 0.6$ of the latter valence 
quark models.
In our opinion, the observed deviation from the valence quark picture
indicates an importance of chiral symmetry as
a generator of ``dynamical sea quark effect'', and the predicted
feature is expected to be confirmed by future measurements of
tensor charges.

  To compare the theoretical first moments of the spin distribution 
functions with the existing data for the longitudinal case and with 
yet-to-be-observed ones for the transversity case, we must take
account of the scale dependence of the relevant moments.
As is well-known, the first moment of
the isovector longitudinal distribution functions, i.e. the isovector 
axial charge is scale independent, i.e. it does not evolve : 
$g_A^{(3)} (Q^2) = g_A^{(3)} (Q_{init}^2)$. This is due to the
conservation of the flavor nonsinglet axial-vector current [51].
This is not generally 
the case for the flavor singlet (isoscalar) axial charge owing to the 
so-called axial anomaly of QCD [46,47].
(Still, one can take a scheme called the
chiral invariant factorization scheme in which the flavor singlet
axial charge is independent of $Q^2$ [48]. Here, we take more
standard gauge invariant factorization scheme [49].)
In the singlet sector, the nth moments of the longitudinally polarized 
distribution functions are coupled with the corresponding gluon 
contributions. The evolution of these nth moments is governed by the 
anomalous dimension matrix
\begin{equation}
   \gamma^{(p)n} \equiv 
   \left(\begin{array}{cc}
   \gamma_{qq}^{(p)n} & \gamma_{qg}^{(p)n} \\
   \gamma_{gq}^{(p)n} & \gamma_{gg}^{(p)n}
   \end{array} \right) .
\end{equation}
where $\gamma^{(0)}$ and $\gamma^{(1)}$ are $1$- and $2$-loop 
contributions to the anomalous dimensions. An analytic solution to this
coupled evolution equation of the NLO is given in the matrix
form [50,51] :
\begin{equation}
   \Gamma^n (Q^2) = 
   \left(
   \begin{array}{c}
   \Delta \Sigma^n (Q^2) \\
   \Delta G^n (Q^2)
   \end{array}
   \right) ,
\end{equation}
\begin{eqnarray}
   \Gamma^n (Q^2)
   &=& \Biggl\{ 
   {\left( \frac{\alpha_s (Q^2)}{\alpha_s (Q_{init}^2)}
   \right)}^{\Lambda_-^n/2 \beta_0}
   \Biggl[ P_{-}^n - \frac{1}{2 \beta_0} 
   \frac{\alpha_s (Q_{init}^2) - \alpha_s (Q^2)}{4 \pi} 
   \,P_{-}^n \gamma^n P_{-}^n \nonumber \\
   &-& \left(
   \frac{\alpha_s (Q_{init}^2)}{4 \pi} - \frac{\alpha_s (Q^2)}{4 \pi}
   {\left(
   \frac{\alpha_s (Q^2)}{\alpha_s (Q_{init}^2)}
   \right)}^{(\lambda_+^n - \lambda_-^n)/2 \beta_0} \right)
   \cdot \frac{P_-^n \gamma^n P_+^n}{2 \beta_0 + 
   \lambda_+^n - \lambda_-^n}
   \Biggr] \nonumber \\
   &+& \hspace{40mm} (+ \leftrightarrow - ) \hspace{40mm} \Biggr\}
   \,\, \Gamma^n (Q_{init}^2) . \ \ \ 
\end{eqnarray}
Here $\alpha_s (Q^2)$ is the QCD running coupling constant at the 
next-to-leading order with $\overline{MS}$ scheme, $\beta_0$ and
$\beta_1$ are the $1$- and $2$-loop QCD beta functions, 
respectively, and
\begin{equation}
   \gamma^n = \gamma^{(1)n} - \frac{\beta_1}{\beta_0} \,
   \gamma^{(0)n} .
\end{equation}
$P_{\pm}^n$ are $2 \times 2$ projection matrices defined by
\begin{equation}
   P_{\pm}^n = \pm \,(\gamma^{(0)n} - \lambda_{\mp}^n \,\hat{1}) 
   / (\lambda_+^n - \lambda_-^n) ,
\end{equation}
with $\hat{1}$ being a $2 \times 2$ unit matrix and with
\begin{equation}
   \lambda_{\pm}^n = \frac{1}{2} \,\left[ \,\gamma_{qq}^{(0)n} + 
   \gamma_{gg}^{(0)n} 
   \pm \sqrt{(\gamma_{qq}^{(0)n} - \gamma_{gg}^{(0)n})^2 
   + 4 \gamma_{qg}^{(0)n} \gamma_{gq}^{(0)n}} \,\,\right] ,
\end{equation}
the eigenvalues of the $1-$loop anomalous dimension matrix
$\gamma^{(0)n}$. Since the necessary anomalous dimension matrices 
are all given in [51], it is easy to calculate the $Q^2$ evolution
of the first moment of the 
flavor singlet longitudinally polarized distribution functions, i.e. 
the isosinglet axial-charge.

Because of its chiral-odd nature, the moments of the transversity 
distributions do not couple with gluons, irrespective of the 
flavor quantum numbers, which especially means that isovector 
and isoscalar tensor charges follow the same  evolution equation.
The anomalous dimension of the transversity distribution at the
leading $1-$loop order was first given by Artru and Mekhfi [52], while
the corresponding $2-$loop contributions have recently been given by
three groups independently [53-55]. Once the relevant anomalous dimensions 
are known, it is easy to obtain an analytical solution of the NLO
evolution equation for the nth moment of transversity distribution.
Here, we use the form given by Hayashigaki et al. [54] as
\begin{equation}
   \frac{\delta q_1^{(n)} (Q^2)}{\delta q_1^{(n)} (Q_{init}^2)} 
   = {\left( 
   \frac{\alpha_s (Q^2)}{\alpha_s (Q_{init}^2)} 
   \right)}^{\gamma_h^{(0)n} / 2 
   \beta_0}
   {\left( \frac{\beta_0 + \beta_1 \frac{\alpha_s (Q^2)}{4 \pi}}
   {\beta_0 + \beta_1 
   \frac{\alpha_s (Q_{init}^2)}{4 \pi}}\right)}^{\frac{1}{2} 
   \bigl( \frac{\gamma_h^{(1)n}}{\beta_1} 
   - \frac{\gamma_h^{(0)n}}{\beta_0} \bigr)}
\end{equation}
where the relevant anomalous dimensions $\gamma_h^{(0)n}$ and 
$\gamma_h^{(1)n}$ are all given in [51].
Fig.15 show the calculated $Q^2$ dependence of the axial and tensor
charges. For obtaining it, we start with the theoretical first moments
given at the initial energy scale $Q_{init}^2 = 0.25 \mbox{GeV}$ :
\begin{eqnarray}
   g_A^{(3)} (Q_{init}^2) &=& 1.41 \\
   g_A^{(0)} (Q_{init}^2) &\equiv& \Delta 
   \Sigma (Q_{init}^2) \ = \ 0.35 \\
   \Delta G (Q_{init}^2) &=& 0 \\
   g_T^{(3)} (Q_{init}^2) &=& 1.22 \\
   g_T^{(0)} (Q_{init}^2) &=& 0.56
\end{eqnarray}
One sees that the $Q^2$ dependence of the flavor singlet axial 
charge is very small (it is almost constant except in the very
low $Q^2$ region). A characteristic prediction of the CQSM for the axial 
charges, i.e. large isovector charge and small isoscalar charge 
appears to be qualitatively consistent with the corresponding 
experimental data at the relevant energy scale. As was pointed out by
many authors [38,52-55], the $Q^2$ dependence of the tensor
charges are sizably large.
Although there is no experimental information for these latter
quantities, this $Q^2$ dependence must be taken seriously when 
comparing the theoretical prediction of low energy models with future 
experimental date. (Note however that that
the ratio $g_T^{(0)} / g_T^{(3)}$ is $Q^2$ independent.)

Because of the coupling between the flavor singlet axial charge 
(the longitudinal quark polarization) and the gluon polarization
in the evolution equation, non-zero gluon polarization
appears at high $Q^2$ even if we have assumed $\Delta G = 0$
at the initial energy scale of $Q_0^2 = 0.25 \,\mbox{GeV}^2$.
We show in Fig.16 the $Q^2$ evolution of $\Delta G$ in comparison 
with that of $\Delta \Sigma = g_A^{(0)}$. One sees that the gluon 
polarization rapidly grows with increasing $Q^2$.
Already at $Q^2 \simeq 2 \,\mbox{GeV}^2$, $\Delta G$ 
is seen to be larger than $\Delta \Sigma$. As explained in [49], 
the growth of the gluon polarization with $Q^2$ can be traced back to 
the positive sign of the anomalous dimension
$\gamma_{qg}^{(0)1}$ 
at the leading order ($\gamma_{qg}^{(0)1} = 2$).
The positivity of this quantity means that a polarized quark is 
preferred to radiate a gluon with helicity parallel to the quark 
polarization. Since the net quark spin component in the proton is 
positive, it follows that $\Delta G > 0$ at least for the gluons 
perturbatively emitted from quarks [49]. It is hoped that the
direct information on $\Delta g(x,Q^2)$ from the di-jet asymmetry
analyses at HERA in conjunction with the precise NLO analyses of
$g_1 (x,Q^2)$ will soon provide us with an accurate determination of
the polarized gluon distribution as well as its first moment [56].

\vspace{4mm}
\section{Summary}

\ \ \ \ \ In summary, we have shown that the CQSM naturally explains
qualitative behavior
of the experimentally measured longitudinally polarized structure
functions of the proton and the neutron. As was shown in our previous
papers, the model also
reproduces an excess of $\bar{d}$ sea over the $\bar{u}$ sea
in the proton very naturally [17-19]. Furthermore, it predicts
qualitative difference between the transversity distribution functions
and longitudinally polarized distribution functions.
For example, in simple valence quark models like the NRQM or the
MIT bag model, the ratios of the isoscalar to isovector charges
are just the same for both of the axial charges and the tensor
charges. On the contrary, in the CQSM or in the lattice gauge theory,
this ratio turns out to be
much smaller for the axial charges than for the tensor charges.
In our viewpoint, what makes this difference is ``dynamical sea quark
effects'' dictated by the spontaneous chiral symmetry breaking of the
QCD vacuum. Another noteworthy prediction of the CQSM is the
opposite (spin) polarization of the $\bar{u}$ and $\bar{d}$ sea quarks,
thereby indicating SU(2) asymmetric sea quark polarization.
These observations then indicates
that nonperturbative QCD dynamics due to the spontaneous
chiral symmetry breaking would {\it survive} and manifest itself in the
isospin (or flavor) dependence of high energy spin observables,
especially in that of the polarized (as well as unpolarized)
{\it antiquark} distribution functions.

\vspace{5mm}
\section*{Acknowledgement}

\ \ \ The authors would like to express their gratitude to T.~Watabe
at Ruhr Universit\"{a}t Bochum for useful discussion on the
importance of nonlocality corrections in time.
Numerical calculation was performed by using the workstations at the
Laboratory of Nuclear Studies, and those at the Research Center for
Nuclear Physics, Osaka University. 

%
%
\vspace{5mm}
\section*{References}
\newcounter{refnum}
\begin{list}%
{[\arabic{refnum}]}{\usecounter{refnum}}
\item EMC Collab., J.~Aschman et al., Phys. Lett. 
{\bf B206}, 364 (1988) ; \\
Nucl. Phys. {\bf B328}, 1 (1989).
\item F.E~Close, {\it An Introduction to Quarks and Partons} \ 
(Academic Press, London, 1979) ;\\
T.~Muta, {\it Foundations of Quantum Chromodynamics} \ 
(World Scientific, Singapore, 1987).
\item M.~Gockeler, H.~Oelrich, P.E.L.~Rakow, G.Schierholz,
R.~Horsley, E.M.~Ilgenfritz, H.~Perlt and A.~Schiller,
J. Phys. {\bf G22}, 703 (1996) ;\\
M.Gockeler, R.~Horsley, L.~Mankiewicz, H.~Perlt, P.~Rakow,
G.~Schierholz and A.~Schiller, Phys. Lett. {\bf B414},
340 (1997) ; \\
C.~Best, M.~Gockeler, R.~Horsley, L.~Mankiewicz, H.~Perlt,
P.~Rakow, A.~Schafer, G.~Schierholz, A.~Schiller, S.~Schramm and
P.~Stephenson, hep-ph / 9706502.
\item D.I.~Diakonov, V.Yu.~Petrov and P.V.~Pobylista, 
Nucl. Phys. {\bf B306}, 809 (1988).
\item M.~Wakamatsu and H.~Yoshiki, Nucl. Phys.
{\bf A524}, 561 (1991).
\item For reviews, see, M.~Wakamatsu, Prog. Theor. Phys. Suppl.
{\bf 109}, 115 (1992) ; \\
Chr.V.~Christov, A.~Blotz, H.-C.~Kim, P.~Pobylitsa, T.~Watabe,
Th.~Meissner, \\
E.~Ruiz Arriola and K.~Goeke, Prog. Part.
Nucl. Phys. {\bf 37}, 91 (1996) ;\\
R.~Alkofer, H.Reinhardt and H.~Weigel, Phys. Rep. {\bf 265}, 139
(1996).
\item A.~Bramon, Riazuddin and M.D.~Scadron, J. Phys.
{\bf G24}, 1 (1998).
\item M.~Wakamatsu, Phys. Lett. {\bf B300}, 152 (1993).
\item J.D.~Sullivan, Phys. Rev. {\bf D5}, 1732 (1972).
\item E.M.~Henley and G.A.~Miller, Phys. Lett. {\bf B251}, 453
(1990).
\item S.~Kumano, Phys. Rev. {\bf D43}, 59 (1991) ;\\
S.~Kumano and J.T.~Londergan, Phys. Rev. {\bf D44}, 717 (1991).
\item H.~Weigel, L.~Gamberg and H.~Reinhardt, Mod. Phys. Lett.
{\bf A11}, 3021 (1996) ;\\
Phys. Lett. {\bf B399}, 287 (1997) ;\\
L.~Gamberg, H.~Reinhardt and H.~Weigel, Phys. Rev. {\bf D58},
054014 (1998).
\item D.I.~Diakonov, V.Yu.~Petrov and P.V.~Pobylista,
M.V.~Polyakov and C.~Weiss,\\
Nucl. Phys. {\bf B480}, 341 (1996). 
\item D.I.~Diakonov, V.Yu.~Petrov and P.V.~Pobylista,
M.V.~Polyakov and C.~Weiss,\\
Phys. Rev. {\bf D56}, 4069 (1997).
\item K.~Tanikawa and S.~Saito, Nagoya Univ. preprint,
DPNU-96-37 (1996).
\item NMC Collab., P.~Amaudruz et al.,
Phys. Rev. Lett. {\bf 66}, 2712 (1991).
\item M.~Wakamatsu and T.~Kubota, Phys. Rev. {\bf D57},
5755 (1998).
\item M.~Wakamatsu, Phys. Rev. {\bf D44}, R2631 (1991) ;
Phys. Lett. {\bf B269}, 394 (1991) ;\\
Phys. Rev. {\bf D46},
3762 (1992).
\item M.~Wakamatsu, in {\it Weak and Electromagnetic Interactions
in Nucei (WEIN-92)},\\
Proceeding of the International Seminar,
Dubna, Russia, 1992, edited by \\
Ts.D.~Vylov (World Scientific, Singapore, 1993).
\item P.V.~Pobylitsa, M.V.~Polyakov, K.~Goeke, T.~Watabe
and C.~Weiss, hep-ph / 9804436.
\item J.C.~Collins and D.E.~Soper, Nucl. Phys. {\bf B194},
445 (1982) ;\\
J.~Kogut and D.~Soper, Phys. REv. {\bf D1}, 2901 (1970).
\item R.L.~Jaffe and X.~Ji, Nucl. Phys. {\bf B375}, 527 (1992).
\item R.L.~Jafe, {\it Spin, Twist, and Hadron Structure in
Deep Inelastic Procceses}, Lectures given at Ettore Majorana
International School of Nucleon Structure (1995) :
hep-ph / 9602236.
\item M.~Wakamatsu, Prog. Theor. Phys. {\bf 95}, 143 (1996).
\item D.I.~Diakonov and V.Yu.~Petrov, Nucl. Phys. {\bf B272},
457 (1986).
\item C.~Weiss and K.~Goeke, hep-ph / 9712447.
\item S.~Kahana and G.~Ripka, Nucl. Phys. {\bf A429}, 462 (1984) ;\\
S.~Kahana, G.~Ripka and V.~Soni, Nucl. Phys. {\bf A415}, 351 (1984).
\item F.~D\"{o}ring, A.~Blotz, C.~Sch\"{u}ren, T.~Meissner,
E.~Ruiz-Arriola and K.~Goeke, \\
Nucl. Phys. {\bf A415}, 351 (1984).
\item M.~Wakamatsu and T.~Watabe, Phys. Lett. {\bf B312}, 184 (1993).
\item Chr.V.~Christov, A.~Blotz, K.~Goeke, P.~Pobylitsa, 
V.Yu.~Petrov, M.~Wakamatsu \\
and T.~Watabe, Phys. Lett. {\bf B325}, 467 (1994).
\item M.~Gl\"{u}ck, E.~Reya, M.~Stratmann and W.~Vogelsang,
Phys. Rev. {\bf D53}, 4775 (1996).
\item M.~Miyama and S.~Kumano, Comput. Phys. Commun. {\bf 94},
185 (1996).
\item M.~Hirai, S.~Kumano and M.~Miyama, Comput. Phys. Commun.
{\bf 108}, 38 (1998).
\item M.~Hirai, S.~Kumano and M.~Miyama, Comput. Phys. Commun.
{\bf 111}, 150 (1998).
\item R.L.~Jaffe, Phys. Lett. {\bf B93}, 313 (1980) ;
Ann. Phys. (NY) {\bf 132}, 32 (1981).
\item L.~Gamberg, H.Reinhardt and H.Weigel, hep-ph/9707352.
\item M.~G\"{u}ck, E.~Reya and A.~Vogt, Z. Phys. {\bf C67},
433 (1995).
\item V.~Barone, T.~Calarco and A.~Drago, Phys. Lett. {\bf B390},
287 (1997).
\item K.~Suzuki and T.~Shigetani, Nucl. Phys. {\bf A626},
886 (1997) ;\\
K.~Suzuki and W.~Weise, Nucl. Phys. {\bf A634}, 141 (1998).
\item E143 Collab., K.~Abe et al., hep-ph / 9802357.
\item E154 Collab., K.~Abe et al., Phys. Rev. Lett. {\bf 79},
26 (1997).
\item H.-C.~Kim, M.~Polyakov and K.~Goeke, Phys. Lett.
{\bf B387}, 577 (1996).
\item Y.~Kuramashi, Nucl. Phys. {\bf A629}, 235c (1998) ;\\
See, also, M.~Fukugita, Y.~Kuramashi, M.~Okawa and A.~Ukawa,\\
Phys. Rev. Lett. {\bf 75}, 2092 (1995) ;\\
S.-J.~Dong, J.-F.~Laga\"{e} and K.-F.~Liu, Phys. Rev. Lett.
{\bf 75}, 2096 (1995) ;\\
S.~Aoki, M.~Doui, T.~Hatsuda and Y.~Kuramashi, Phys. Rev.
{\bf D56}, 433 (1997).
\item C.~Caso et al. (Particle Data Group), Europian Physical
Journal {\bf C3}, 1 (1998).
\item J.~Ellis and M.~Karliner, Phys. Lett. {\bf B341}, 397 (1995).
\item G.~Altarelli and G.G.~Ross, Phys. Lett. {\bf B212},
391 (1988).
\item R.D.~Carlitz, J.C.~Collins and A.H.~Mueller,
Phys. Lett. {\bf B214}, 229 (1988).
\item G.T.~Bodwin and J.~Qiu, Phys. Rev. {\bf D41}, 2755 (1990).
\item For review, see, H.-Y.~Cheng, {\it Status of the Proton
Spin Problem}, Lectures given at \\
Xth Spring School on
Particle and Fields (1996) : hep-ph / 9607254.
\item W.~Furmanski and R.~Petronzio, Z. Phys. {\bf C11},
293 (1982).
\item M.~Gl\"{u}ck, E.~Reya and A.~Vogt, Z. Phys. {\bf C48},
471 (1990).
\item X.~Artru and M.~Mekhfi, Z. Phys. {\bf C45}, 669 (1990).
\item S.~Kumano and M.~Miyama, Phys. Rev. {\bf D56}, 2504 (1997).
\item A.~Hayashigaki, Y.~Kanazawa and Y.~Koike, Phys. Rev.
{\bf D56}, 7350 (1997).
\item W.~Vogelsang, Phys. Rev. {\bf D57}, 1886 (1998).
\item A.~De Roeck, A.~Deshpande, V.W.~Huges, J.~Lichtenstadt
and G.~R\"{a}del, \\
hep-ph / 9801300.
\end{list}
\vspace{8mm}
\newpage
\begin{flushleft}
\large\bf{Figure caption} \\
\end{flushleft}
\ \\
\begin{minipage}{2cm}
Fig. 1.
\end{minipage}
\begin{minipage}[t]{13cm}
The $O (\Omega^0)$ contributions to the isovector longitudinally
polarized distribution functions $\Delta u(x) - \Delta d(x)$
(solid curves) and $\Delta \bar{u} (x) - \Delta \bar{d} (x)$
(dashed curves). Here the three figures (a), (b), and (c)
correspond to the contributions of the discrete valence level,
those of the Dirac sea quarks, and their sums, respectively.
\end{minipage}
\ \\
\vspace{6mm}
\ \\
\begin{minipage}{2cm}
Fig. 2.
\end{minipage}
\begin{minipage}[t]{13cm}
The $O (\Omega^1)$ contributions to the isovector longitudinally
polarized distribution functions $\Delta u(x) - \Delta d(x)$
(solid curves) and $\Delta \bar{u} (x) - \Delta \bar{d} (x)$
(dashed curves). The meaning of the three figures (a), (b),
and (c) is the same as in Fig.1.
\end{minipage}
\ \\
\vspace{6mm}
\ \\
\begin{minipage}{2cm}
Fig. 3.
\end{minipage}
\begin{minipage}[t]{13cm}
The $O (\Omega^1)$ contributions to the isoscalar longitudinally
polarized distribution functions $\Delta u(x) + \Delta d(x)$
(solid curves) and $\Delta \bar{u} (x) + \Delta \bar{d} (x)$
(dashed curves). The meaning of the three figures (a), (b),
and (c) is the same as in Fig.1.
\end{minipage}
\ \\
\vspace{6mm}
\ \\
\begin{minipage}{2cm}
Fig. 4.
\end{minipage}
\begin{minipage}[t]{13cm}
The final predictions of the CQSM for the longitudinally
polarized distribution functions $\Delta u(x) - \Delta d(x)$
and $\Delta \bar{u} (x) - \Delta \bar{d} (x)$ given as the sums
of the $O (\Omega^0)$ and $O (\Omega^1)$ contributions,
in comparison with those for the isoscalar longitudinally
polarized distribution functions $\Delta u(x) + \Delta d(x)$
and $\Delta \bar{u} (x) + \Delta \bar{d} (x)$ comong
from the $O (\Omega^1)$ terms.
\end{minipage}
\ \\
\vspace{6mm}
\ \\
\begin{minipage}{2cm}
Fig. 5.
\end{minipage}
\begin{minipage}[t]{13cm}
The theoretical predictions for the longitudinally polarized
distribution functions, $x (\Delta u(x) + \Delta \bar{u} (x) + 
\Delta d(x) + \Delta \bar{d} (x))$ and $x (\Delta u(x) + \Delta
\bar{u} (x) - \Delta d(x) - \Delta \bar{d} (x))$, are compared
with the corresponding semi-phenomenological parametrization
given by Gl\"{u}ck, Reya, Stratmann and Vogelsang [31].
Of the two theoretical
curves in each figure, the solid curve is the answer of the
present calculation, whereas the dashed curve is obtained by
using the old treatment used in [17], which amounts to dropping
some of the nonlocality effects in time.
\end{minipage}
\ \\
\vspace{6mm}
\ \\
\begin{minipage}{2cm}
Fig. 6.
\end{minipage}
\begin{minipage}[t]{13cm}
The predictions of the CQSM for the polarized antiquark distributions
$x \Delta \bar{u} (x)$ and $x \Delta \bar{d} (x)$ are compared with the
corresponding GRSV parametrization, which assumes SU(2) symmetric
sea quark polarization, $x \Delta \bar{u} (x) = x \Delta \bar{d} (x) \,\,
( \equiv x \Delta \bar{q} (x) )$.
\end{minipage}
\ \\
\vspace{6mm}
\ \\
\begin{minipage}{2cm}
Fig. 7.
\end{minipage}
\begin{minipage}[t]{13cm}
The $O (\Omega^0)$ contributions to the isovector transversity
distribution functions $\delta u(x) - \delta d(x)$
(solid curves) and $\delta \bar{u} (x) - \delta \bar{d} (x)$
(dashed curves). Here the three figures (a), (b), and (c)
correspond to the contributions of the discrete valence level,
those of the Dirac sea quarks, and their sums, respectively.
\end{minipage}
\ \\
\vspace{6mm}
\ \\
\begin{minipage}{2cm}
Fig. 8.
\end{minipage}
\begin{minipage}[t]{13cm}
The $O (\Omega^1)$ contributions to the isovector transversity
distribution functions $\delta u(x) - \delta d(x)$
(solid curves) and $\delta \bar{u} (x) - \delta \bar{d} (x)$
(dashed curves). The meaning of the three figures (a), (b),
and (c) is the same as in Fig.1.
\end{minipage}
\ \\
\vspace{6mm}
\ \\
\begin{minipage}{2cm}
Fig. 9.
\end{minipage}
\begin{minipage}[t]{13cm}
The $O (\Omega^1)$ contributions to the isoscalar transversity
distribution functions $\delta u(x) + \delta d(x)$
(solid curves) and $\delta \bar{u} (x) + \delta \bar{d} (x)$
(dashed curves). The meaning of the three figures (a), (b),
and (c) is the same as in Fig.1.
\end{minipage}
\ \\
\vspace{6mm}
\ \\
\begin{minipage}{2cm}
Fig. 10.
\end{minipage}
\begin{minipage}[t]{13cm}
The final predictions of the CQSM for the transversity
distribution functions $\delta u(x) - \delta d(x)$
and $\delta \bar{u} (x) - \delta \bar{d} (x)$ given as the sums
of the $O (\Omega^0)$ and $O (\Omega^1)$ contributions,
in comparison with those for the isoscalar transversity
distribution functions $\delta u(x) + \delta d(x)$
and $\delta \bar{u} (x) + \delta \bar{d} (x)$ coming
from the $O (\Omega^1)$ terms.
\end{minipage}
\ \\
\vspace{6mm}
\ \\
\begin{minipage}{2cm}
Fig. 11.
\end{minipage}
\begin{minipage}[t]{13cm}
The solid curve represents the theoretical distribution functions
$\Delta u(x) - \Delta d(x)$, whereas the dashed curve is a modified
one obtained from it by multiplying a $x$-dependent cutoff factor
$(1 - x^{10})$.
\end{minipage}
\ \\
\vspace{6mm}
\ \\
\begin{minipage}{2cm}
Fig. 12.
\end{minipage}
\begin{minipage}[t]{13cm}
The theoretical predictions for the twist-2 spin dependent quark
distribution functions before and after $Q^2$-evolution.
Here $\Delta u(x)$ and $\delta u(x)$ (in (a)) respectively
stand for the longitudinal and transversity distributions
of $u$-quark, while $\Delta d(x)$ and $\delta d(x)$ (in (b))
are the corresponding quantities for $d$-quark.
\end{minipage}
\ \\
\vspace{6mm}
\ \\
\begin{minipage}{2cm}
Fig. 13.
\end{minipage}
\begin{minipage}[t]{13cm}
The theoretical predictions for the twist-2 spin dependent
antiquark distribution functions before and after $Q^2$-evolution.
Here $\Delta \bar{u} (x)$ and $\delta \bar{u} (x)$ (in (a))
respectively stand for the longitudinal and transversity
distributions of $\bar{u}$-quark, while $\Delta \bar{d} (x)$
and $\delta \bar{d} (x)$ (in (b)) are the corresponding
quantities for $\bar{d}$-quark.
\end{minipage}
\ \\
\vspace{6mm}
\ \\
\begin{minipage}{2cm}
Fig. 14.
\end{minipage}
\begin{minipage}[t]{13cm}
The theoretical predictions for the proton and neutron spin
structure functions $g_1^p (x,Q^2)$ and $g_1^n (x,Q^2)$
at $Q^2 = 4 \,\mbox{GeV}^2$ in comparison with the
corresponding SLAC data. The solid and dashed curves in (a)
respectively stand for the prediction of the CQSM and that
of the naive MIT bag model for $g_1^p (x,Q^2)$, whereas the
black circles are the E143 data. The corresponding theoretical
predictions for the $g_1^n (x,Q^2)$ are shown in (b) together
with the E154 data.
\end{minipage}
\ \\
\vspace{6mm}
\ \\
\begin{minipage}{2cm}
Fig. 15.
\end{minipage}
\begin{minipage}[t]{13cm}
The scale dependence of the axial and tensor charges.
The evolution equations at the next-to-leading order are solved
under the initial conditions $g_A^{(3)} (Q_{init}^2) = 1.41$,
$g_A^{(0)} (Q_{init}^2) \equiv \Delta \Sigma (Q_{init}^2) = 0.35$,
$g_T^{(3)} (Q_{init}^2) = 1.22$, $g_T^{(0)} (Q_{init}^2) = 0.56$,
and $\Delta G (Q_{init}^2) = 0$ at $Q_{init}^2 = 0.25 \,\mbox{GeV}^2$.
\end{minipage}
\ \\
\vspace{6mm}
\ \\
\begin{minipage}{2cm}
Fig. 16.
\end{minipage}
\begin{minipage}[t]{13cm}
The scale dependence of the flavor singlet axial charge (or the
quark polarization) and the gluon polarization. The initial
conditions for the evolution equation is the same as given in Fig.15.
\end{minipage}
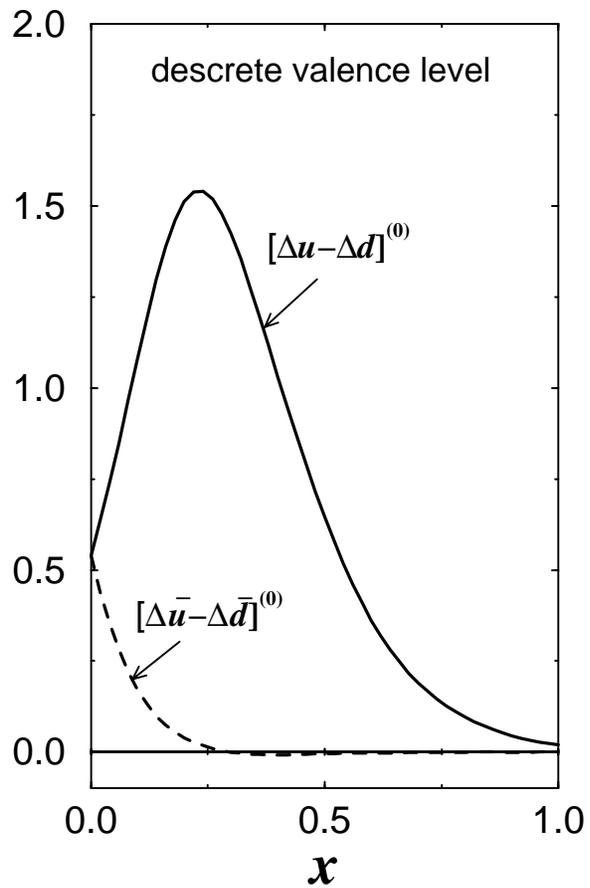
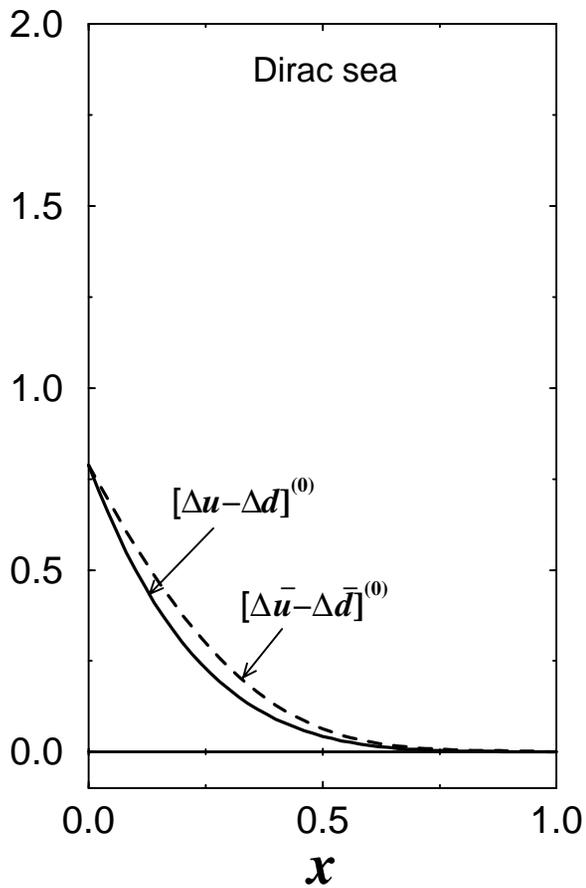
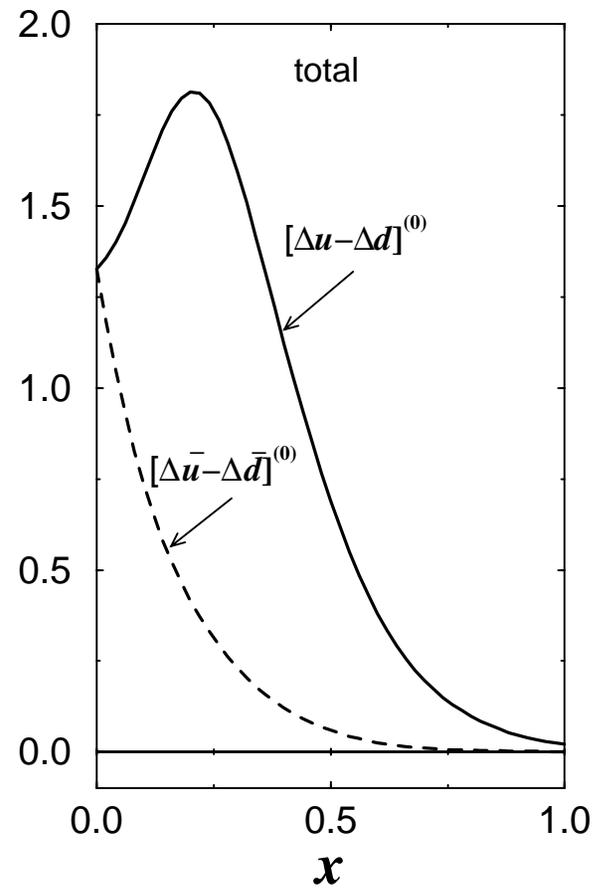
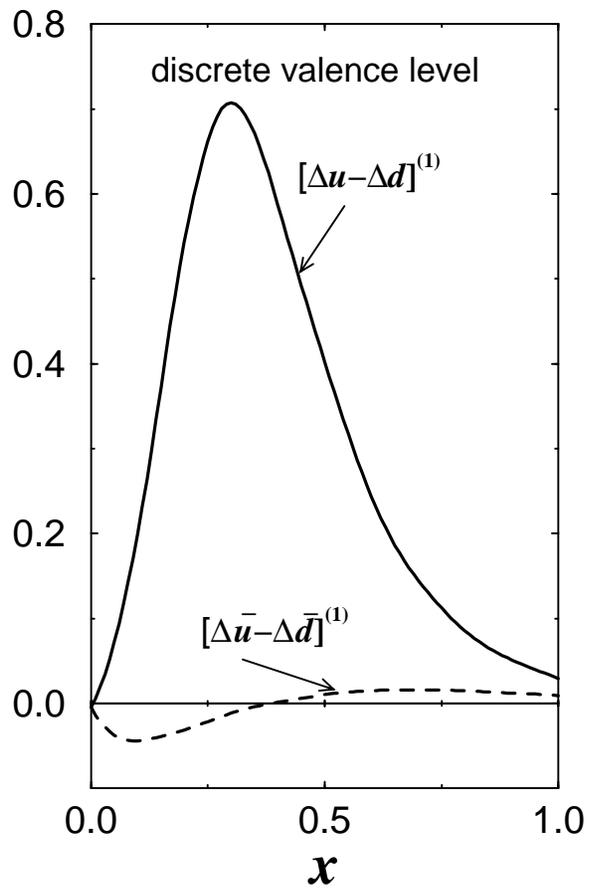
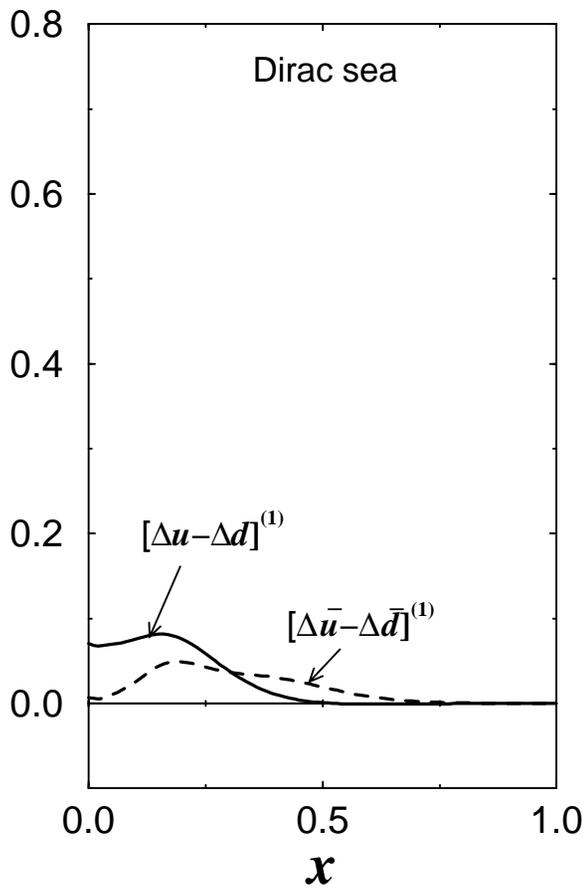
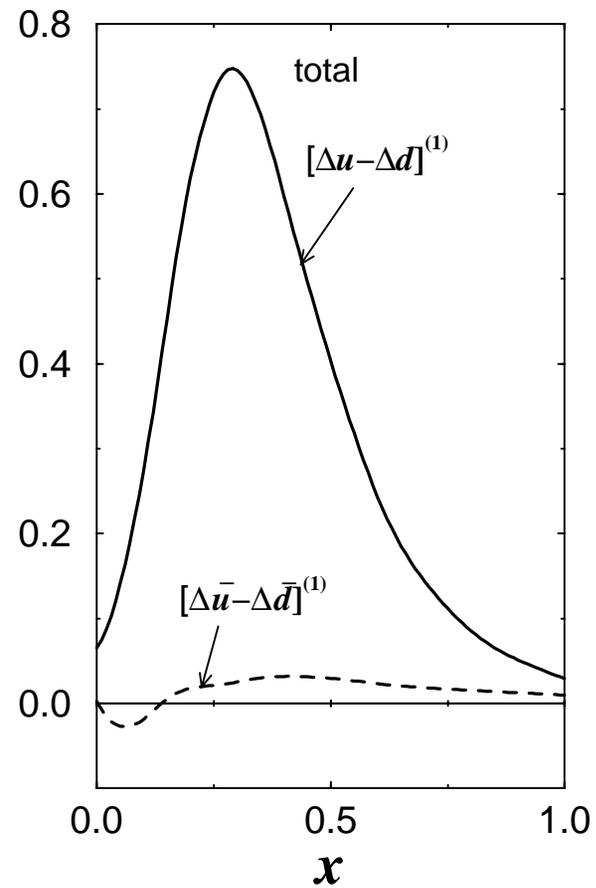
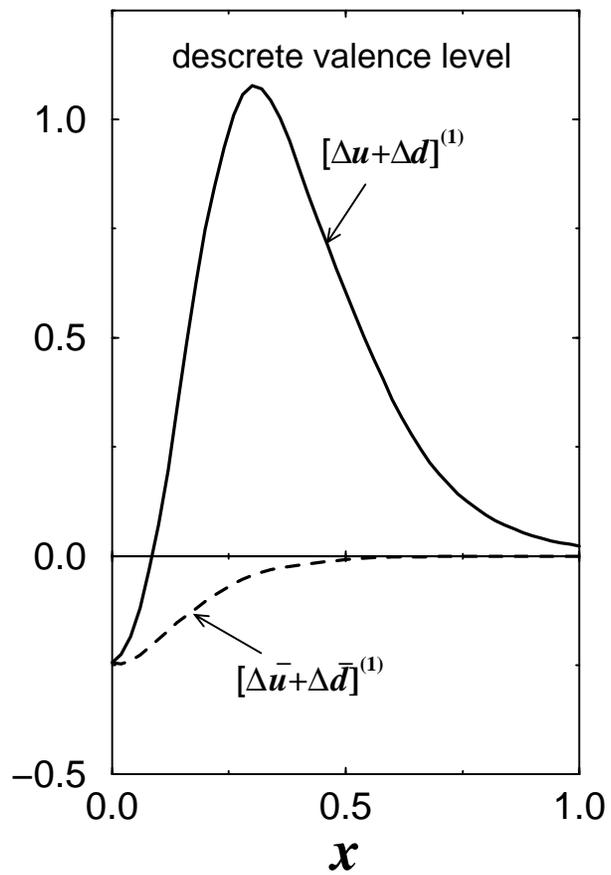
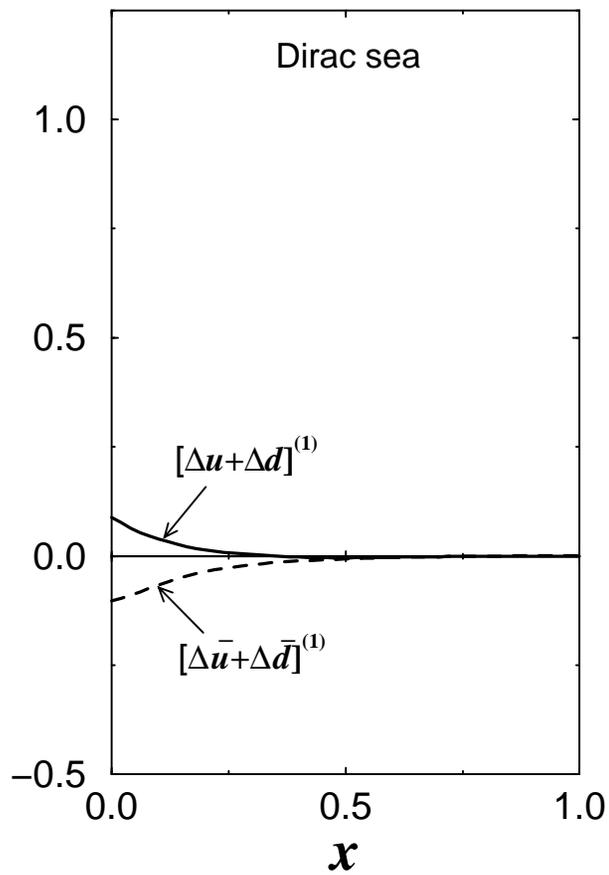
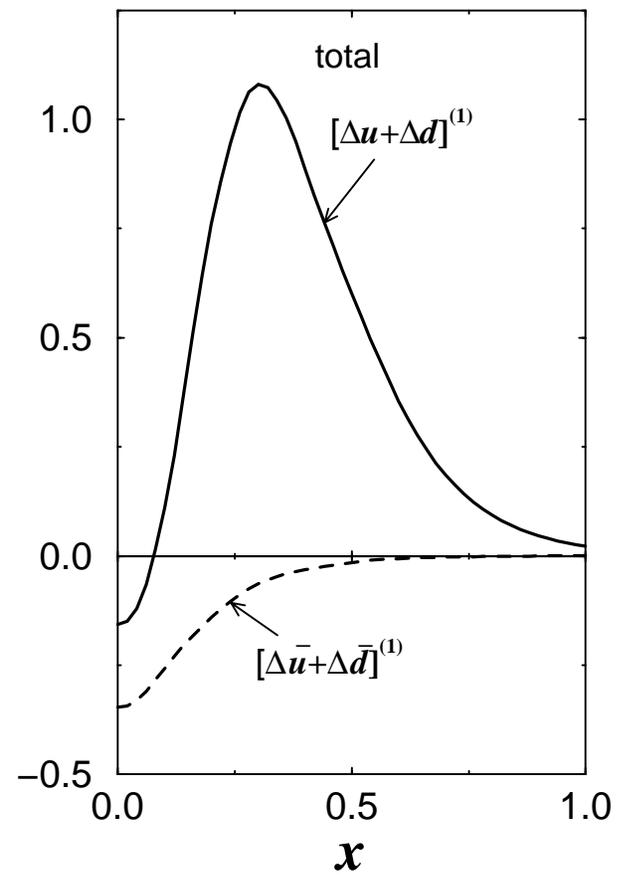
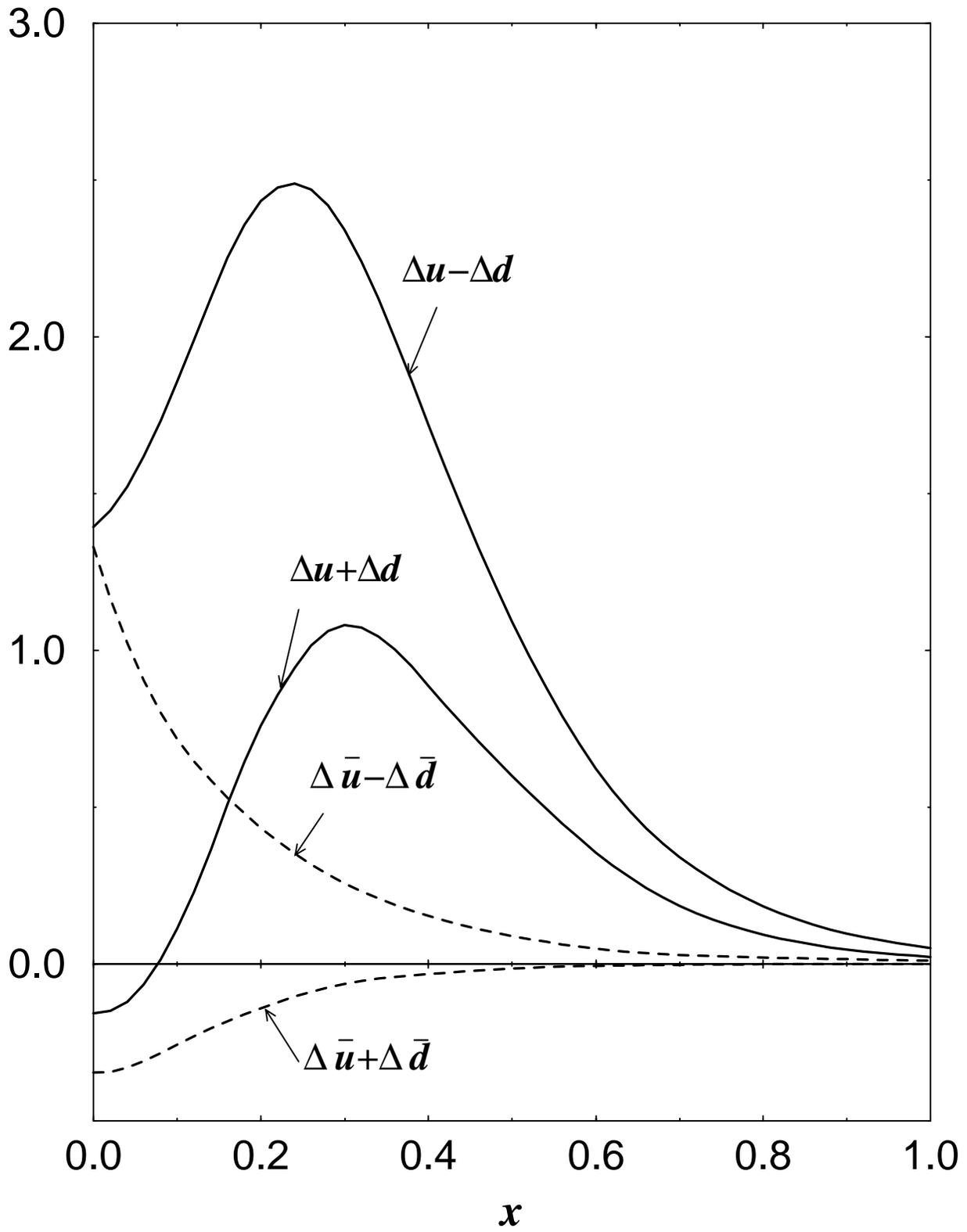
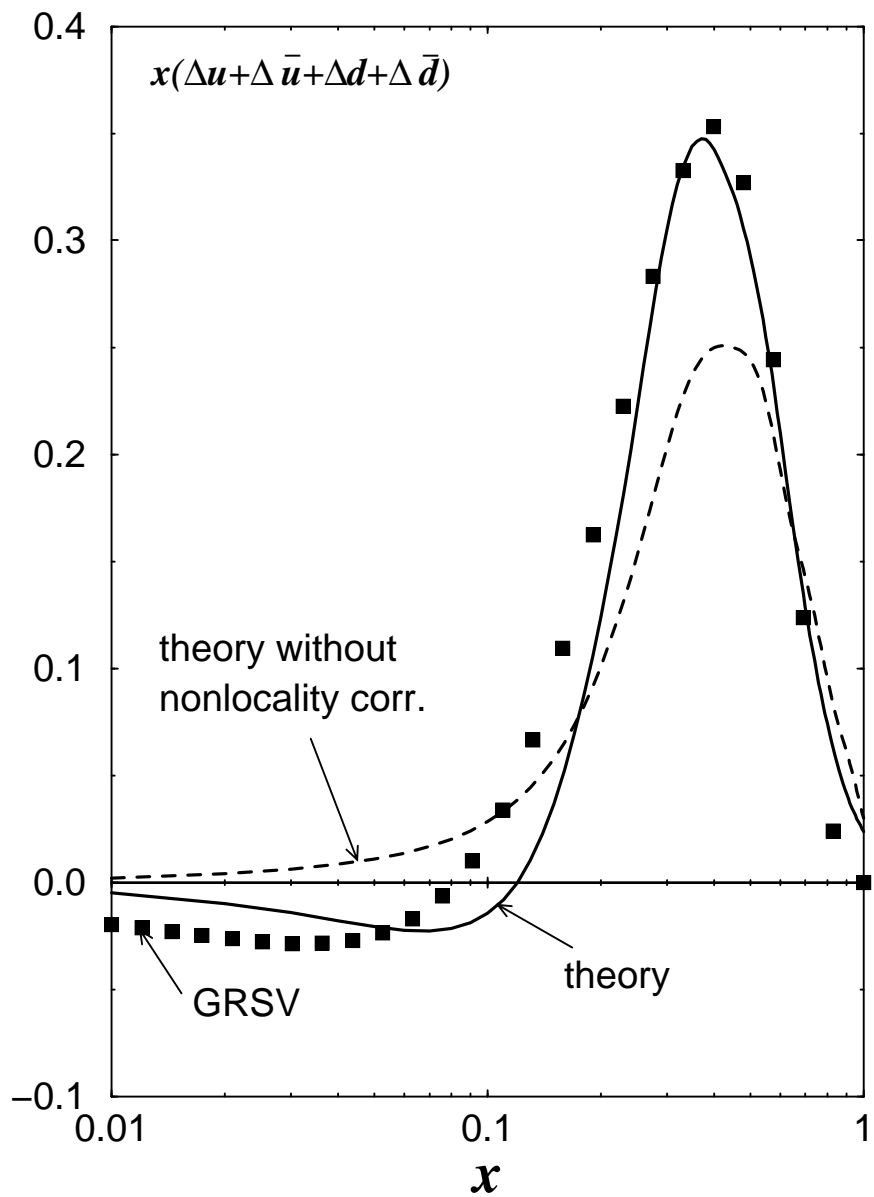
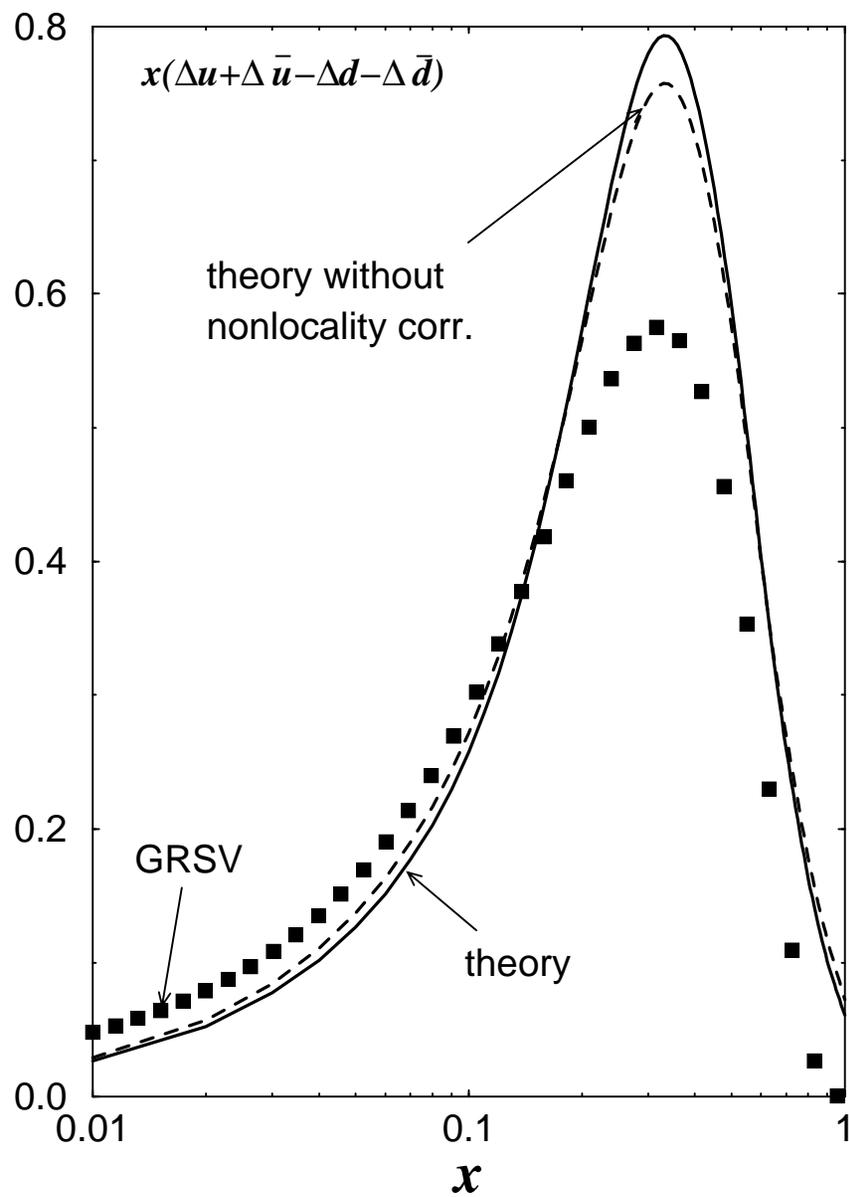
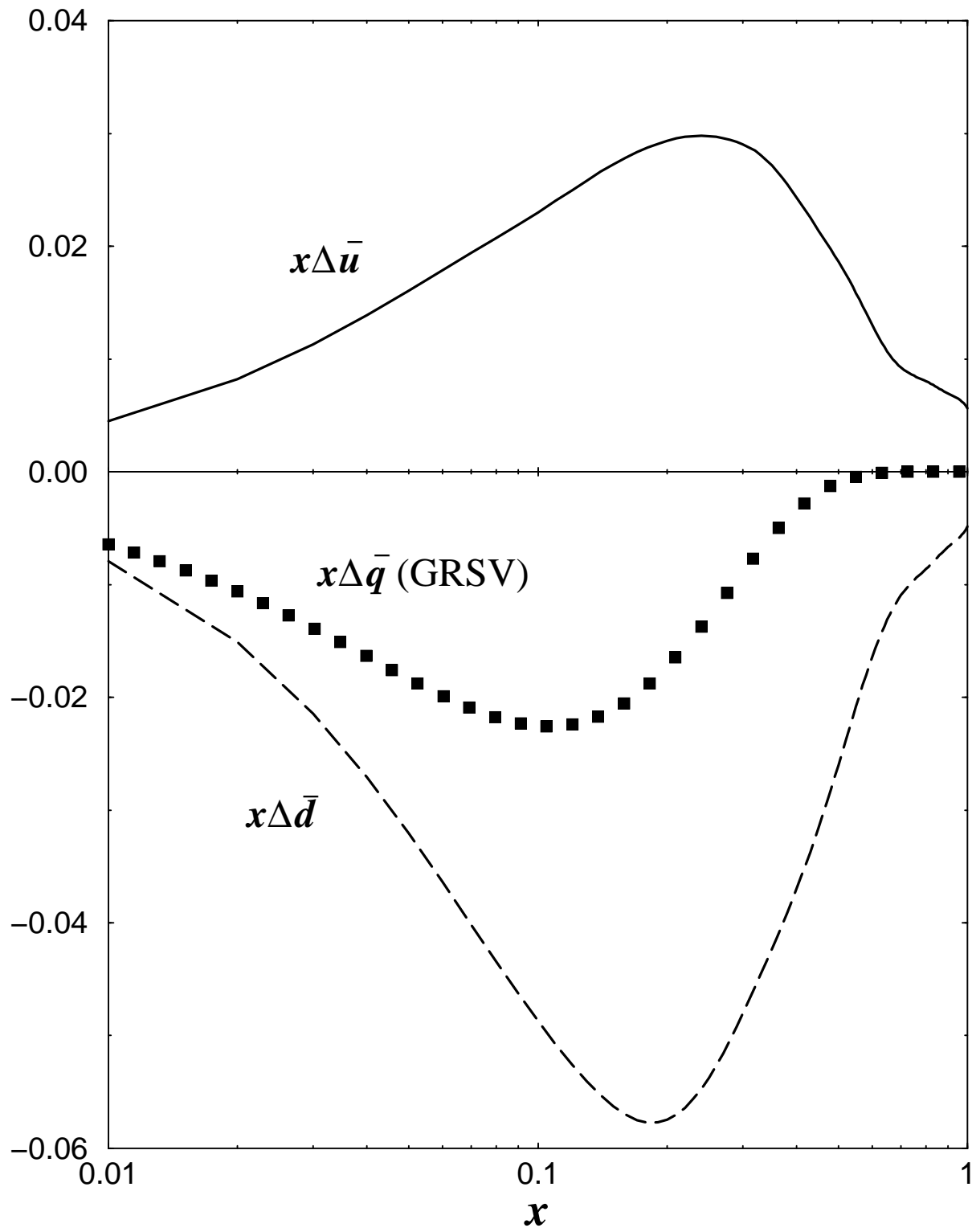
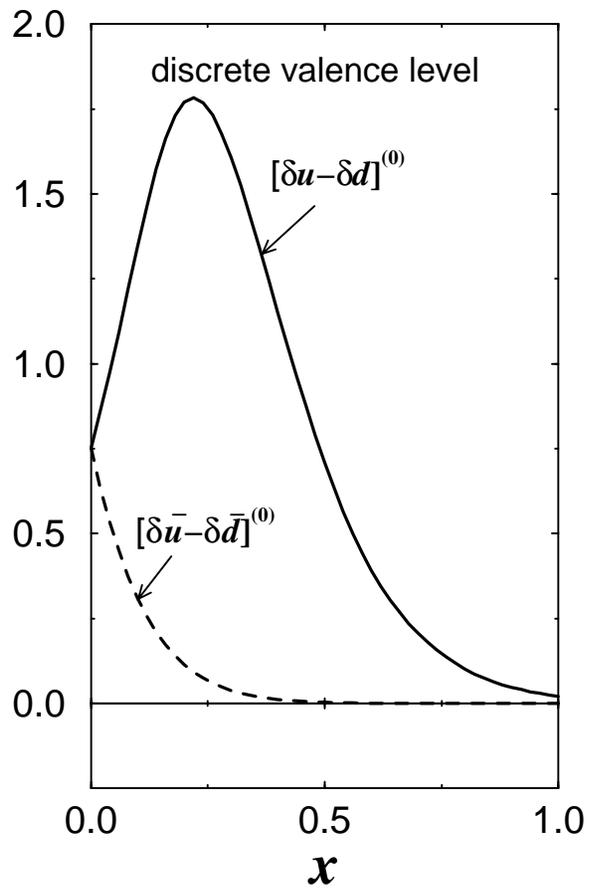
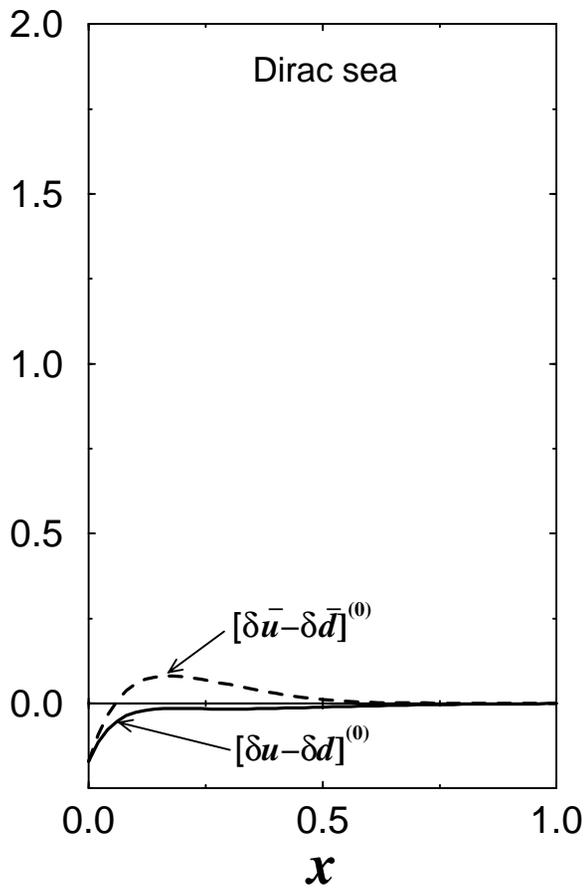
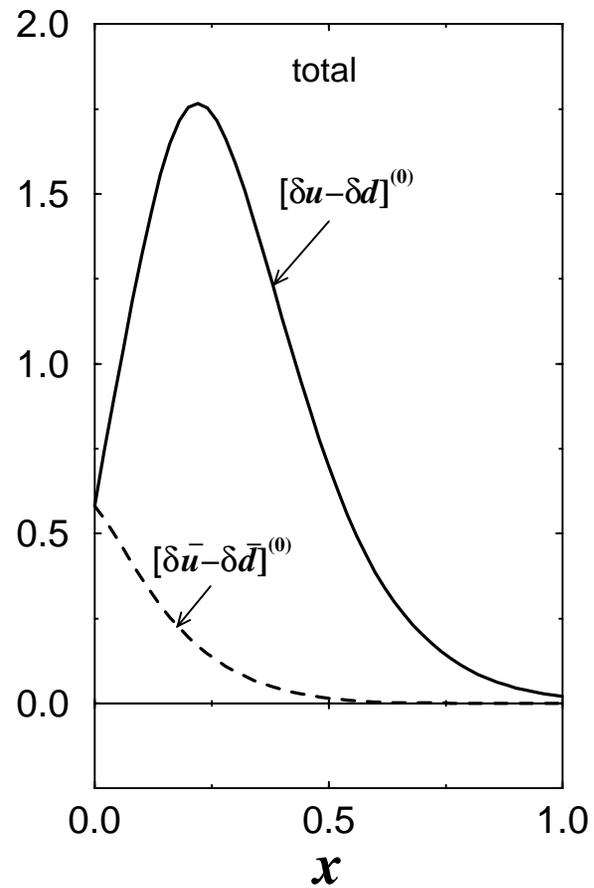
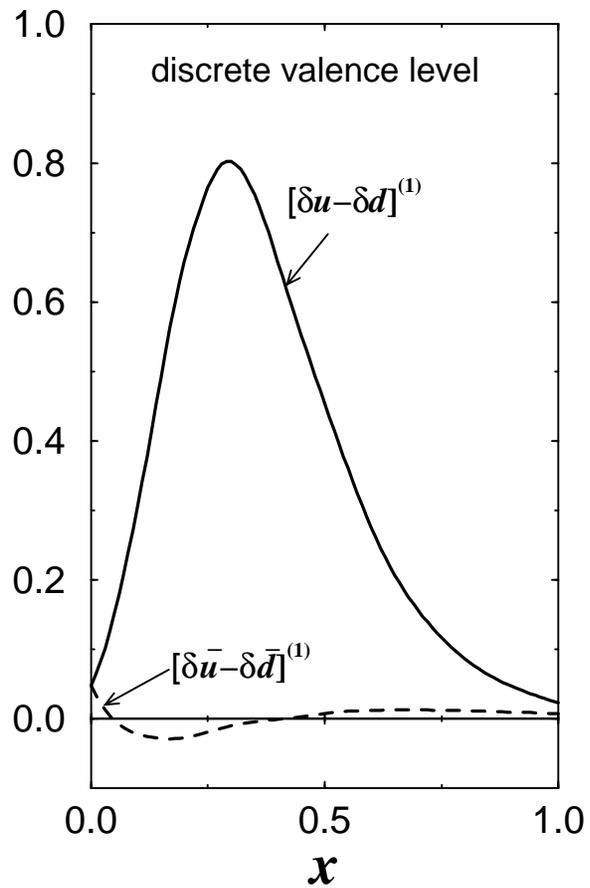
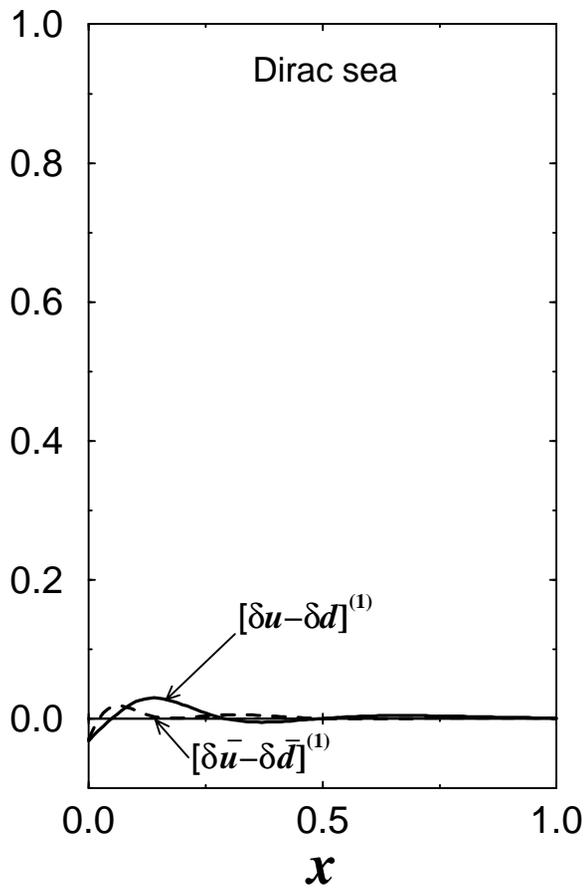
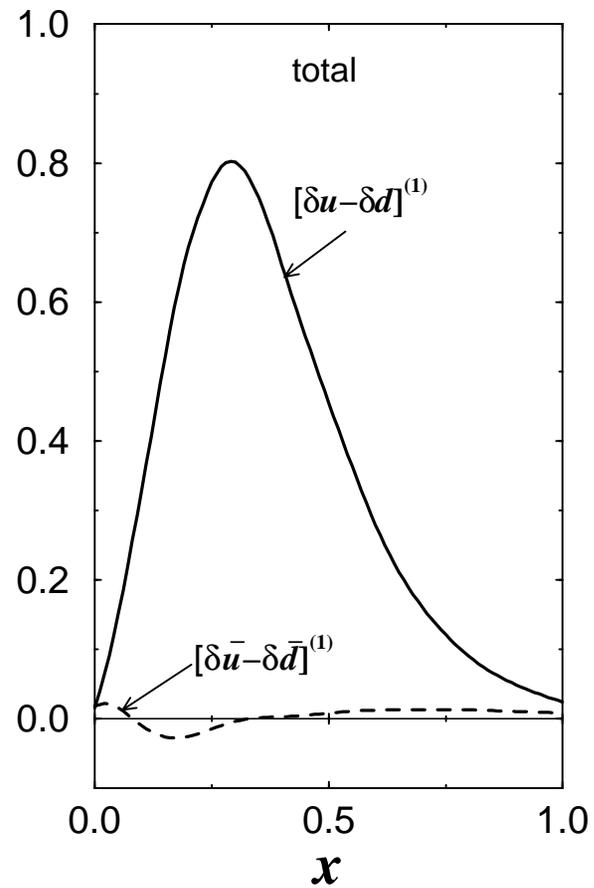
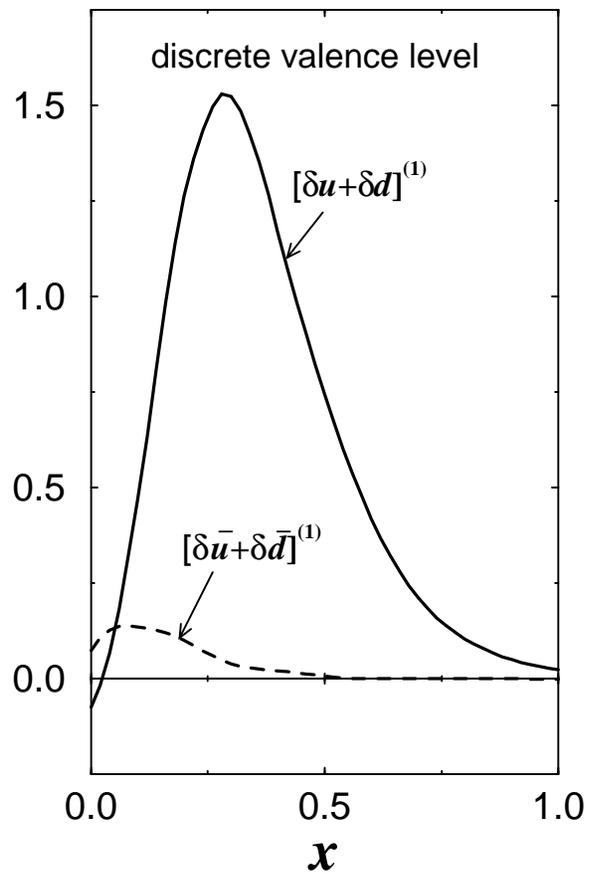
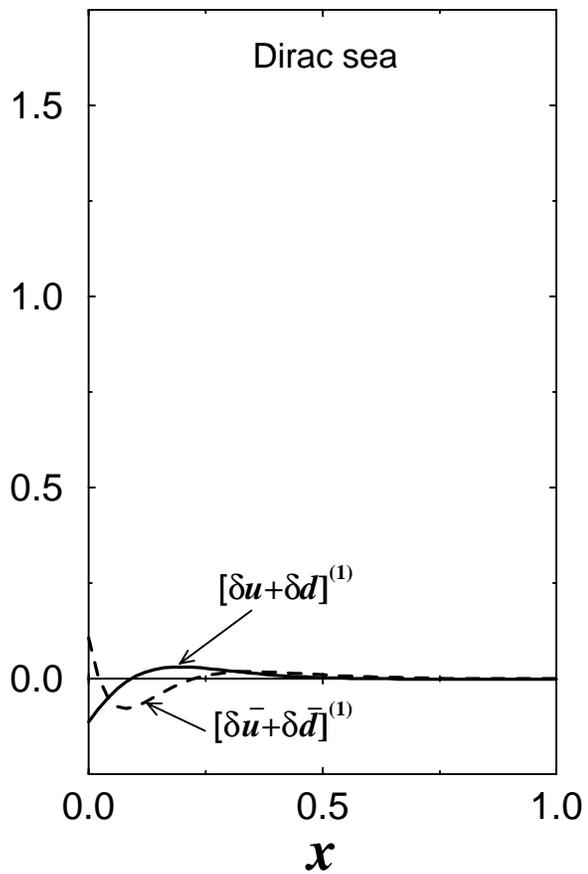
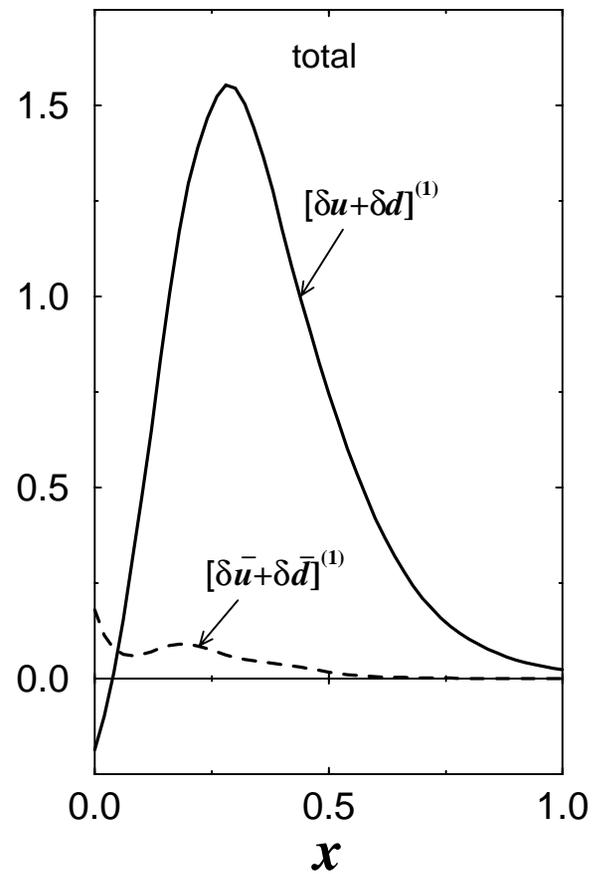
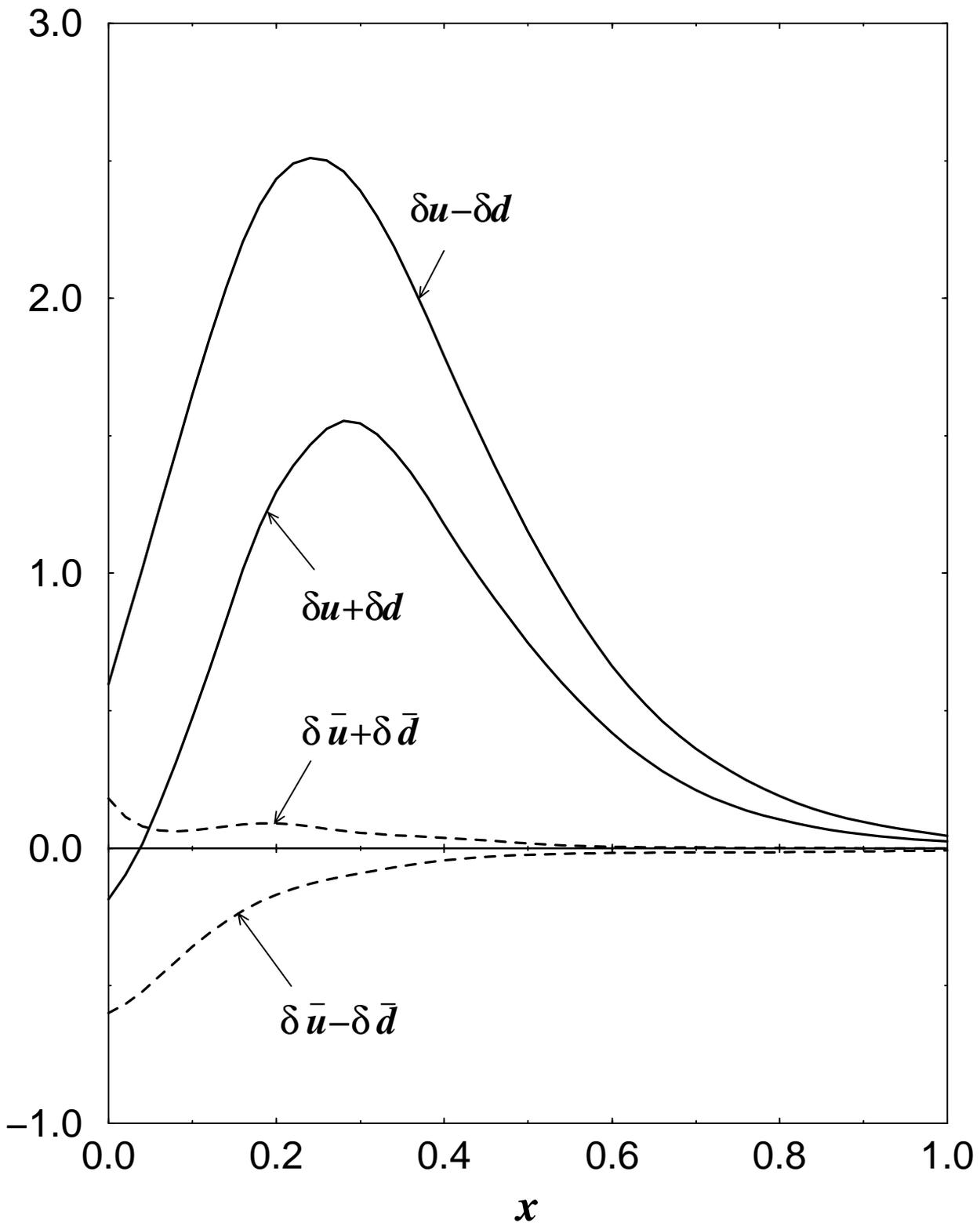
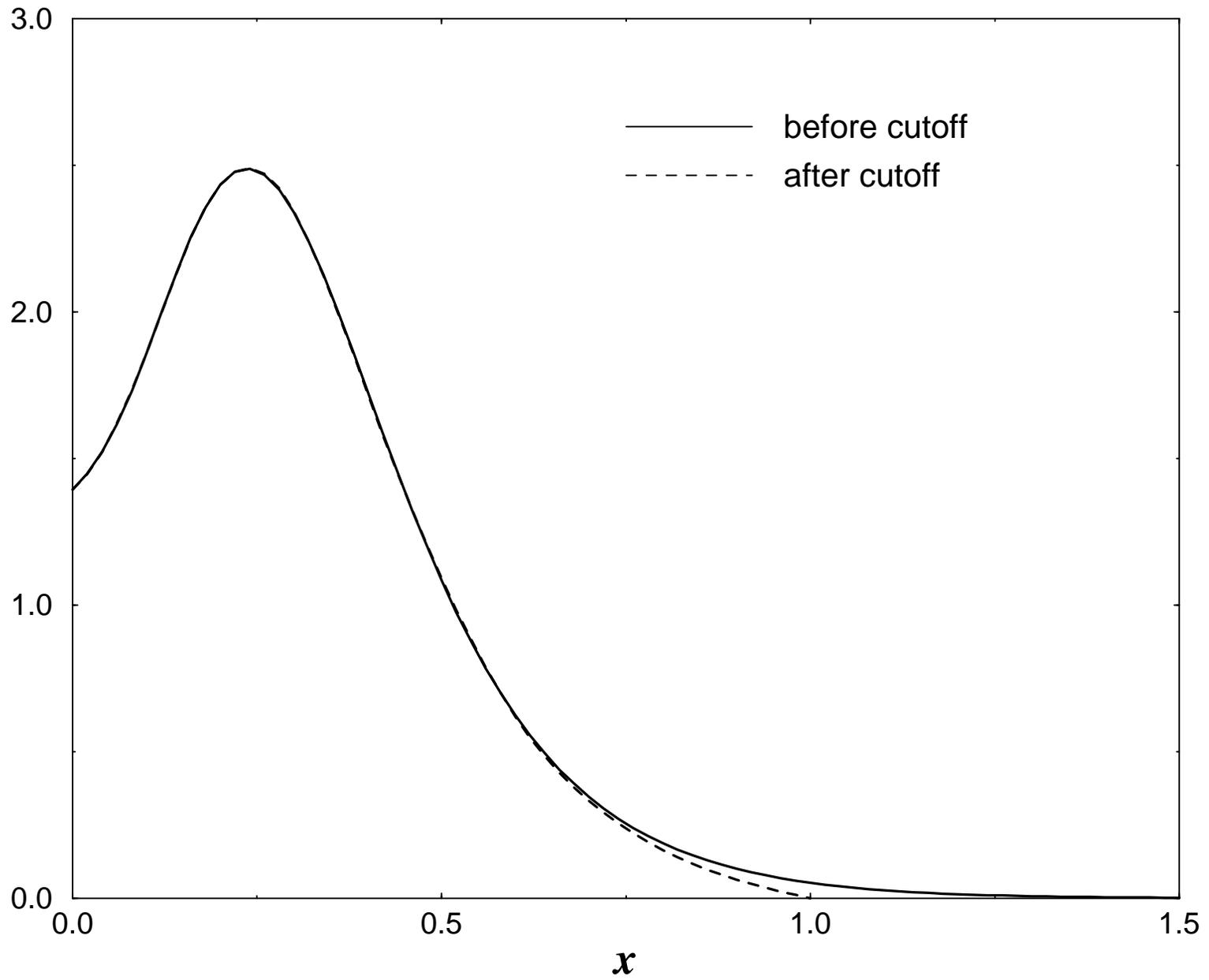
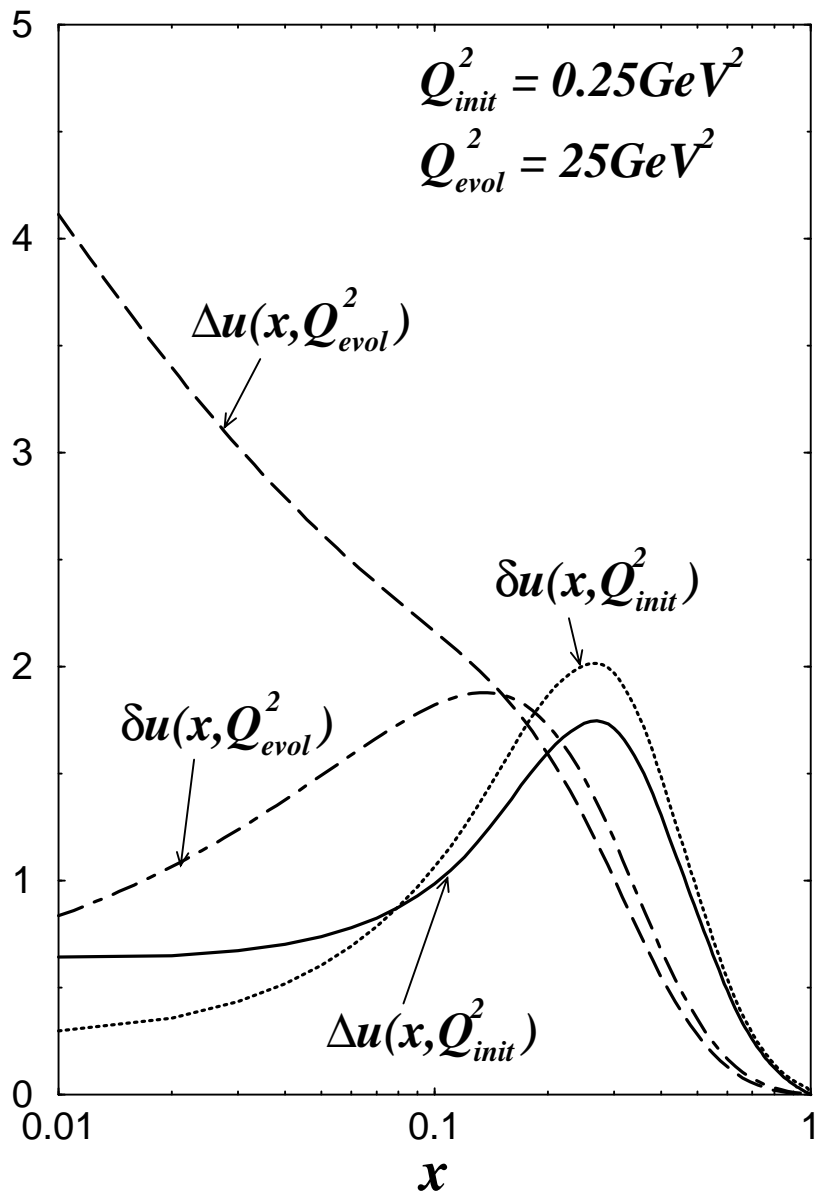
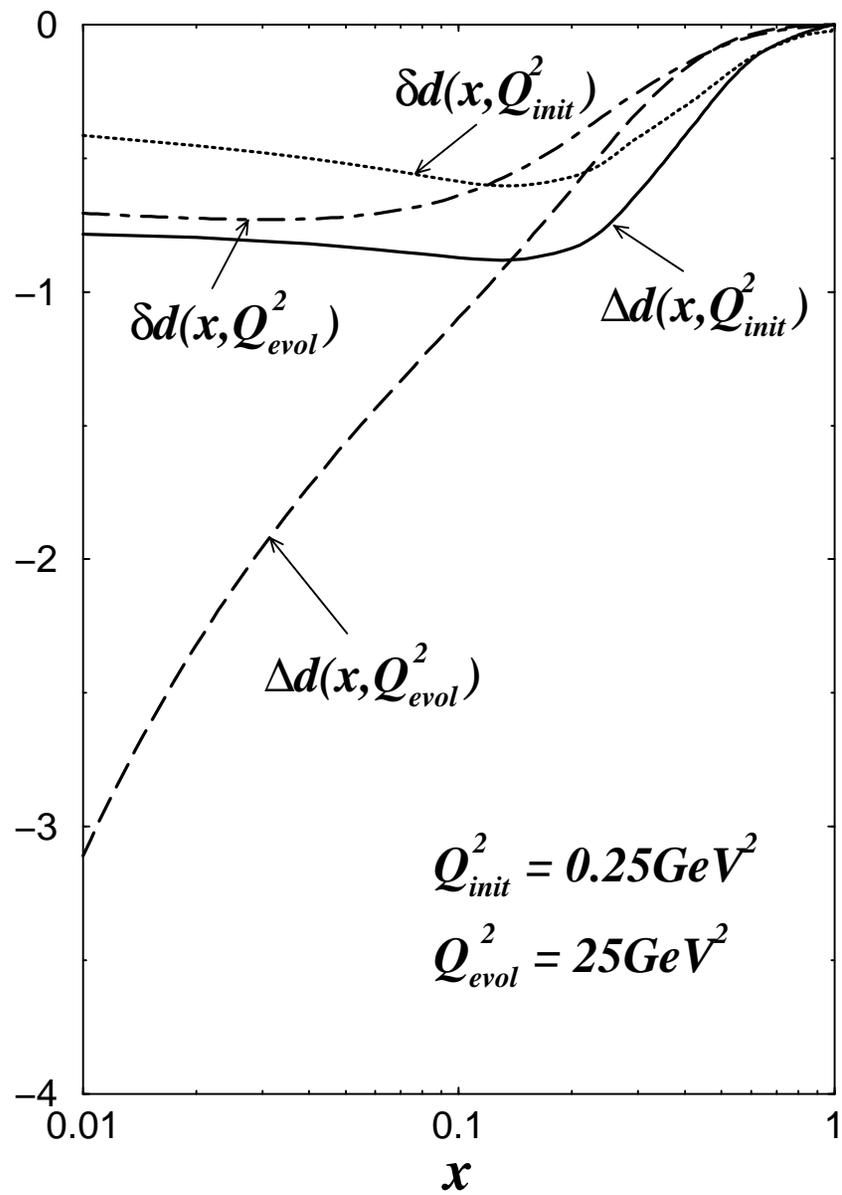
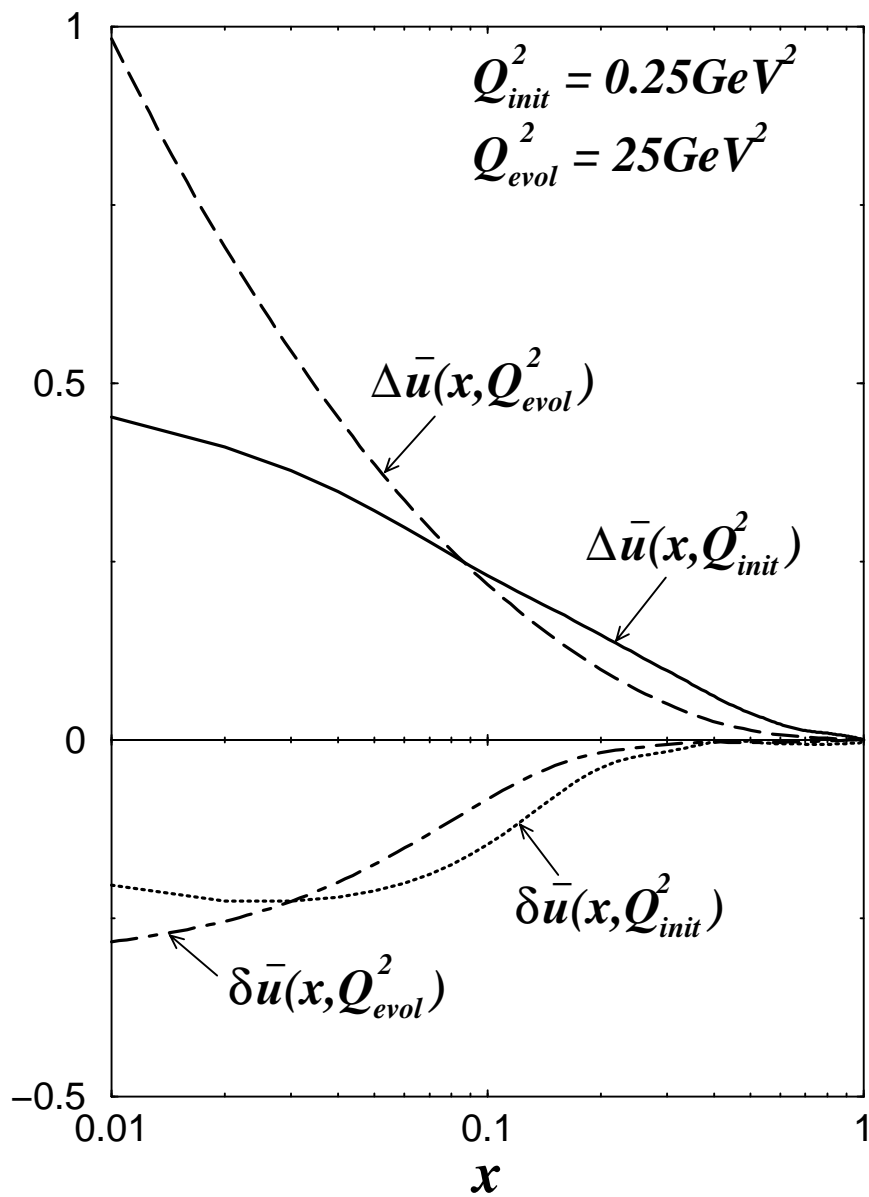
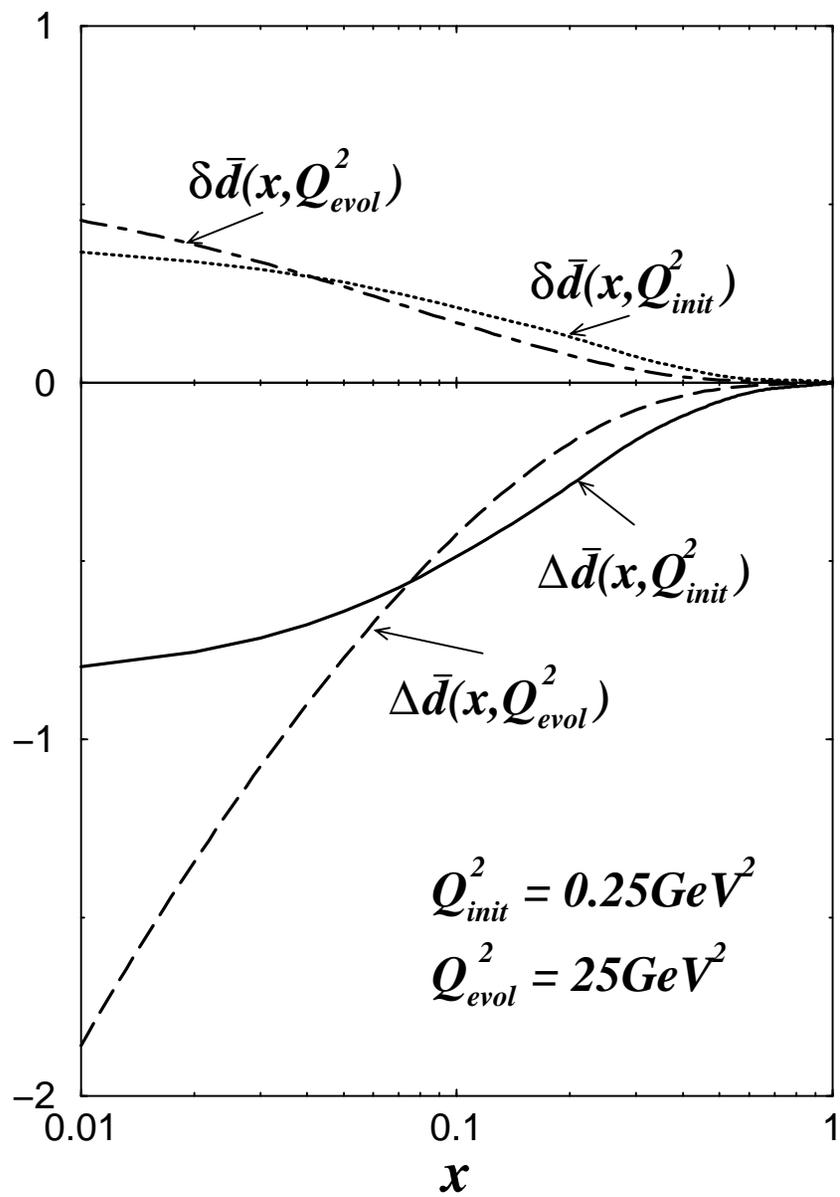
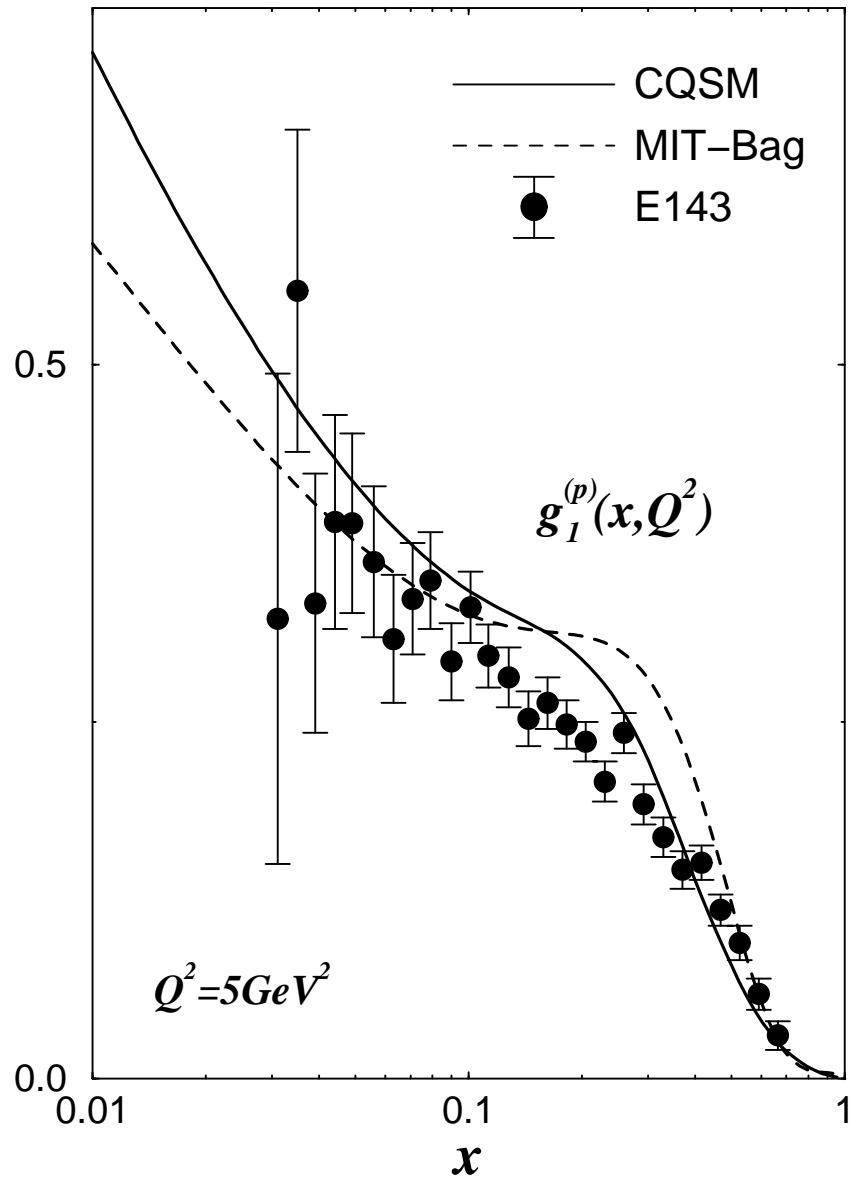
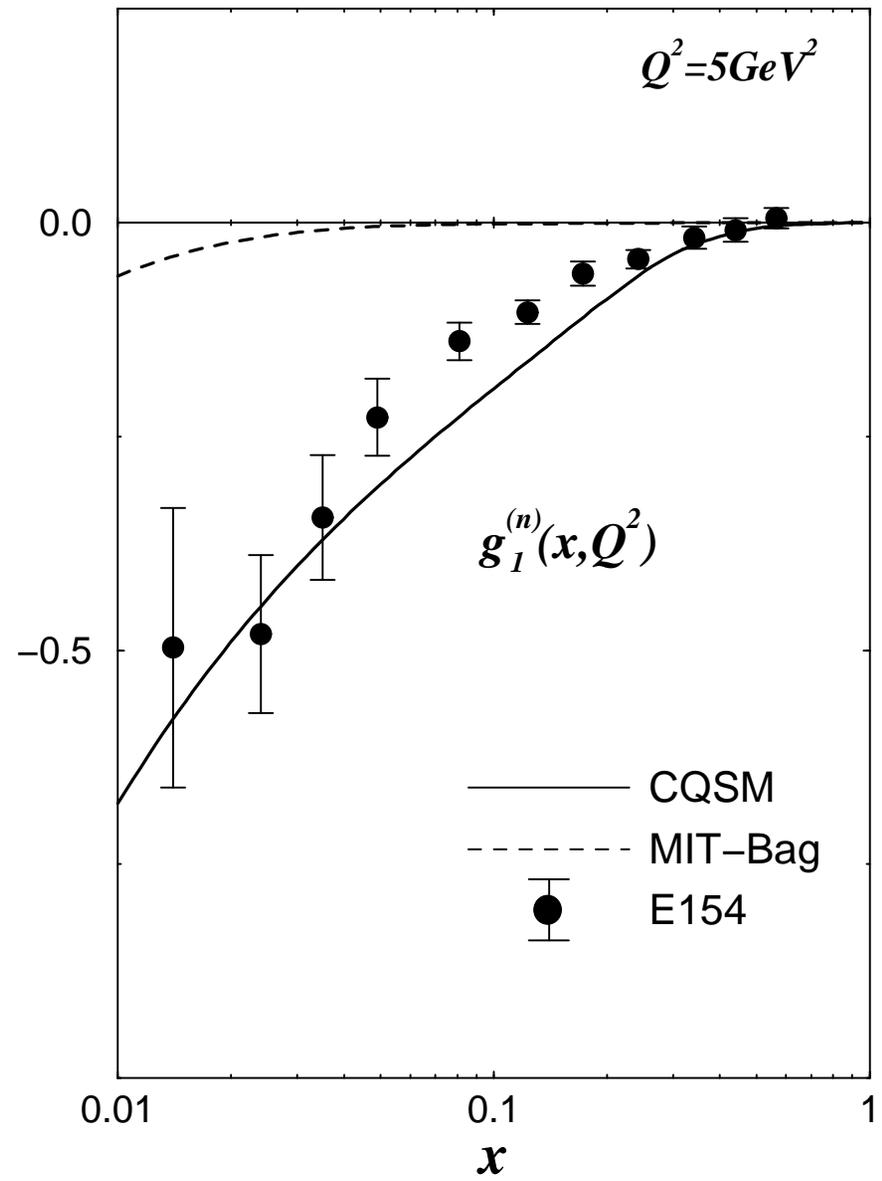
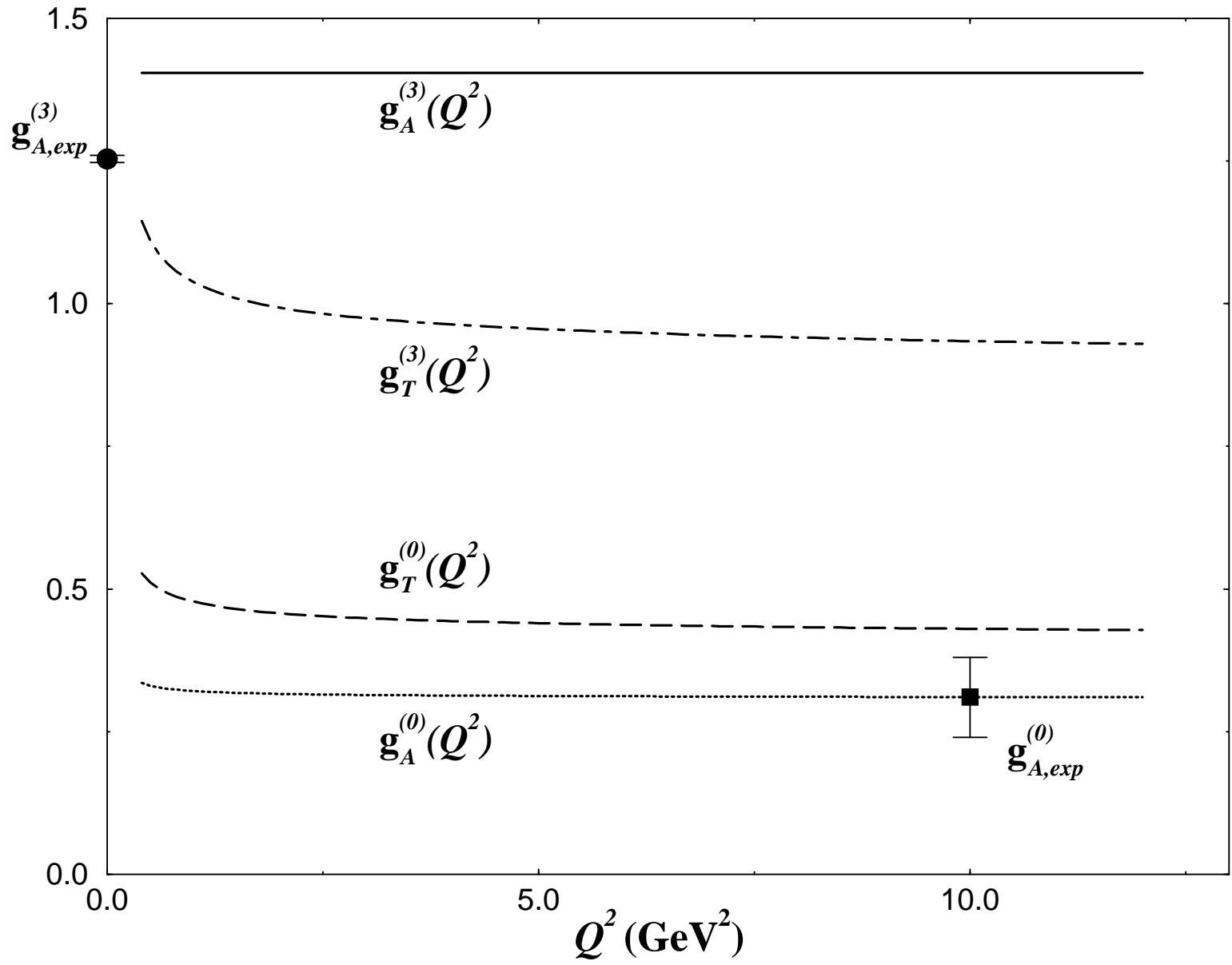
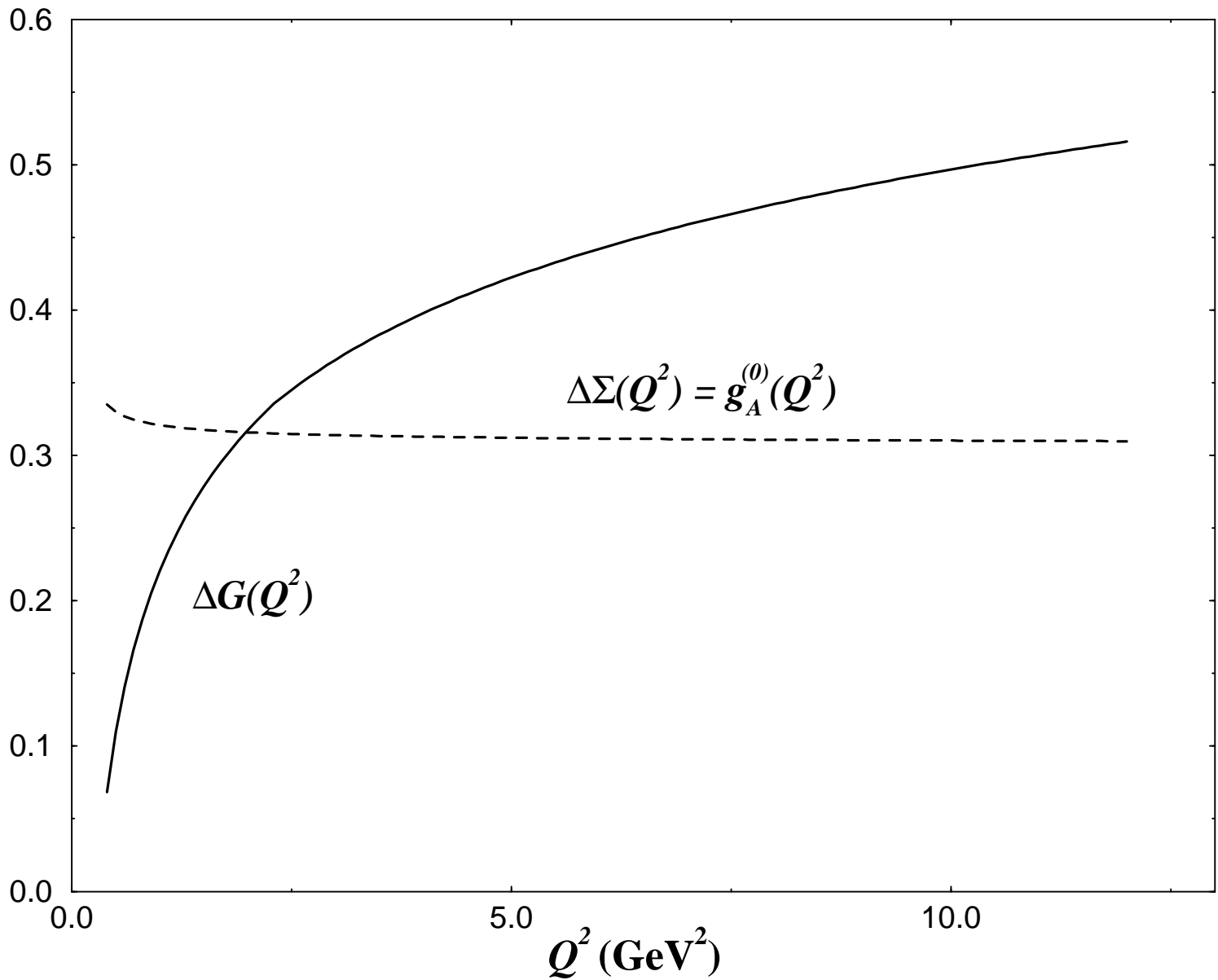
\end{document}